\documentclass{tech}

\usepackage{cite}
\usepackage{amsmath,amssymb,amsfonts}
\usepackage{graphicx}
\usepackage{textcomp}
\usepackage{xcolor}
	
\usepackage{setspace}	
\usepackage{balance} 

\usepackage[font=small]{caption}

\usepackage{tikz}
\usepackage{listings}
\usepackage{caption}
\newcommand{\ignore}[1]{}
\usepackage{graphicx}
\usepackage{multirow}
\DeclareGraphicsExtensions{.jpg,.JPG,.png,.pdf,.eps}

\newcommand{\includesvg}[2]{
	\immediate\write18{inkscape -z -D --file=#2.svg --export-pdf=#2.pdf && pdfcrop #2.pdf #2.pdf }
	\centering
	\includegraphics[width=#1]{#2.pdf}
}

\newcommand{\includepdfcrop}[2]{
	\immediate\write18{pdfcrop #2.pdf #2.pdf}
	\includegraphics[width=#1]{#2.pdf}
}

\usepackage{float}

\floatstyle{ruled}
\newfloat{pseudocode}{h}{lop}
\floatname{pseudocode}{Snippet}

\usepackage{algorithmicx}
\usepackage{algorithm}
\usepackage{algpseudocode}
\usepackage{enumitem}

\algnewcommand\algorithmicinput{\textbf{Input:}}
\algnewcommand\Input{\item[\algorithmicinput]}

\algnewcommand\algorithmicoutput{\textbf{Output:}}
\algnewcommand\Output{\item[\algorithmicoutput]}

\algnewcommand\algorithmicfornoend{\textbf{For }}
\algnewcommand\ForNoEnd{\item[\algorithmicfornoend]}

\algnewcommand\algorithmicwhilenoend{\textbf{While }}
\algnewcommand\WhileNoEnd{\item[\algorithmicwhilenoend]}

\algnewcommand\algorithmicifnoend{\textbf{If }}
\algnewcommand\IfNoEnd{\item[\algorithmicifnoend]}

\usepackage{boxedminipage}
\usepackage{tabularx}
\usepackage{tabu}
\usepackage{pifont}
\usepackage{amsmath}
\RequirePackage[english=usenglishmax]{hyphsubst}

\usepackage{color}

\definecolor{dred}{rgb}{0.6,0.0,0}
\definecolor{dgreen}{rgb}{0.0,0.5,0}
\definecolor{dblue}{rgb}{0.6,0.0,0}
\definecolor{orange}{rgb}{0.9,0.5,0.2}

\usepackage{nicefrac}

\definecolor{lgray}{gray}{0.85}

\newcommand{\dimtree} {{PIM-Tree}}
\newcommand{\imtree} {{IM-Tree}}

\newcommand{\bptree} {{B\textsuperscript{+}-Tree}}

\newcommand{\IBWJ} {{IBWJ}}

\usepackage{balance}

\graphicspath{{Diagram}}

\begin{document}

\title{Parallel Index-based Stream Join on a Multicore CPU}

\numberofauthors{1} 
\author{
	\alignauthor
	Amirhesam Shahvarani, Hans-Arno Jacobsen\\
	\affaddr{Technical University of Munich}\\
	\email{shahvara@in.tum.de}
}
\maketitle

\begin{abstract}
	
	There is increasing interest in using multicore processors to
	accelerate stream processing. For example, indexing sliding window
	content to enhance the performance of streaming queries is greatly
	improved by utilizing the computational capabilities of a multicore
	processor. However, designing an effective concurrency control
	mechanism that addresses the problem of concurrent indexing in highly
	dynamic settings remains a challenge.
	In this paper, we introduce an index data structure, called the
	\textit{Partitioned In-memory Merge-Tree}, to address the
	challenges that arise when indexing highly dynamic data, which are common
	in streaming settings. To complement the index, we design an
	algorithm to realize a parallel index-based stream join that exploits
	the computational power of multicore processors. Our experiments 
	using an octa-core processor show that our parallel stream
	join achieves up to 5.5
	times higher throughput than a single-threaded
	approach.
	
\end{abstract}

\vspace*{-0.2cm}
\section{Introduction}
\vspace*{-0.0cm}

For a growing class of data management applications, such as social
network analysis~\cite{gao2015parallel}, fraud
detection~\cite{zhang2008detecting}, algorithmic
trading~\cite{montana2008data}, and real-time data
analytics~\cite{stonebraker20058}, an information source is available
as a transient, in-memory, real-time, and continuous sequence of
tuples (also known as a \textit{data stream}) rather than as a persistently
disk-stored dataset~\cite{publication-15241}. 
In these applications, processing is mostly performed
using long-running queries known as \textit{continuous
  queries}~\cite{babu2001continuous}. Although its size is steadily 
increasing, the limited capacity of system memory is a general obstacle to
processing potentially infinite data streams. To address this
problem, the scope of continuous queries is typically limited to a
\textit{sliding window} that limits the number of tuples to process
at any one point in time. The window is either defined over a fixed
number of tuples (\textit{count based}) or is a function of time
(\textit{time based}).

Indexing the content of the sliding window is necessary to eliminate
memory-intensive scans during searches and to enhance the performance of
window queries, as in conventional databases~\cite{golab2004indexing}. 
In terms of indexing data structures, hash
tables are generally faster than tree-based data structures for both
update and search operations. However, hash-based indexes are
applicable only for operations that use equality predicates since the
logical order of indexed values is not preserved by a hash table.
Consequently, tree-based indexing is essential for applications
that analyze continuous variables and employ nonequality
predicates~\cite{zhang2015memory}. Thus, in this paper, we focus on 
tree-based indexing approaches, which are also applicable
to operators that use nonequality predicates.

{\color{black} Due to the distinct characteristics of the data flow in streaming
settings, the indexing data structures designed for conventional databases,
such as \bptree, are not efficient for indexing streaming data.} Data in streaming
settings are highly dynamic, and the underlying indexes must be 
continuously updated. In contrast to indexing in conventional databases,
where search is among the most frequent and critical operations,
support for an efficient index update is vital in a streaming
setting. 
Moreover, tuple movement in sliding windows follows a
specific pattern of arrival and departure that could be utilized to
improve indexing performance.  

In addition to the index maintenance overhead arising from data
dynamics, proposing a concurrency control scheme for multithreaded
indexing that handles frequent updates is also a challenging
endeavor. In conventional databases, the index update rate is lower
than the index lookup rate, and concurrency control schemes are designed
accordingly. Therefore, these approaches are suboptimal for
indexing high\-ly dynamic data, such as sliding windows, for which they have
not been designed. Thus, dedicated solutions are desired to
coordinate dynamic workloads with highly concurrent index updates.
These issues will be further exacerbated because the continued
leveraging of the computational power of multicore processors is
becoming inevitable in high-performance stream processing.
The shift in processor design from the single-core to the multicore
paradigm has initiated widespread efforts to leverage parallelism in
all types of applications to enhance performance, and stream processing
is no exception~\cite{gepner2006multi}.

In terms of the underlying hardware, stream processing systems (SPSs)
are divided into two categories, multi-node and single-node.
Single-node SPSs are designed to exploit the computation power of a
single high-performance machine and are optimized for
\textit{scale-up} execution, such as Trill \cite{trill}, StreamBox~\cite{streambox}
and Saber~\cite{saber}.  In contrast, multi-node SPSs are intended to
exploit a multi-node cluster and are optimized for \textit{scale-out}
execution, such as Storm~\cite{storm}, Spark~\cite{spark} and Flink~\cite{flink}.  In
general, a multi-node SPS relies on massive parallelism in the
workload and the producer-consumer pattern to distribute tasks among
nodes.  As a consequence, multi-node SPSs achieve sub-optimal single
node performance and require a large cluster to match the performance
of a scale-up optimized solution using a single machine.  With
advances in modern single-node servers, scale-up optimized solutions
become an interesting alternative for high-throughput and low-latency
stream processing for many applications~\cite{Zeuch:2019:AES:3303753.3316441}.

Thus, in this paper, we address the challenges of parallel tree-based
sliding window indexing, which is designed to exploit a multicore processor
on the basis of uniform memory access.
The distinct characteristics of streaming data motivated us to reconsider 
how to parallelize a stream index and design a novel
mechanism dedicated to a streaming setting. 
{\color{black}
We propose a two-stage data structure based on two known techniques,
data partitioning and delta update, called the \textit{Partitioned In-memory Merge-Tree} 
(\dimtree), that consists of a mutable component 
and an immutable component to address the challenges inherent to
concurrent indexing in highly dynamic settings.} 
The mutable component in \dimtree\ is partitioned into multiple 
disjoint ranges which are dynamically adapt to the range of the streaming tuple values.
This multi-partition design enables \dimtree\ to benefit from the queries' distribution to
reduce potential conflicts among queries and to support parallel index
lookup and update through a simple and low-cost concurrency control method.
{\color{black} Moreover, leveraging a coarse-grained tuple disposal scheme based on this two-stage 
design, \dimtree\ reduces the amortized cost of sliding window updates 
significantly compared to individual tuple updates in conventional indexes such as a \bptree.}
By combining these two techniques \dimtree\ outperforms state-of-the-art indexing 
approaches in both single- and multi-threaded settings.

To validate our indexing approach, we evaluate it in the context of
performing a window band join. 
Stream join is a fundamental operation for performing real-time analytics 
by correlating the tuples of two streams, 
and it is among the most computationally
intensive operations in stream processing systems. 
Nonetheless, our indexing approach is generic and applies equally
well to other streaming operations.

To complement our data structure, we develop a parallel window band
join algorithm based on dynamic load balancing and shared sliding 
window indexes.
These features enable our join algorithm to perform a parallel window 
join using an arbitrary number of available threads.
Thus, the number of threads assigned for a join operation can 
be adjusted at run time based on the workload and the hardware available.
Moreover, our join
algorithm preserves the order of the result tuples such that if tuple
$t_1$ arrives before $t_2$, the join result of tuple $t_1$ will be
propagated into the output stream before that for $t_2$.

\begin{figure}
	\includesvg{0.9\columnwidth}{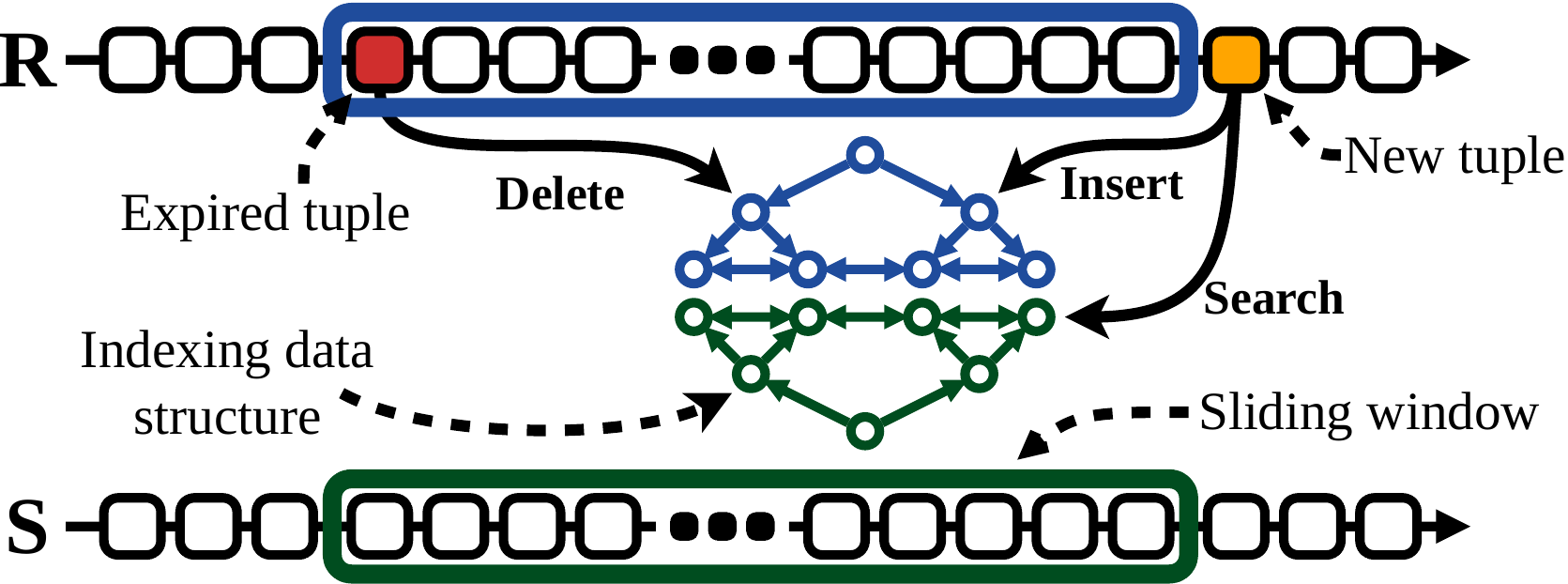}
	\vspace*{-0.1cm}
	\caption{Index-based window join.}
	\label{fig:IBJ}
	\vspace*{-0.6cm}
\end{figure}

The evaluation results indicate that utilizing an octa-core processor, 
our multithreaded join algorithm using \dimtree\ achieves up to 5.6 
times higher throughput than our single-threaded implementation.
Moreover, a single-threaded stream band join using \dimtree\ is 
60\% faster on average than that using \bptree, which demonstrates 
the efficiency of our data structure for stream indexing applications.
Compared with a stream band join using the state-of-the-art 
parallel indexing tree index Bw-Tree~\cite{levandoski2013bw},
using \dimtree\ improves the system performance by a factor of 2.6 on average.

In summary, the contributions of this paper are fourfold:
(1) We propose \dimtree, which is a novel two-stage data structure 
designed to address the challenges of indexing highly dynamic data,
{\color{black} which outperforms state-of-the-art indexing methods
in the application of window join in both single- and multi-threaded settings.}
(2) We develop an analytical model to compare the costs of 
window joins using the indexing approaches studied in this
paper {\color{black}in order to provide a better insight about our design decisions.}
(3) We propose a parallel index-based window join (IBWJ) 
algorithm that addresses the challenges arising from using 
a shared index in a concurrent manner.
(4) We conduct an extensive experimental study of IBWJ 
employing \dimtree\ and provide a detailed quantitative 
comparison with state-of-the-art approaches.

\vspace*{-0.cm}

\vspace*{-0.2cm}
\section{Index-Based Window Join} \label{section:IBJ}

In this section, we define the stream join operator semantics and
study Index-Based Window Join (IBWJ) using three existing indexing approaches, 
including \bptree , chain-index and round-robin partitioning, 
in order to point out the challenges of sliding window indexing
and the shortcoming of existing methods.
We also provide an analytical comparison of processing a tuple using each
approach to provide a more clear insight about each mechanism and to highlight 
their differences to our approach. The notation that we use in
this paper is as follows.

\vspace*{0.2cm}
{
	
	\footnotesize
	\setlength{\parindent}{0cm}
	\hspace*{-0.1cm}
	\begin{tabular} { p{0.1cm} p{0.01cm}  p{0.95\columnwidth} }
		$w$&:&\hspace*{-0.2cm} Size of sliding window.													\\
		$\tau_c$&:&\hspace*{-0.2cm} Time complexity of comparing two tuples.												\\
		$\sigma$&:&\hspace*{-0.2cm} Join selectivity ($0 \leq \sigma \leq 1$).												\\
		$\sigma_s$&:&\hspace*{-0.2cm} Match rate ($w \times \sigma$).			\\	
		$f_{T}$&:&\hspace*{-0.2cm} Inner node fan-out of a tree of type 'T'.		\\
		$\lambda^O_{T}$&:&\hspace*{-0.2cm} Time complexity of performing an operation (O: \underline{I}nsert, \\
		&&\hspace*{-0.1cm} \underline{S}earch, \underline{D}elete) on a node of a tree of type 'T'.		\\
	\end{tabular}
	\begin{tabular} { p{0.15\columnwidth}  p{0.75\columnwidth} }
	\end{tabular}
	\vspace*{-0.4cm}
}\\

\vspace*{-0.cm}

Throughout the remainder of this paper, $\lambda_b^s$, $\lambda_b^i$ and
$\lambda_b^d$ denote the time complexities of search, insert and delete
operations at each node of \bptree, respectively, and $f_b$ denotes the 
inner node fan-out of \bptree.

\vspace*{-0.1cm}

\subsection{Window Join}

The common types of sliding windows are \textit{tuple-based} and
\textit{time-based} sliding windows. The former defines the window boundary based on
the number of tuples, also referred to as the count-based window
semantic, and the latter uses time to delimit the window. We present
our approach based on tuple-based sliding windows, although there is
no technical limitation for applying our approach to time-based sliding
windows.

We denote a two-way window $\theta$-join as $W_R \bowtie_\theta
W_S$, where $W_R$ and $W_S$ are the sliding windows of
streams $R$ and $S$, respectively. The join result contains all pairs
of the form $(r, s)$ such that $r \in W_R$ and $s \in W_S$,
where $\theta (r, s)$ evaluates to $true$.
A join operator processes a
tuple $r$ arriving at stream $R$ as follows.
(1) Lookup $r$ in $W_S$ to determine matching tuples and propagate the
results into the output stream. (2) Delete expired tuples from $W_R$.
(3) Insert tuple $r$ into $W_R$. 
The cost of each step depends on the choice of the join algorithm and
index data structure used. To simplify the time complexity analysis
for different join implementations, we assume that the lengths of the
sliding windows of both streams, $R$ and $S$, are identical, denoted
by $w$. Additionally, we ignore the cost of the sliding window update in our
analysis since it is identical when using different join algorithms and
indexing approaches.
Let $C_S, C_D$ and $C_I$ represent
the time complexities of search, delete, and insert operations,
respectively; then, the time complexity of processing a single tuple
($C_T$) is given by Equation~\ref{equ:joincost}.

\vspace*{-0.3cm}
\begin{equation} \label{equ:joincost}
\footnotesize
\begin{split}
C_T = & \overbrace{ C_{S} }^\text{Step 1} + \overbrace{ C_{D} }^\text{Step 2} + \overbrace{ C_{I} }^\text{Step 3}
\end{split}
\end{equation}
\vspace*{-0.2cm}

\vspace*{-0.1cm}
\subsection{Index-Based Window Join}

IBWJ accelerates window lookup by
utilizing an index data structure.  Although maintaining an extra data
structure along the sliding window increases the update cost, the
performance gain achieved during lookup offsets this extra cost and
results in higher overall throughput.  The general idea of IBWJ is
illustrated in Figure~\ref{fig:IBJ}.  Tuples in $W_R$ and $W_S$ are
indexed into two separate index structures called $I_R$ and $I_S$,
respectively. Upon the arrival of a new tuple $r$ into stream $R$,
IBWJ searches $I_S$ for matching tuples. In addition, $I_R$ must be
updated based on the changes in the sliding window. Here, we examine
IBWJ using \bptree, chained index and context-insensitive
partitioning.
\vspace*{-0.1cm}
\subsubsection{IBWJ using B\textsuperscript{+}-Tree}

We now derive the time complexity of IBWJ based on \bptree. Let $H_b$
be the height of the \bptree{} storing $w$ records ($H_b \approx
\log^w_{f_b}$). The join algorithm processes a given tuple $r$ from
stream $R$ as follows. (1) Search $I_S$ to reach a leaf node
($H_b \cdot \lambda_b^s$); then, linearly scan the leaf node to determine
all matching tuples ($\sigma_s \cdot \tau_c$).  (2) Delete the expired tuple
from $I_R$ ($H_{b} \cdot \lambda_{b}^{d}$).  (3) Insert the new tuple, $r$,
into $I_R$ ($H_{b} \cdot \lambda_{b}^{i}$).  
The time complexity of
processing a tuple using IBWJ based on a \bptree{} ($C_{BJ}$) is given
in Equation~\ref{equ:BTREE_COST}.

\vspace*{-0.3cm}
\begin{equation} \label{equ:BTREE_COST}
\footnotesize
\begin{split}
C_{BJ} = & \overbrace{ H_{b} \cdot \lambda_{b}^{s}  + \sigma_s \cdot \tau_c }^\text{Step 1} + \overbrace{ H_{b} \cdot \lambda_{b}^{d}}^\text{Step 2} + \overbrace{ H_{b} \cdot \lambda_{b}^{i}}^\text{Step 3}
\end{split}
\end{equation}
\vspace*{-0.3cm}

\subsubsection{IBWJ using Chained Index}

\begin{figure}
	\includesvg{0.95\columnwidth}{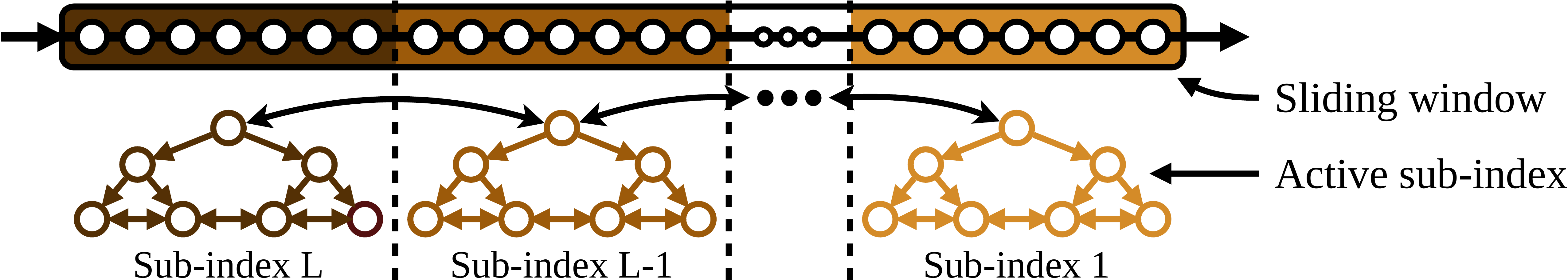}
	\vspace*{-0.1cm}
	\caption{Chained index.}
	\label{fig:chain_index}
	\vspace*{-0.55cm}
\end{figure}

Lin et al.~\cite{Lin:2015:SDS:2723372.2746485} and Ya-xin et
al.~\cite{ya2006indexed} proposed \textit{chained index} to accelerate
stream join processing.  The basic idea of chained index is to
partition the sliding window into discrete intervals and construct a
distinct index per each interval.  Figure~\ref{fig:chain_index}
depicts the basic idea of chained index.  As new tuples arrive into
the sliding window, they are inserted into the active subindex until
the size of the active subindex reaches its limit. When this situation occurs,
the active subindex is archived and pushed into the subindex chain,
and an empty subindex is initiated as a new active subindex.  Using
this method, there is no need to delete expired tuples incrementally;
rather, the entire subindex is released from the chain when it
expires.

We now derive the time complexity of IBWJ when both $I_R$ and $I_S$
are set to a chain index of length $L$ ($L \geq 2$) and all
subindexes are \bptree s.  Let $H_c$ be the height of each subindex
($H_c \approx H_b - \log^L_{f_b}$; we also considered the height of
the active subindex being equal to that of archived subindexes to simplify
the equations).  The join algorithm processes a given tuple $r$ from
stream $R$ as follows.  (1) Search all subindexes of $I_S$ to their
leaf nodes ($L \cdot H_c \cdot \lambda_b^s$) and linearly scan leaf nodes to find
matching tuples and filter out expired tuples during the scan. The
number of expired tuples that need to be removed from the result set
is $\sigma_s/(2 \cdot (L-1))$ on average.  (2) Check whether the latest
subindex of $I_R$ is expired and discard the entire subindex. The
cost of this step is negligible, and we consider it to be zero.  (3)
Insert the new tuple, $r$, into the active subindex of $I_R$
($H_{c}.\lambda_{b}^{i}$).  
The time complexity of processing a tuple
using IBWJ on a chained index ($C_{CJ}$) is given in
Equation~\ref{equ:Chain-BTREE_COST}.

\vspace*{-0.2cm}
\begin{equation} \label{equ:Chain-BTREE_COST}
\footnotesize
\begin{split}
C_{CJ} = & \overbrace{L \cdot H_{c} \cdot \lambda_{b}^{s}  + \sigma_s \cdot \tau_c (1 + \frac{1}{2 \cdot (L-1)}) }^\text{Step 1} + \\
& \underbrace{0}_\text{Step 2} + \underbrace{ H_{c} \cdot \lambda_{b}^{i}}_\text{Step 3}
\end{split}
\end{equation}
\vspace*{-0.2cm}

Comparing the cost of the index operations using chained index and \bptree{}
indicates that using chained index to index
sliding windows is more efficient in terms of index update costs than
using a single \bptree{}, whereas range queries are more costly using
chained index because it needs to search multiple individual
subindexes.

\vspace*{-0.1cm}
\subsubsection{IBWJ using Round-robin Partitioning}

{\color{black}
A group of parallel stream join solutions, such as handshake
join~\cite{roy2014low}, SplitJoin \cite{196251} and 
BiStream~\cite{Lin:2015:SDS:2723372.2746485}, are based on
context-insen\-sitive partitioning.  In all these mentioned approaches,
a sliding window is divided into disjoint partitions using 
round-robin partitioning which is based on the arrival order of tuples 
rather than tuple values, 
and each join-core is associated with a single window partition. }
To accelerate the lookup
operation, each thread may maintain a local index for its associated
partition.  Because indexes are local to each thread, there is no need
for a concurrency control mechanism to access indexes. In fact, the
parallelism in these approaches is achieved by dividing a tuple
execution task into a set of independent subtasks rather than
utilizing a shared index data structure and distributing tuples among
threads.  As a drawback of approaches based on context-insensitive
partitioning, it is required to have all joining threads available to
generate the join result of a single tuple because each thread can
only generate a portion of the join result.

\begin{figure}
	\includesvg{0.95\columnwidth}{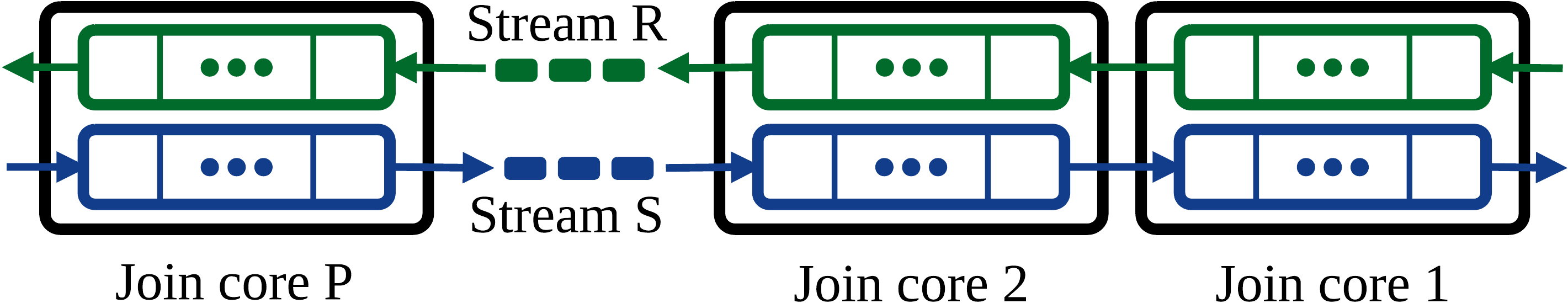}
	\vspace*{-0.1cm}
	\caption{Low-latency handshake join.}
	\label{fig:handshake}
	\vspace*{-0.55cm}
\end{figure}

Here, we explain the cost of IBWJ using the low-latency variant of
handshake join (LHS) employing $P$ threads.
Figure~\ref{fig:handshake} illustrates the join-core arrangement and
the flow of streams in LHS.  In LHS, join-cores are linked as a linear chain
such that each thread only communicates with its two neighbors, and
data streams $R$ and $S$ propagate in two opposite directions.  In
the original handshake join, tuples arrive and leave each join-core in
sequential order, and tuples may have to queue for a long period of
time before moving to the next join-core.  This results in significant
latency in join result generation and in higher computational
complexity because all tuples are required to be inserted and deleted
from each local index.  In LHS, however, tuples are fast forwarded
toward the end of the join-core chain to meet all join-cores faster.
Moreover, each tuple is only indexed by a single join-core, which is
assigned in a round-robin manner.  Consequently, LHS results in higher
throughput and lower latency than the original handshake join.
  
We now derive the time complexity of the index operations required to
process a single tuple using round-robin partitioning with $P$ join-cores.  Let all
join-cores use \bptree{} as local indexes and $H_p$ be the height of
each local index ($H_p \approx H_b - \log^P_{f_b}$). The cost of
processing a given tuple $r$ from stream $R$ is as follows.  (1) Tuple
$r$ is propagated among all join-cores, and all cores search their
local $I_S$ until the leaf nodes ($P \cdot H_p \cdot \lambda_b^s$) and linearly
scan leaf nodes to find matching tuples ($\sigma_s.\tau_c $).  (2) The
join-core that is assigned to index tuple $r$ deletes the expired
tuple from its $I_R$ ($H_{p} \cdot \lambda_{b}^{d}$).  (3) The same
join-core as in Step~2 inserts the new tuple, $r$, into its $I_R$
($H_{p} \cdot \lambda_{b}^{i}$).  
The time complexity of processing a tuple
using round-robin partitioning ($C_{RRJ}$) is given in Equation~\ref{equ:RR_COST}.

\vspace*{-0.2cm}
\begin{equation} \label{equ:RR_COST}
\footnotesize
\begin{split}
C_{RRJ} = & \overbrace{ P \cdot H_{p} \cdot \lambda_{b}^{s}  + (\sigma_s \cdot \tau_c) }^\text{Step 1} + \overbrace{(H_{p} \cdot \lambda_{b}^{d})}^\text{Step 2} + \overbrace{(H_{p} \cdot \lambda_{b}^{i})}^\text{Step 3}
\end{split}
\end{equation}
\vspace*{-0.1cm}

Comparing the cost of the index operations using round-robin partitioning 
with the cost of IBWJ using \bptree{} results in the following:
Using round-robin partitioning is more efficient for inserting 
or deleting a tuple from sliding window than using a single \bptree{} because
the heights of the local indexes for each partition are less than a single
\bptree{} indexing $w$ tuples ($H_p < H_b$). However, because it is
necessary to search multiple local indexes using round-robin partitioning to find matching
tuples, using a single \bptree{} is more efficient in terms of range
querying.  Generally, as the number of join-cores increases, the total
cost of searching local indexes using round-robin partitioning also increases, which is a
consequence of context-insensitive window partitioning.  This
redundant index search limits the efficiency of approaches based on 
round-robin partitioning in the application of IBWJ.

\vspace*{-0.2cm}

\section{Concurrent Window Indexing} 

In this section, we present the design of our
indexing data structures for join processing. 

\vspace*{-0.11cm}
\subsection{Overview} 

We propose a novel two-stage indexing mechanism to accelerate parallel 
stream join by combining two previously known techniques, 
\textit{delta merging} and \textit{data partitioning} resulting in a highly 
efficient indexing solution for both single- and multi-threaded sliding 
window indexing. 
Our indexing solution consists of a
mutable component and an immutable component. The mutable component is an
insert-efficient indexing data structure in which all the new tuples are
initially inserted. The immutable component is a search-efficient
data structure where updates are applied using delta merging. Utilizing the
strength of each indexing component and a coarse-grained tuple
disposal method, our two-stage data structure results in more 
efficient sliding window indexing compared with a single-component indexing
data structure. Moreover, we extend our indexing solution
by splitting the mutable component into multiple mutable partitions,
where partitions are assigned to disjoint ranges. Consequently,
operations on different value ranges can be performed concurrently.
This technique enables our indexing solution to leverage the queries'
distribution to support efficient task parallelism with a lightweight
concurrency control mechanism.

Throughout this section, we first study the effect of delta merging 
in the application of sliding window indexing and then we extend the 
delta merging method with index partitioning to support parallel 
sliding window indexing.

In this work, we use two different \bptree{} designs that have
distinct performance characteristics. The first design is the classic \bptree{} design, 
where each node explicitly stores the references to
its children. This design, which we simply refer to as
\textit{\bptree}, supports efficient incremental updates. In
contrast, as an immutable data structure, \bptree{} nodes can be
arranged into an array in a breadth-first fashion. In this
representation, given a node position, it is possible to retrieve the
location of its children implicitly without needing to store actual
references. By eliminating child references, more space is available
in inner nodes for keys, and it is feasible to achieve a higher
fan-out and decrease the tree depth. Therefore, lookup operations in
this design, which we call \textit{immutable \bptree}, are faster than
in the classic design based on node referencing. As a drawback, it is
inefficient to perform individual updates in an immutable
\bptree{} since the entire tree must be reconstructed; however, this
type of access is not required in our use of the index.

Throughout this paper, $\lambda_{ib}^s$ denotes the time 
complexity of search at each node of the immutable \bptree, and 
$f_{ib}$ denotes the inner node fan-out of immutable \bptree.

\vspace*{-0.15cm}
\subsection{In-memory Merge-Tree}

We now describe our \textit{In-memory Merge-Tree} (\imtree), which is
designed to accelerate sliding window indexing.  \imtree\ consists of
two separate indexing components ($T_I$ and $T_S$).  $T_I$ is a
regular \bptree{} that is capable of performing individual updates, and
$T_S$ is an immutable \bptree{} that is only efficient for bulk updates.
All new tuples are initially indexed by $T_I$.  When the size of $T_I$
reaches a predefined threshold, the entire $T_I$ is merged into $T_S$,
and simultaneously, all expired tuples in $T_S$ are discarded.  The
merging threshold is defined as $m \times w$, where $m$ is a parameter
between zero and one ($0 < m \leq 1$), referred to as the
\textit{merge ratio}.  To query a range of tuples, it is necessary to
search both components, $T_I$ and $T_S$, separately.  Additionally, it
is necessary to filter out expired tuples of $T_S$ from the result
set.  When a tuple expires, it is flagged in the sliding window as
expired but not eliminated.  To drop expired tuples from the index
search results, every result tuple is checked in the sliding window to
determine whether it is flagged as expired.  At the end, all expired
tuples are eliminated from both the sliding window and the index data
structure during the merge operation.

Both chained index and \imtree\ utilize a coarse-grained tuple
disposal technique to alleviate the overhead of tuple removal, but the
tuple disposal techniques differ between these indexing approaches.
Chained index disposes of an entire subtree, whereas
\imtree\ eliminates expired tuples periodically during the merge
operation. The periodic merge enables \imtree\ to maintain all indexed
tuples in only two index components and to provide better search
performance than chained index.

{\color{black}
Although both LSM-Tree~\cite{Neil1996} and \imtree{} are multi component
indexing solutions which use the delta update mechanism 
to transfer data between their components, the two data structures 
are designed differently to tackle distinct problems.
Components in LSM-Tree are configured to be used 
in different storage media and LSM-Tree applies delta update to 
alleviate the cost of write operations in low bandwidth storage media.
In contrast, \imtree{} consists of two in-memory components specialized 
for different operations and \imtree{} applies periodic merges 
in order to enhance the performance of range queries.
Moreover, LSM-Tree is based on incremental merging between its components
which is not applicable on immutable data structures such as 
immutable \bptree{} used in our \imtree .
}

\vspace*{-0.1cm}
\subsubsection{IBWJ using IM-Tree}

Let $H_I$ and $H_S$ be the heights of $T_I$ and $T_S$, respectively.
The time complexity of processing a tuple $s$ arriving at stream
$S$ for IBWJ using \imtree\ is as follows. (1) Search both $T_I$ and
$T_S$ of the opposite stream to the leaf nodes ($H_I \cdot \lambda_b^s +
H_S \cdot \lambda_{ib}^s$) and perform a linear scan of the leaf node to
determine matching tuples ($\sigma_s \cdot \tau_c$) and filter out expired tuples
($\sigma_s \cdot \tau_c \cdot \frac{m}{2}$). (2)
Tuples in IM-Tree are deleted in a batch during a $T_I$ and $T_S$ merge.
Let $M$ be the time complexity of the merge; then, the average cost per
tuple is $M/{(m \cdot w)}$. (3) Insert the new tuple into the index of
stream $S$ ($H_I \cdot \lambda_b^i$). 
The time complexity of
processing a single tuple using IBWJ based on an
\imtree\ ($C_{MJ}$) is given by Equation~\ref{equ:IM_COST}.

The stepwise comparison between the window join using \bptree{} and IM-Tree
is controlled by the merge ratio $m$. Assigning a proper value for
$m$ is subject to various trade-offs. A late merge creates a larger
$T_I$ on average and results in a more expensive insert and
search of $T_I$. Additionally, it increases the average number of expired 
tuples in $T_S$ and results in an inefficient lookup in $T_S$. 
Meanwhile, merge operations are costly, and overdoing such operations results in
a significant performance loss. Generally, increasing the
value of $m$ causes the costs of Steps~1 and~3 to increase and the cost of
Step~2 to decrease.

\vspace*{-0.3cm}
\begin{equation} \label{equ:IM_COST}
\footnotesize
\begin{split}
C_{MJ} = & \overbrace{H_S \cdot \lambda_{ib}^s + H_I \cdot \lambda_b^s + \sigma_s \cdot \tau_c \cdot (1+\frac{m}{2}) }^\text{Step 1} + \\
& \underbrace{M/{(m \cdot w)}}_\text{Step 2} + \underbrace{ H_I \cdot \lambda_b^i}_\text{Step 3}
\end{split}
\end{equation}
\vspace*{-0.2cm}

\begin{figure}[]
	\includesvg{\columnwidth}{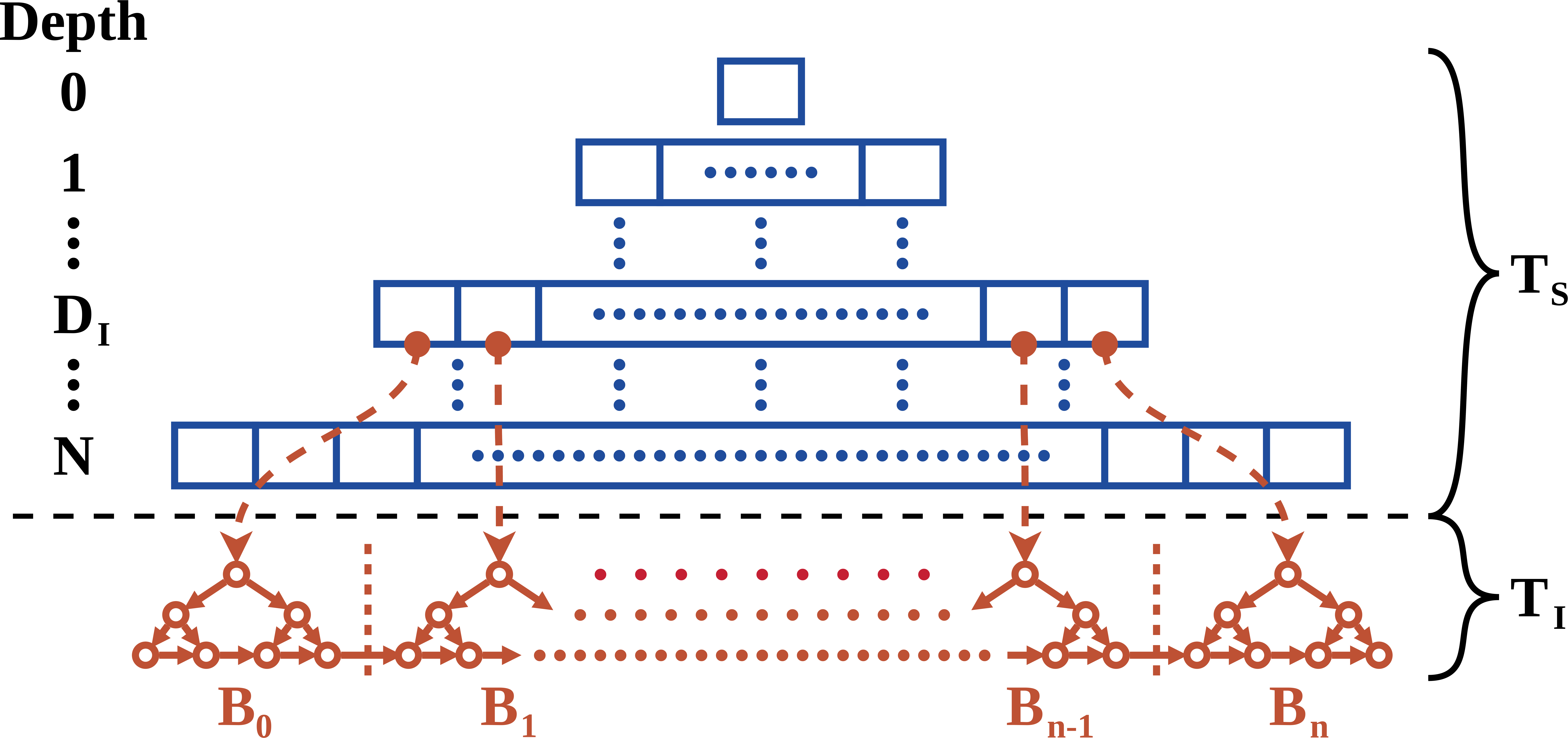}
	\vspace*{-0.1cm}
	\caption{Structure of \dimtree\ (blue and red sections are $T_S$ and $T_I$ components, respectively).}
	\label{fig:mtree}
	\vspace*{-0.5cm}
\end{figure}

\vspace{-0.0cm}
\subsection{Partitioned In-memory Merge-Tree}
Partitioned In-memory Merge-Tree (\dimtree) is an extended variant of
\imtree\ that is designed to address the challenges of parallel
sliding window indexing.  Similar to \imtree, \dimtree\ is also
composed of two components in which recently inserted tuples are
periodically merged into a lookup-efficient index.  In fact, the key
difference is in the design of the insert-efficient component
$T_I$. Rather than using a single \bptree{} for all incoming tuples, we
opt to use a set of \bptree s that are associated with disjoint tuple
value ranges. To provide a uniform workload among trees, these ranges
periodically adapt to the distribution of values in the sliding
window. Each \bptree{} is associated with a lock that allows only a
single thread to access the tree to handle parallel updates and
lookups. {\color{black} Unlike approaches that target resolving concurrency at the
tree node level, such as Bw-tree~\cite{levandoski2013bw} or 
B-link~\cite{lehman1981efficient}, parallelism in \dimtree\ is based on concurrent
operations over disjoint partitions and relies on the distribution of
incoming tuples.}  An advantage of our approach is that the routines
for performing operations are as efficient as those of the
single-threaded approach, and their only overhead is to obtain a
single lock per each tree traversal.

\vspace*{-0.1cm}
\subsubsection{\dimtree{} Structure}

Figure~\ref{fig:mtree} provides an overview of the
\dimtree\ structure.  \dimtree\ consists of two separate components,
$T_S$ and $T_I$. $T_S$ is an immutable \bptree; it is similar to our
IM-Tree, which stores static data. $T_I$ represents a set of
\textit{subindexes} named $B_0, .. ,B_n$ attached to $T_S$ at depth
$D_I$ (\textit{insertion depth}), where each $B_i$ is associated with
the same range of values as the $i^{th}$ node of $T_S$ at the
insertion depth. Each $B_i$ is an independent \bptree, where the tail
leaf node of each $B_i$ $(0 \leq i < n)$ is connected to the head leaf
node of the successor \bptree{} ($B_{i+1}$) to create a single sorted
linked list of all elements in $T_I$.

To insert a new record, the update routine first searches $T_S$ until
the depth of $D_I$ to identify the matching $B_i$ that is associated
with the range that includes the given value. Then, the routine
inserts the record into $B_i$ using the \bptree{} insert algorithm.
Similar to IM-Tree, the two components of \dimtree\ need to be
periodically merged for maintenance. This maintenance occurs when the
total number of tuples in $T_I$ equals $m \times w$. Merging
eliminates expired tuples in $T_S$ and arranges the remaining tuples to
be combined with those from $T_I$ into a sorted array that is taken as
the last level of the new $T_S$. Subsequently, $T_S$ is built from the
bottom up, and every $B_i$ is initialized as an empty \bptree.

\vspace*{-0.1cm}
\subsubsection{IBWJ Using \dimtree}

Let $H^\prime_I$ be the average height of $B_i, 0 \leq i \leq n$. The
join algorithm processes a given tuple $r$ from stream $R$ as follows.
(1) Search the index of stream $S$ to identify matching tuples, which
requires first searching $T_S$ ($H_S \cdot \lambda_{ib}^{s}$) and the
corresponding $B_i$ ($H^\prime_I \cdot \lambda_{b}^{s}$) to the leaf nodes
and then performing a leaf node scan to determine matching tuples and
filter out expired tuples ($\sigma_s \cdot \tau_c \cdot (1+m/2)$).  (2) Similar to
\imtree, tuples are deleted in a batch during the merge of $T_I$ and
$T_S$; thus, the average cost per tuple is $M^\prime/{(m \cdot w)}$,
where $M^\prime$ is the cost of merging $T_I$ and $T_S$ in
\dimtree. (3) Insert the new tuple, $r$, into $T_I$, which requires
first traversing $T_S$ to depth $D_I$ ($D_I \cdot \lambda_{ib}^s$) and then
inserting the tuple into the corresponding $B_i$
($H^\prime_I \cdot \lambda_b^i$). 
The total cost of IBWJ using \dimtree\ per
tuple ($C_{PJ}$) is given by Equation~\ref{equ:PIM_cost}.

\vspace*{-0.2cm}
\begin{equation} \label{equ:PIM_cost}
\footnotesize
\begin{split}
C_{PJ} = & \overbrace{ H_S \cdot \lambda_{ib}^{s} + H^\prime_I \cdot \lambda_{b}^{s} +  \sigma_s \cdot \tau_c \cdot (1+\frac{m}{2}) }^\text{Step 1} + \\
& \underbrace{M^\prime/{(m \cdot w)}}_\text{Step 2} + \underbrace{D_I \cdot \lambda_{ib}^s + H^\prime_I \cdot \lambda_b^i}_\text{Step 3}
\raisetag{3.5\baselineskip}
\end{split}
\end{equation}
\vspace*{-0.2cm}

{\color{black}
Comparing the costs of IBWJ using IM-Tree and \dimtree, we obtain the
following. Searching in \dimtree\ is faster because the average height
of a subindex in \dimtree\ is less than $T_I$ in IM-Tree.  The costs
for merging $T_I$ and $T_S$ in both trees are almost identical ($M =
M^\prime$), and consequently, the overall cost of tuple deletion is
the same in both trees.  The insertion costs in \dimtree\ and IM-Tree
are controlled by the number of tuples in $T_I$. Let the number of
tuples in $T_I$ be represented by $|T_I|$. For $|T_I| = 0$ (after
merge), the constant overhead of traversing $T_S$ to depth $D_I$ in
\dimtree\ is dominant and results in slower insertion in \dimtree. As
$|T_I|$ increases, the cost of insertion in IM-Tree increases faster
and eventually surpasses the insertion cost in \dimtree.}

\vspace*{-0.1cm}
\subsubsection{Concurrency Control in \dimtree} 

To protect the \dimtree\ structure during concurrent indexing,
each subindex ($B_i$) is associated with a lock that coordinates the 
accesses of the threads to the subindex.
Moreover, a searching thread may move from a $B_i$ to its successor
($B_{i+1}$) during the leaf node scan to determine matching tuples. 
To address this issue, the last leaf node of each $B_i$ is
flagged such that the searching thread recognizes the movement from one
subindex to another. In this case, the searching thread releases the
lock and acquires the one associated with the successor.

Traversing $T_S$ is completely lock-free since its structure never
changes, and there is no need for a concurrency control mechanism to
avoid race conditions. 

\vspace*{-0.2cm}

\section{Parallel Stream Join using \\ Shared Indexes} 

In this section, we present our parallel window join algorithm which
addresses the challenges of using shared indexes in a multi-threaded setting. 
During concurrent join, tuples might be inserted into indexes in an order 
different to their arrival order depending on the threads' scheduling in the system.
We design a join algorithm which is aware of the indexing status of tuples 
in order to avoid duplicated or missing results.
Moreover, our join algorithm is based on an asynchronous 
parallel model which enables threads to join or leave the operator dynamically 
depending on the system load.

\begin{figure}
	\includesvg{0.95\columnwidth}{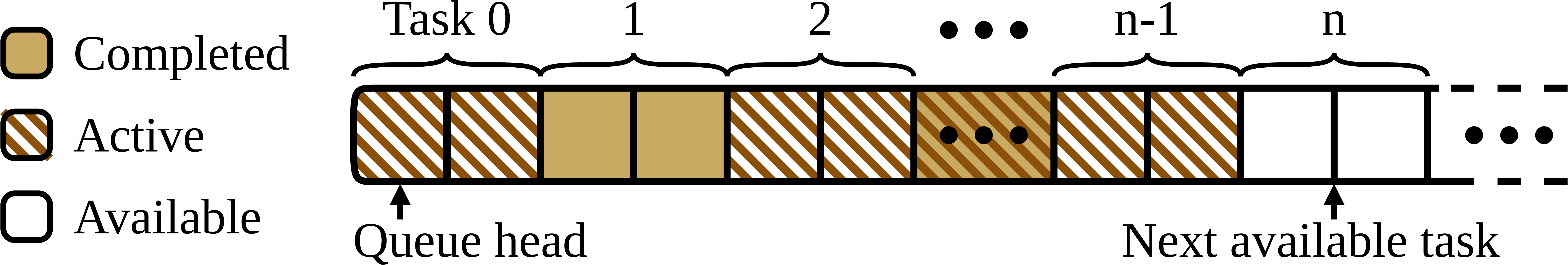}
	\vspace*{-0.1cm}
	\caption{Shared task queue with task size of 2.}
	\label{fig:queue}
	\vspace*{-0.4cm}
\end{figure}

\subsection{Concurrent Stream Join Algorithm} 

Our parallel join algorithm processes incoming tuples in four steps:
(1) task acquisition, (2) result generation, (3) index update, and (4)
result propagation.

\textbf{Task acquisition} -- A task represents a unit of
work to schedule, which is a set of incoming tuples. The task size
is the number of tuples assigned to a thread per each task
acquisition round, which determines the trade-off between maximizing
throughput and minimizing response time. Large tasks reduce
scheduling and lock acquisition overhead but simultaneously
increase system response time, whereas small tasks result in the
opposite.  In our join algorithm, tasks are distributed among
threads based on dynamic scheduling; thus, a thread is assigned with
a task whenever the thread is available.  This method enables our
join algorithm to utilize an arbitrary number of threads and not
stall because threads are unavailable.

We arrange incoming tuples into a shared work queue according to their
arrival order, regardless of which stream they belong to; and 
we protect the accesses to this queue using a shared mutex. Each tuple in the
work queue is assigned a status flag: \textit{available}
indicates that the tuple is ready to be processed but not yet assigned to any
thread, \textit{active} indicates that the tuple is assigned to a
thread but the join results are not ready, and \textit{completed}
indicates that processing of the tuple is completed and the join results
are ready but the results are not propagated. 
When a tuple arrives in the queue, its status is
initialized to available. Figure~\ref{fig:queue} illustrates the
status of the work queue during a window join with a task size of 2.

During a concurrent stream join, sliding windows must store all tuples
that are required to process active tuples of the opposite stream, 
which generally results in windows larger than $w$. 
In the case of a time-based sliding window, it is possible to filter out
unrelated tuples using timestamps; however, for count-based sliding
windows, it is necessary to record the boundaries of the opposite
window at the point in time when a tuple is assigned to a thread. We
refer to these boundaries as $t_l$ (latest tuple) and $t_e$ (earliest
tuple). When a thread acquires a task, it changes the status of the
tuples to active and saves $t_l$ and $t_e$ for each tuple.

\begin{figure}
	\includesvg{0.95\columnwidth}{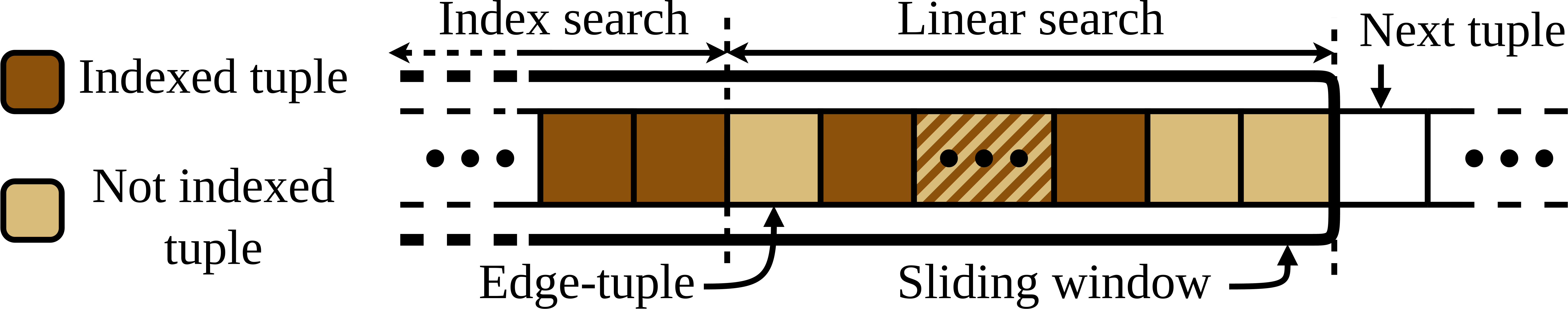}
	\vspace*{-0.15cm}
	\caption{Sliding window during parallel stream join.}
	\label{fig:window}
	\vspace*{-0.65cm}
\end{figure}
\vspace*{-0.cm}

\textbf{Result generation} -- To avoid duplicate or missing results,
we keep references to the earliest nonindexed tuple of each sliding
window, referred to as the \textit{edge tuple}. This tuple declares
that all tuples before it are already indexed, whereas the statuses of
the subsequent tuples are undetermined. When a thread starts to
process a tuple, it stores the position of the edge tuple in a local
variable since the value might be updated during processing.  Using an
old value of the edge tuple might increase the computational cost
slightly, but it is safe in terms of result correctness. The lookup
algorithm determines matching tuples in two steps. First, it queries
the index for matching tuples and filters out those after the edge
tuple or before $t_l$. Second, it linearly searches the sliding window
from the edge tuple to $t_e$ and adds any results to the previously
found results. Figure~\ref{fig:window} illustrates the sliding window
during the join operation.  When a thread finishes processing a tuple,
it stores the results in shared memory and updates the task status to
\textit{completed} in the shared queue but does not yet propagate the
results into the output stream at this step.

\textbf{Index update} -- After a thread generates the join results for
a tuple, it inserts the tuple into the index and marks the tuple in
the sliding window as indexed. Subsequently, the thread attempts to update
the edge tuple accordingly. To avoid a race condition, a
shared mutex coordinates write accesses to the edge tuple. Using a
test-and-set operation, the thread checks whether the mutex is held by
another thread. If so, it avoids the edge tuple update and continues
to the next step. Otherwise, it increments the edge tuple to the next
nonindexed tuple in the sliding window and releases the mutex.

\textbf{Result propagation} -- In the final step, a thread attempts to
propagate the results of completed tuples. Similar to the edge tuple
update routine, a shared mutex coordinates threads during result
propagation. The thread checks the status of the mutex. In the case that the
mutex is already held by another thread, the thread skips this step
and begins to process another task. Otherwise, it verifies whether
the results for the tuple at the work queue head are completed. If
so, it propagates the results into the output stream and removes the
tuple from the work queue. This routine is repeated until the status
of the tuple at the work queue head is either active or
available. Finally, the thread releases the mutex and starts to
process another task.

\subsection{Nonblocking Merge and Indexing}	

Performing merging as a blocking operation negatively impacts system
availability and latency, which are both often critical concerns for
stream processing applications. To address this challenge, we propose
a nonblocking merge method. Our approach enables the stream join
processing threads to continue the join without significant
interruption during merge processing.  Figure~\ref{fig:merging}
illustrates the overall scheme of performing a nonblocking merge.  The
operation consists of two phases: first, creating a new \dimtree, and
second, applying pending updates.   

Whenever merging is needed, a thread called the
\textit{merging thread} is assigned to perform the merge operation.
At the beginning of each stage, the merging thread blocks the assignment
of new tasks until all active threads finish their currently processing tasks.
During the first phase, the merging thread creates a new
\dimtree\ without modifying the previous index tree. Concurrently,
other threads resume performing tasks without an index update.
When the merging thread finishes creating the updated \dimtree, it 
starts the next phase.
At the beginning of the second phase, the merging thread swaps the old index 
with the new one before it unblocks the task assignment process.
During the second phase, the merging thread applies pending updates and
other threads begin to perform the join operation with index update.
When the pending updates are finished, the merging thread leaves merge 
operation and begins to perform the join operation.

During the first phase of nonblocking merge, the index
data are not updated; therefore, the position of the edge tuple
does not change during this phase.  Consequently, the linear search
in the nonindexed portion of the sliding window becomes more
expensive.

\begin{figure}
	\includesvg{0.95\columnwidth}{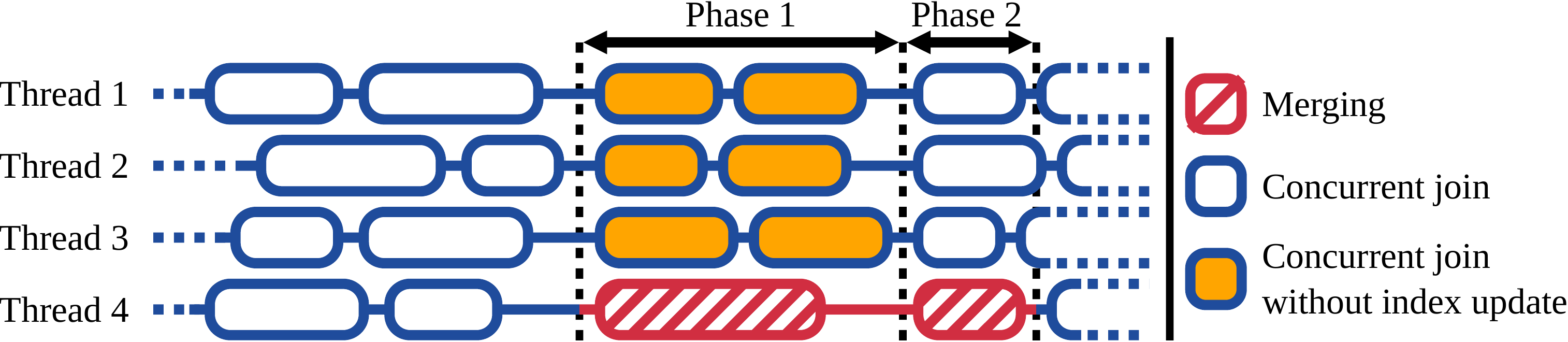}
	\vspace*{-0.1cm}
	\caption{Nonblocking merge.}
	\label{fig:merging}
	\vspace*{-0.4cm}
\end{figure}

\begin{figure*}[]
	\begin{minipage}[t]{0.28\textwidth}
		\captionsetup{labelformat=empty,labelsep=none}
		\centering
		\includepdfcrop{\columnwidth}{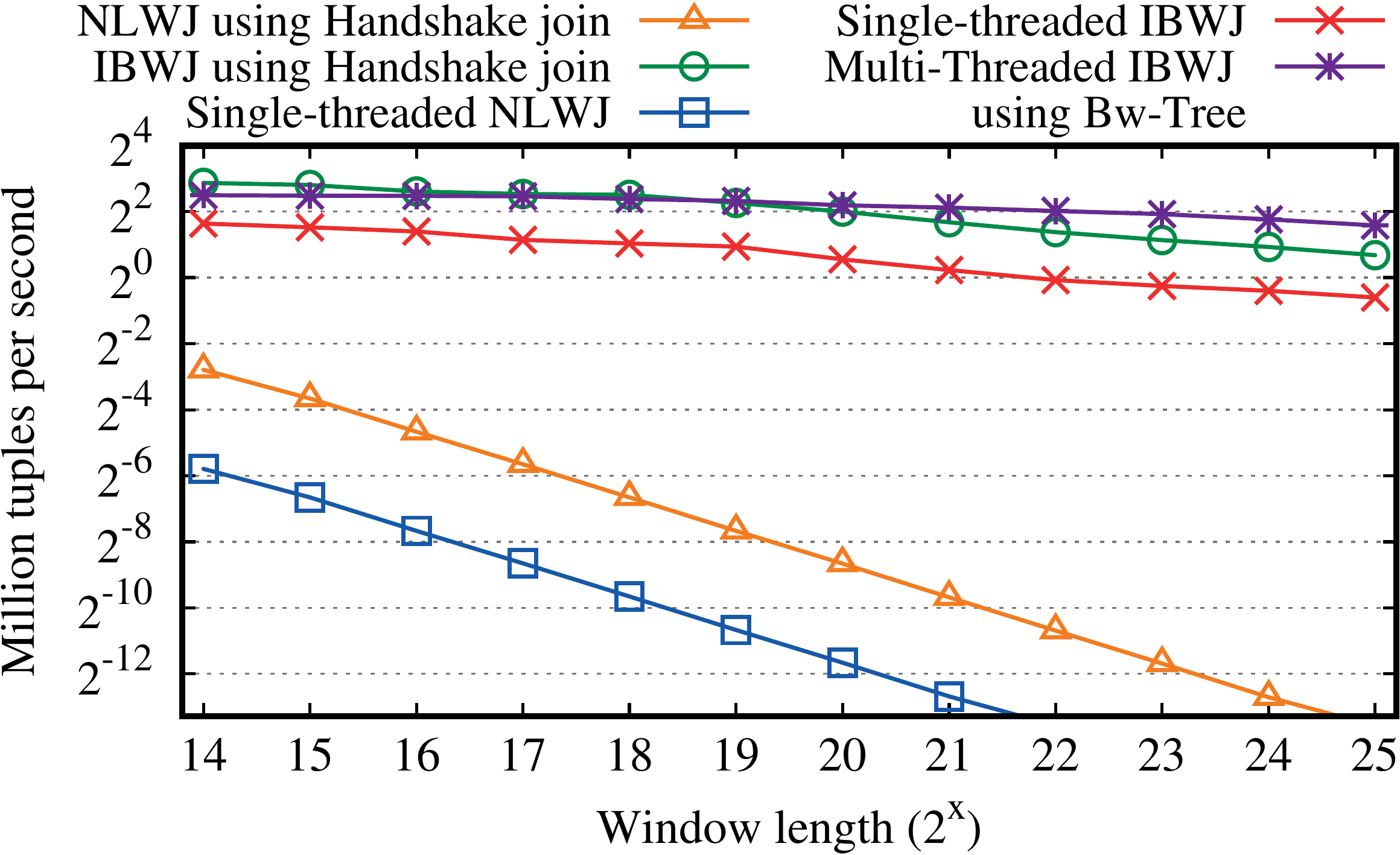}
		{\vspace*{-0.95cm} \caption{\hspace*{-3.9cm}\footnotesize(a)}}
	\end{minipage}
	\hfill
	\begin{minipage}[t]{0.23\textwidth}
		\captionsetup{labelformat=empty,labelsep=none}
		\centering
		\includepdfcrop{\columnwidth}{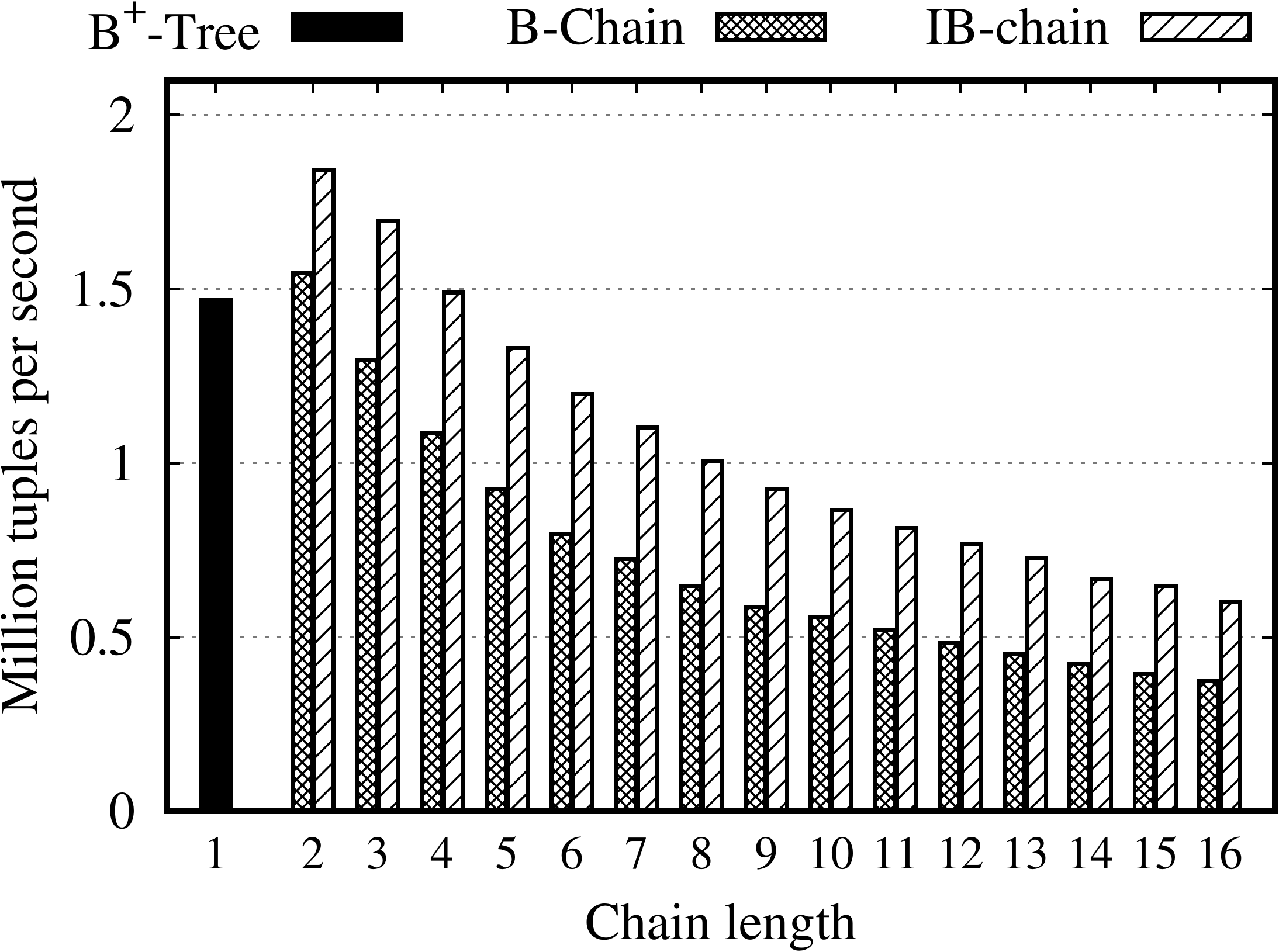}
		{\vspace*{-0.95cm} \caption{\hspace*{-3.4cm}\footnotesize(b)}}
	\end{minipage}
	\hfill
	\begin{minipage}[t]{0.23\textwidth}
		\captionsetup{labelformat=empty,labelsep=none}
		\centering
		\includepdfcrop{\columnwidth}{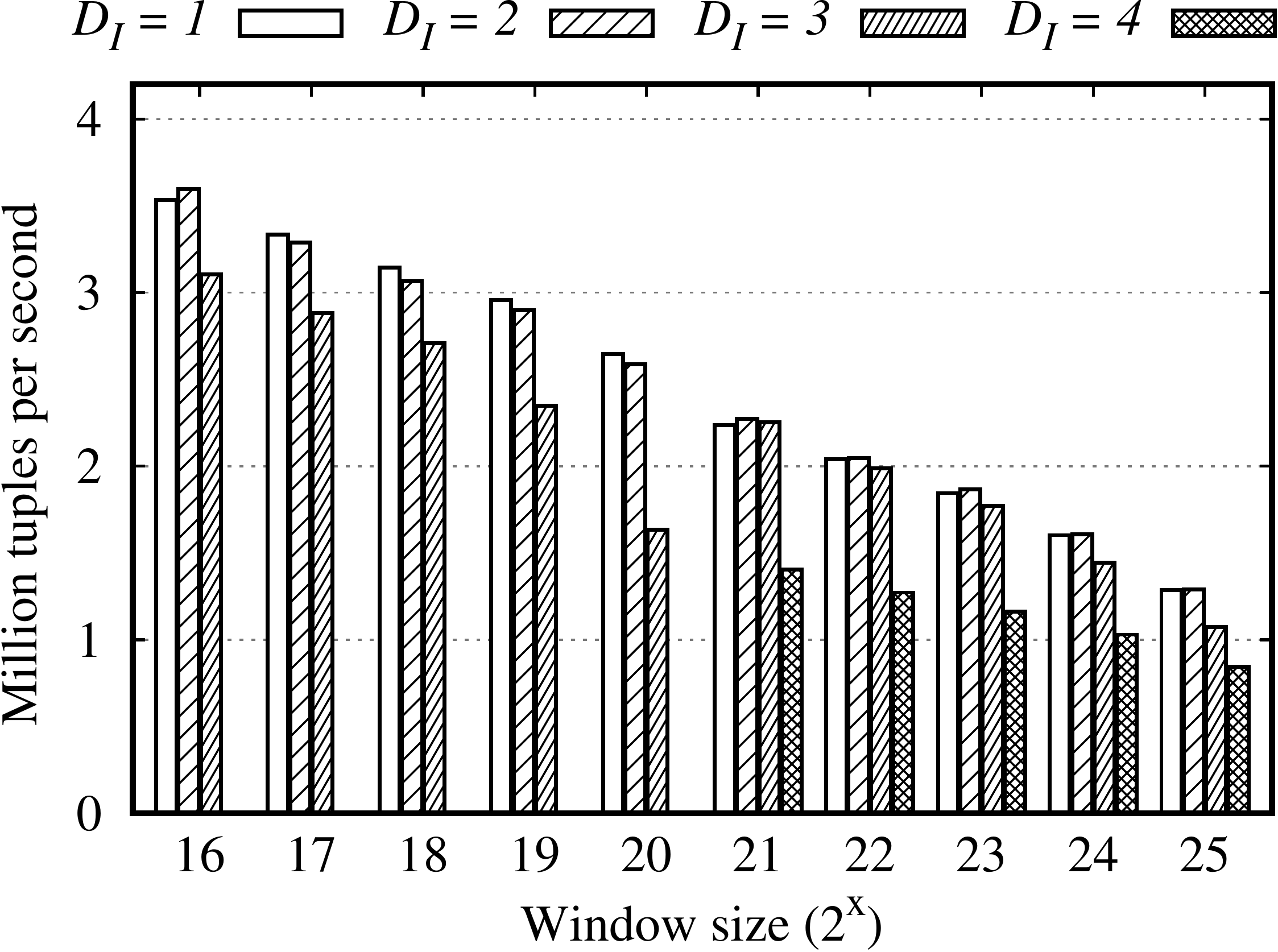}
		{\vspace*{-0.95cm} \caption{\hspace*{-3.4cm}\footnotesize(c)}}
	\end{minipage}
	\hfill
	\begin{minipage}[t]{0.23\textwidth}
		\captionsetup{labelformat=empty,labelsep=none}
		\centering
		\includepdfcrop{\columnwidth}{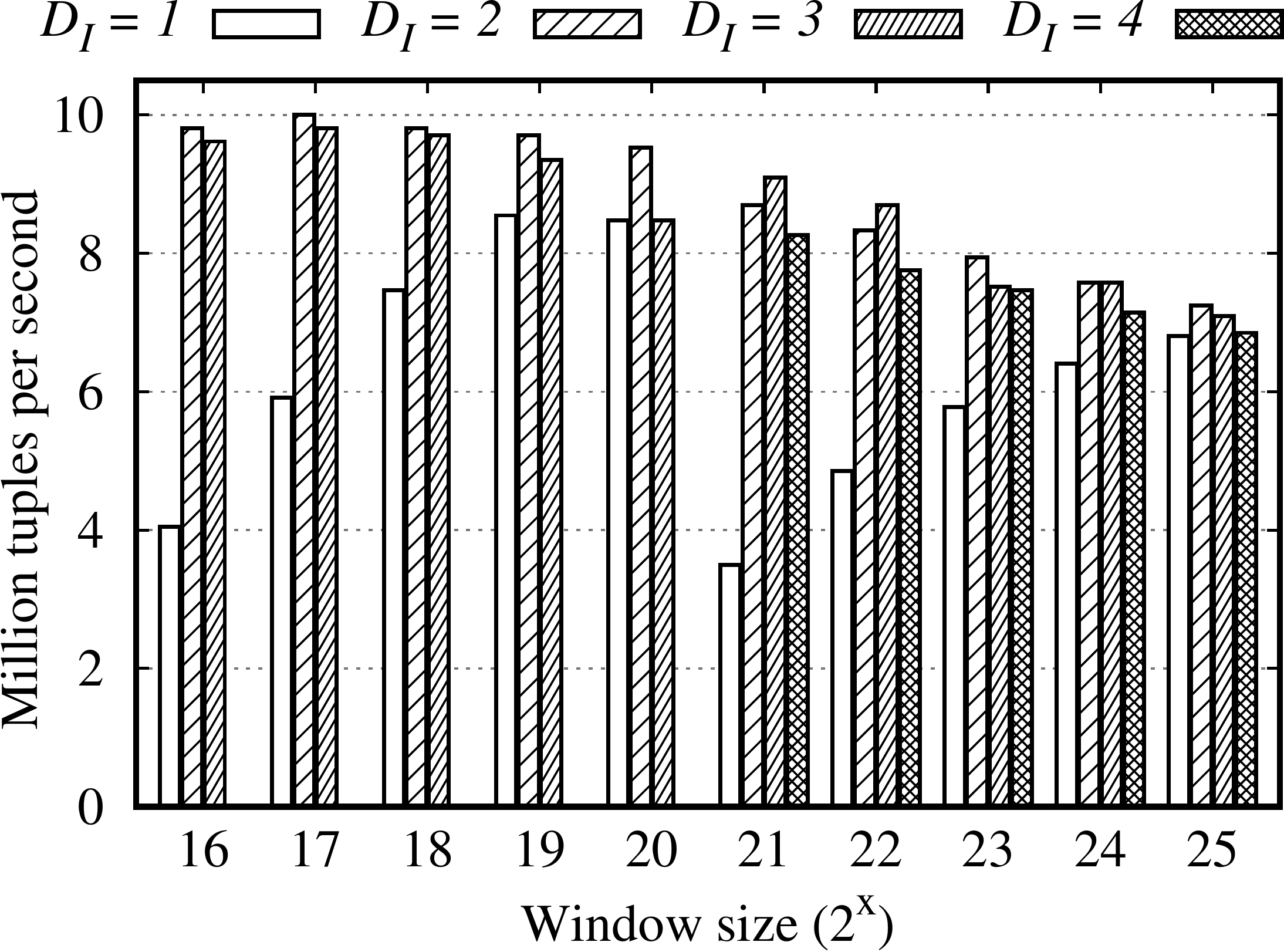}
		{\vspace*{-0.95cm} \caption{\hspace*{-3.4cm}\footnotesize(d)}}
	\end{minipage}
	\vspace{-0.2cm}
	\setcounter{figure}{7}
	\caption{
		a) Performance evaluation of multi-threaded window join using round-robin partitioning.
		b) Throughput comparison of IBWJ using chained index and \bptree{} ($w = 2^{20}$).
		c) Throughput vs. insertion depth ($D_I$) for single-threaded IBWJ using \dimtree.
		d) Throughput vs. insertion depth ($D_I$) for parallel IBWJ using \dimtree.
	}
	\label{plot:NCH}
	\vspace{-0.3cm}
\end{figure*}

\vspace{-0.1cm}

\section{Evaluation}

In this section, we present a set of experiments to benchmark the
efficiency of the approaches introduced in this paper and 
empirically determine the corresponding parameters, such as
merge ratio and insertion depth.
Moreover, we study the influence of join selectivity
and skewed value distribution on the performance of our 
parallel window join design.
As the query workload, two streams, $R$ and $S$, are joined 
via the following band join.

\vspace*{-0.0cm}
\lstset{language=SQL}
\begin{lstlisting}[basicstyle=\footnotesize\ttfamily,numbers=none,]  
SELECT * FROM R, S
WHERE ABS(R.x - S.x) <= diff
\end{lstlisting} 
\vspace*{-0.0cm} 

The join attributes ($R.x$ and $S.x$) are assumed to be random 
integers generated according to a uniform distribution 
and the input rates of streams $R$ and $S$ are symmetric unless 
otherwise stated.

Because we evaluate each experiment for different window lengths
($w$), considering a fixed value for \textit{diff} results in various
join match rates (i.e., the match rate of band join with $w=2^{25}$
will be $2^{15}$ times higher than that with $w=2^{10}$), which
influences the overall join performance.  For a more comprehensive
comparison, the value of \textit{diff} is adjusted according to the
window length such that the match rate ($\sigma_s$) is always two 
except for the one that exclusively studies the influence of join selectivity.

We used two forms of band join: two-way join and self-join.
In the former, $R$ and $S$ are two distinct streams, 
and in the latter, an identical stream is used as both $R$ and $S$. The
experiments are generally based on two-way join, except for those 
where we explicitly declare that self-join is used. 

We evaluate our approaches on an octa-core (16 CPU threads, 
hyper-threading enabled) Intel Xeon E5-2665.
For all multithreaded experiments, we utilized all 16 threads 
unless otherwise stated.
{\color{black}We employ the STX-\bptree{} implementation, which is a set of C++ template
classes for an in-memory \bptree~\cite{bingmann2008stx}, and we used our own
CSS-Tree implementation as immutable \bptree~\cite{Rao:1999:CCI:645925.671362}.}

\vspace{-0.1cm}
\subsection{Comparison of Existing Approaches}

{\color{black}
\textbf{Round-robin window partitioning} --
The purpose of this experiment is to study the efficiency of 
round-robin partitioning based approaches, such as low-latency handshake join, 
SplitJoin and BiStream, 
in the application of index-accelerated stream join. 
We evaluate five implementations of
the window join: (1) single-threaded Nested-Loop Window Join (NLWJ), (2) multithreaded NLWJ
based on round-robin partitioning, (3) single-threaded IBWJ using \bptree , (4)
multithreaded IBWJ based on round-robin partitioning, and (5)
multithreaded IBWJ using Bw-tree. Figure~\ref{plot:NCH}a presents the
results for varying window sizes.

Comparing the join algorithms, we observe that NLWJ is more vulnerable to
the sliding window size because its performance linearly decreases as the
window size increases. In contrast, the performance of IBWJ is less
sensitive to the sliding window size.  Multithreaded join using round-robin partitioning
improves the performances of NLWJ and IBWJ by factors of 8 and 2.5,
respectively.  This result implies that although approaches based on
round-robin window partitioning are effective for NLWJ,
these approaches cannot efficiently exploit the computational power of
multicore processors for IBWJ.

Moreover, the performance result of parallel IBWJ using Bw-Tree
indicates that the efficiency of concurrent operations in Bw-Tree
improves as the size of Bw-Tree increases.  The larger the indexing
tree, the lower is the probability of accessing the same node by different
threads at the same time; consequently, the multithreading
efficiency increases.  For the smallest sliding window size ($w =
2^{14}$), parallel IBWJ using Bw-Tree results in 65\% lower throughput
than parallel IBWJ using round-robin partitioning, but for the
largest window size ($w = 2^{25}$) evaluated, parallel IBWJ using
Bw-Tree outperforms round-robin based method and results in 75\%
higher throughput.
}
\vspace{0.1cm}

\textbf{Chained index} --
Figure~\ref{plot:NCH}b shows the throughput of
IBWJ using chained index~\cite{Lin:2015:SDS:2723372.2746485} 
for varying chain lengths. We propose and
evaluate two different designs for chained index, referred to as
\textit{\bptree{} chain (B-chain)} and \textit{Immutable \bptree{} chain
	(IB-chain)}. In the former design, all subindexes are \bptree s,
including the active subindex (the one where newly arriving tuples
are inserted) and all archived subindexes. In the latter design,
only the active subindex is a \bptree, and before archiving an active
subindex, it is converted into an immutable \bptree; thus, all
archived subindexes are immutable \bptree s.

We observe that the IB-chain results in 50\% higher throughput than
the B-chain on average, which indicates that the immutable
\bptree{} vastly outperforms the regular \bptree{} for search queries in
this scenario. For both the B-chain and IB-chain, the shortest chain
length, which is two, results in the best throughput. However, the
performance noticeably decreases when the chain length increases. The
main drawback of chained index is the higher search complexity, which
increases almost linearly with the chain length. Although the index
chain reduces the overhead of tuple removal using coarse-grained data
discarding, the higher search overhead degrades its overall
performance.

\begin{figure*}
	\begin{minipage}[t]{0.24\textwidth}
		\captionsetup{labelformat=empty,labelsep=none}
		\centering
		\includepdfcrop{\columnwidth}{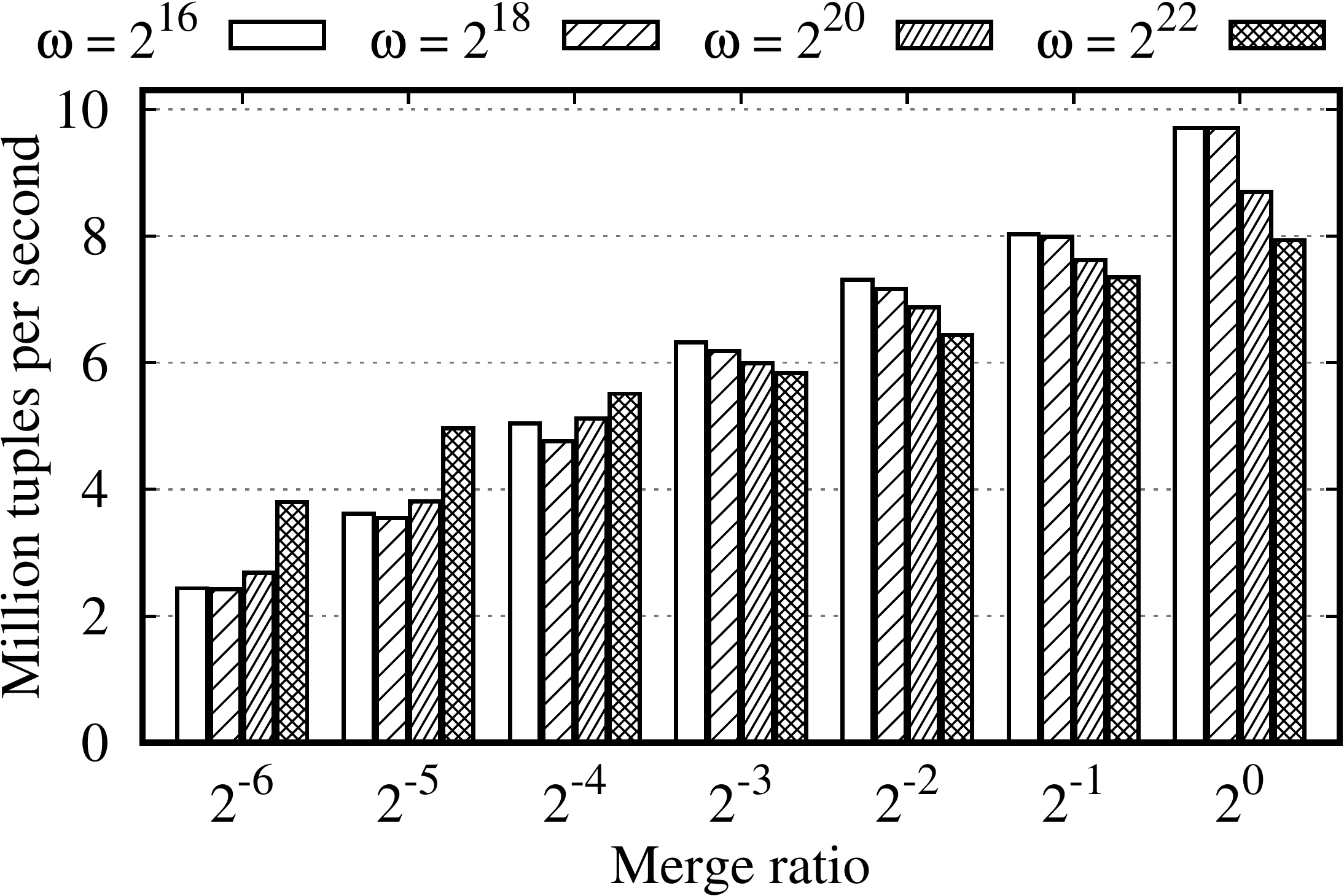}
		{\vspace*{-0.95cm} \caption{\hspace*{-3.9cm}\footnotesize(a)}}
	\end{minipage}
	\hfill
	\begin{minipage}[t]{0.24\textwidth}
		\captionsetup{labelformat=empty,labelsep=none}
		\centering
		\includepdfcrop{\columnwidth}{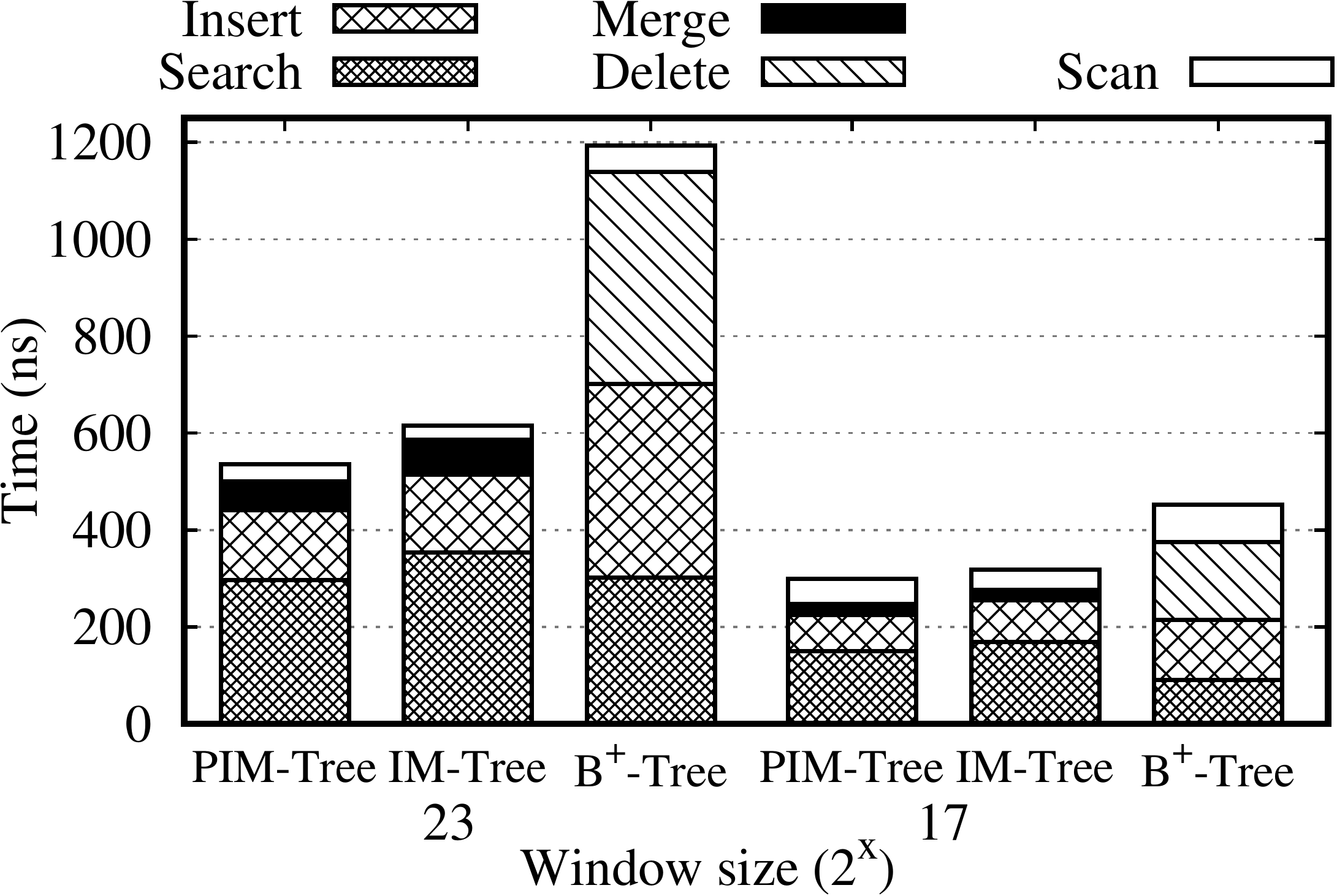}
		{\vspace*{-0.95cm} \caption{\hspace*{-3.6cm}\footnotesize(b)}}
	\end{minipage}
	\hfill
	\begin{minipage}[t]{0.25\textwidth}
		\captionsetup{labelformat=empty,labelsep=none}
		\centering
		\includepdfcrop{\columnwidth}{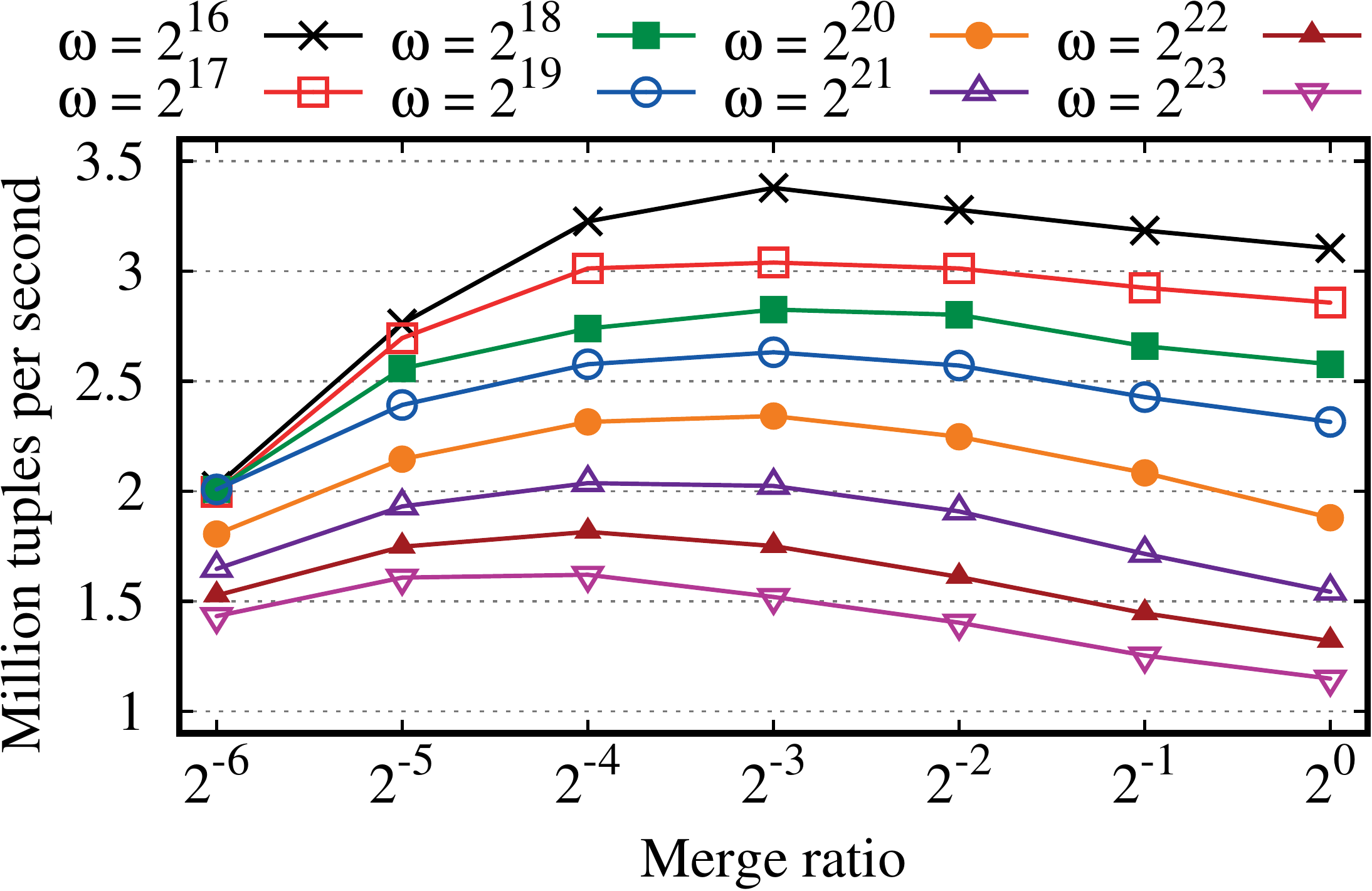}
		{\vspace*{-0.95cm} \caption{\hspace*{-3.9cm}\footnotesize(c)}}
	\end{minipage}
	\hfill
	\begin{minipage}[t]{0.25\textwidth}
		\captionsetup{labelformat=empty,labelsep=none}
		\centering
		\includepdfcrop{\columnwidth}{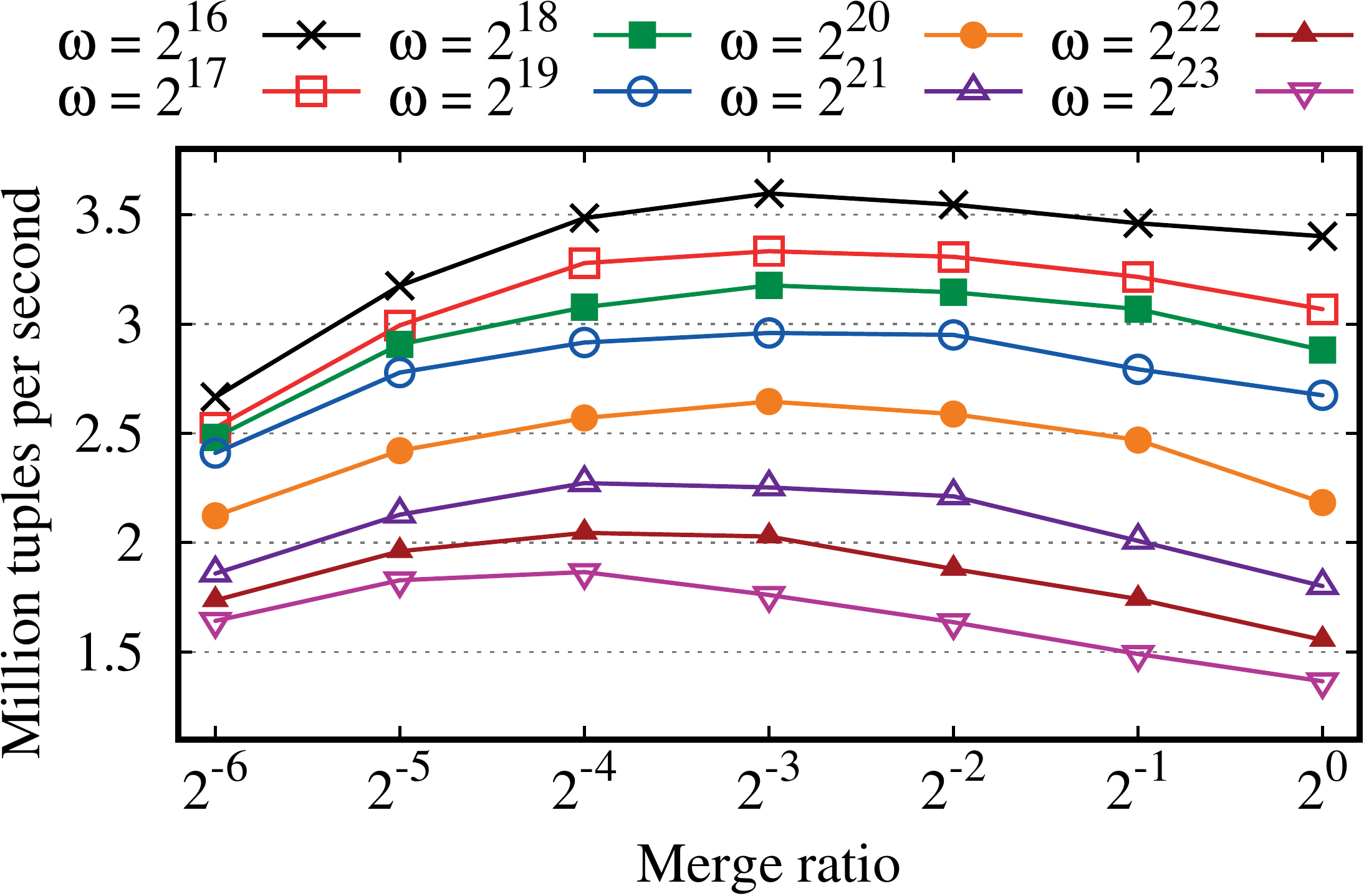}
		{\vspace*{-0.95cm} \caption{\hspace*{-3.9cm}\footnotesize(d)}}
	\end{minipage}
	\setcounter{figure}{8}
	\vspace{-0.45cm}
	\caption{
		a) Throughput vs. merge ratio for parallel IBWJ using \dimtree.
		b) Cost comparison of the different steps of IBWJ for a single tuple using various indexing data structures. 
		c) Throughput vs. merge ratio for IBWJ using \imtree.
		d) Throughput vs. merge ratio for IBWJ using \dimtree.
	}
	\label{plot:HRR}
	\vspace{-0.1cm}
\end{figure*}

\begin{figure*}
	\begin{minipage}[t]{0.24\textwidth}
		\captionsetup{labelformat=empty,labelsep=none}
		\centering
		\includepdfcrop{\columnwidth}{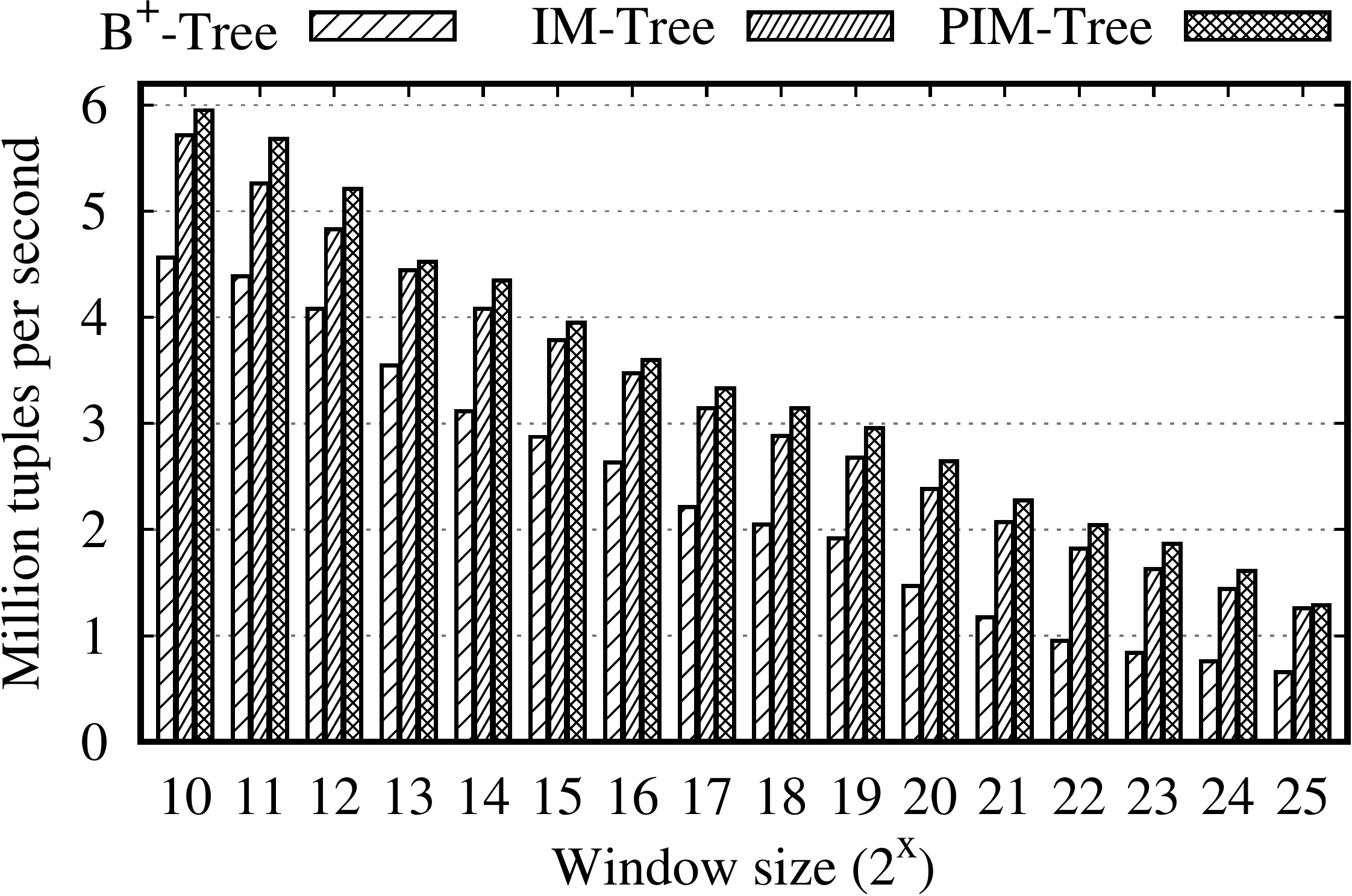}
		{\vspace*{-0.95cm} \caption{\hspace*{-3.9cm}\footnotesize(a)}}
	\end{minipage}
	\hfill
	\begin{minipage}[t]{0.24\textwidth}
		\captionsetup{labelformat=empty,labelsep=none}
		\centering
		\includepdfcrop{\columnwidth}{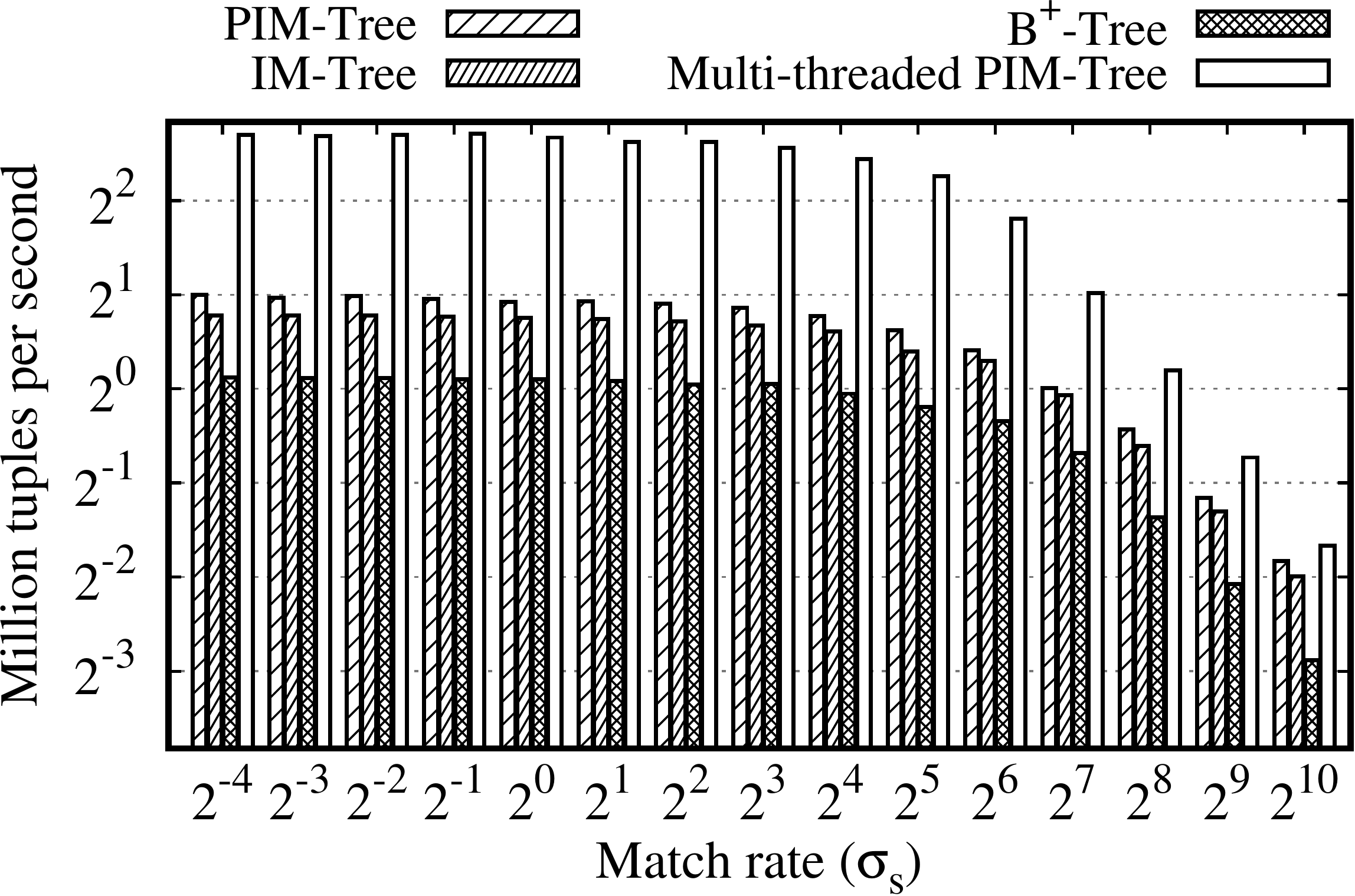}
		{\vspace*{-0.95cm} \caption{\hspace*{-3.9cm}\footnotesize(b)}}
	\end{minipage}
	\hfill
	\begin{minipage}[t]{0.24\textwidth}
		\captionsetup{labelformat=empty,labelsep=none}
		\centering
		\includepdfcrop{\columnwidth}{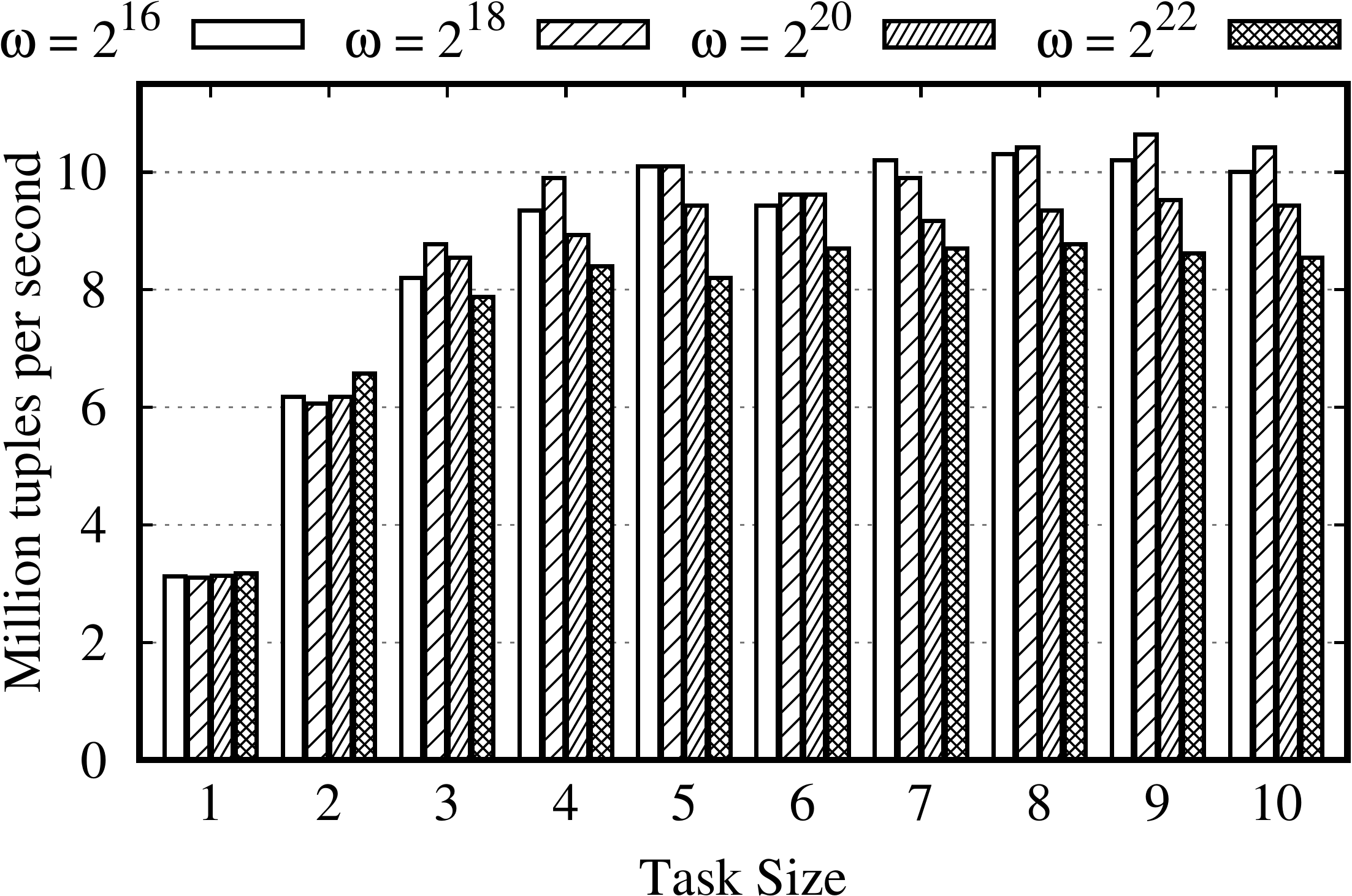}
		{\vspace*{-0.95cm} \caption{\hspace*{-3.9cm}\footnotesize(c)}}
	\end{minipage}
	\hfill
	\begin{minipage}[t]{0.24\textwidth}
		\captionsetup{labelformat=empty,labelsep=none}
		\centering
		\includepdfcrop{\columnwidth}{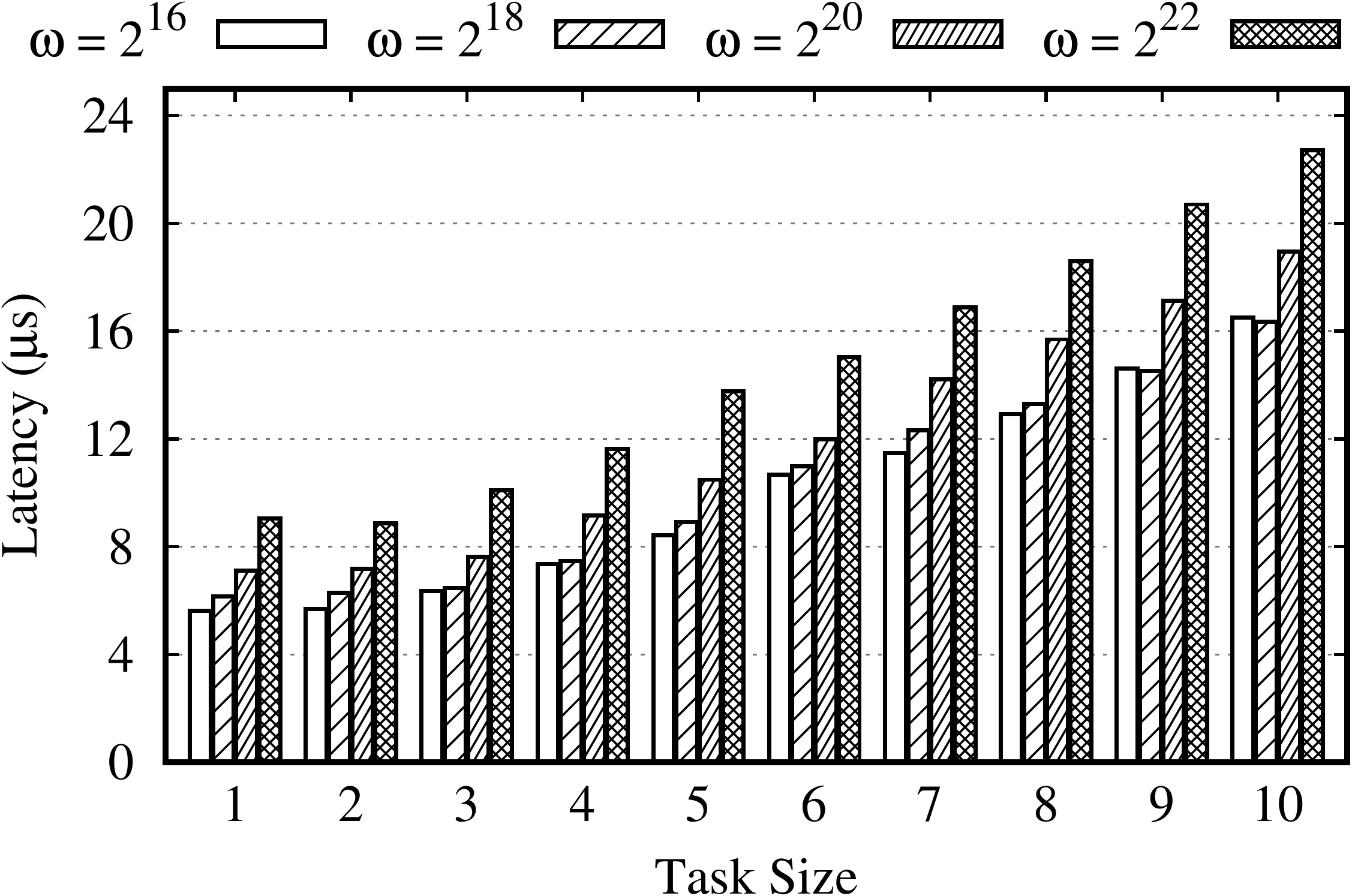}
		{\vspace*{-0.95cm} \caption{\hspace*{-3.9cm}\footnotesize(d)}}
	\end{minipage}
	\vspace{-0.2cm}
	\setcounter{figure}{9}
	\caption{
		a) Performance comparison of single-threaded IBWJ using different indexing data structures. 
		b) Throughput vs. match rate for IBWJ ($w = 2^{20}$). 
		c) Throughput vs. task size for parallel IBWJ using \dimtree. 
		d) Latency vs. task size for parallel IBWJ using \dimtree. 
	}
	\label{plot:SRS}
	\vspace{-0.6cm}
\end{figure*}

\subsection{IBWJ using \dimtree\ and \imtree}

\textbf{Insertion depth} -- In this experiment, we study the impact of
the insertion depth ($D_I$) on the performance of \dimtree.
Increasing $D_I$ results in smaller subindexes ($B_i$s), which
accelerates the operation on subindexes, and it simultaneously 
increases the overhead of searching $T_S$ to find the
corresponding $B_i$. Figures~\ref{plot:NCH}c and \ref{plot:NCH}d 
show the throughputs of single-threaded and parallel IBWJ, respectively, 
using \dimtree\ for different $D_I$s ranging from
one to four, considering that the root node is at a depth of zero.
For the window sizes of $2^{16}$ to $2^{19}$, there are only four levels of
inner nodes (including the root node); thus, the maximum feasible
$D_I$ is three.

The results for $D_I = 1$ reveal that the number of inner nodes at
depth $D_I$ highly influences the performance of parallel IBWJ.  If
the number of subindexes in $T_I$ (which is equal to the number of
inner nodes at depth $D_I$) is not sufficient, then the performance
significantly decreases due to the high partition locking congestion.   
From $w = 2^{16}$ to $2^{20}$, the system throughput rapidly increases
since the number of inner nodes at $D_I = 1$ also increases.  At $w =
2^{21}$, the number of inner nodes at $D_I = 1$ decreases since the
tree depth is incremented by one, which also causes a decrease in the
IBWJ throughput.  For larger values of $D_I$ (three and four), the
IBWJ throughput does not improve, which suggests that the
multithreading is no longer bounded by the number of subindexes.

For the case of single-threaded IBWJ, the achieved throughput for
different $D_I$s is less dependent on the window size.  However,
setting $D_I$ to the highest feasible value results in a higher
overhead for searching $T_S$ and lowers the overall performance.

\textbf{Merge ratio ($m$)} -- To determine the empirically optimal
merge ratio for \imtree\ and \dimtree, we conduct an experiment for
each data structure. Figures~\ref{plot:HRR}c and \ref{plot:HRR}d
illustrate the throughputs of single-threaded IBWJ using \imtree\ and \dimtree,
respectively, with merge ratios ranging from $2^{-6}$ to $1$.
The results for both data structures follow a similar pattern, but the
average throughput employing \dimtree\ is higher than that using
\imtree. Additionally, the system does not perform efficiently for
either very low or very high values of the merge ratio. This
underperformance is a consequence of the excessive overhead imposed
by the frequent merge when the merge ratio is set very low and by the
inefficient insert and search operations when the merge ratio is set
very high.  The results suggest that the choice of the merge ratio is
more influential for smaller sliding windows, and the empirical
optimal ratio is not identical for all window sizes. Over the largest
evaluated sliding window ($2^{23}$), setting the merge ratio to
$1/2^{4}$ results in the highest throughput, whereas for the smallest
one ($2^{16}$), $1/2^{3}$ is the best merge ratio.

Figure~\ref{plot:HRR}a illustrates the throughput of the parallel IBWJ
using \dimtree\ for varying merge ratios ranging from $2^{-6}$ to $1$. 
In contrast to the single-threaded implementation, setting
the merge ratio to the highest value always results in the best
performance in the multithreaded setting, regardless of the window
size.  This result indicates that the cost of merge operations during
a parallel window join is higher than the cost in a single-threaded
setup. Hence, minimizing the number of merges results in the highest
throughput. We also observe that the choice of the merge ratio is more
influential for smaller window sizes. Henceforth, we set the value of
the merge ratio for the multithreaded setup to one.

\begin{figure*}
	\begin{minipage}[t]{0.24\textwidth}
		\captionsetup{labelformat=empty,labelsep=none}
		\centering
		\includepdfcrop{\columnwidth}{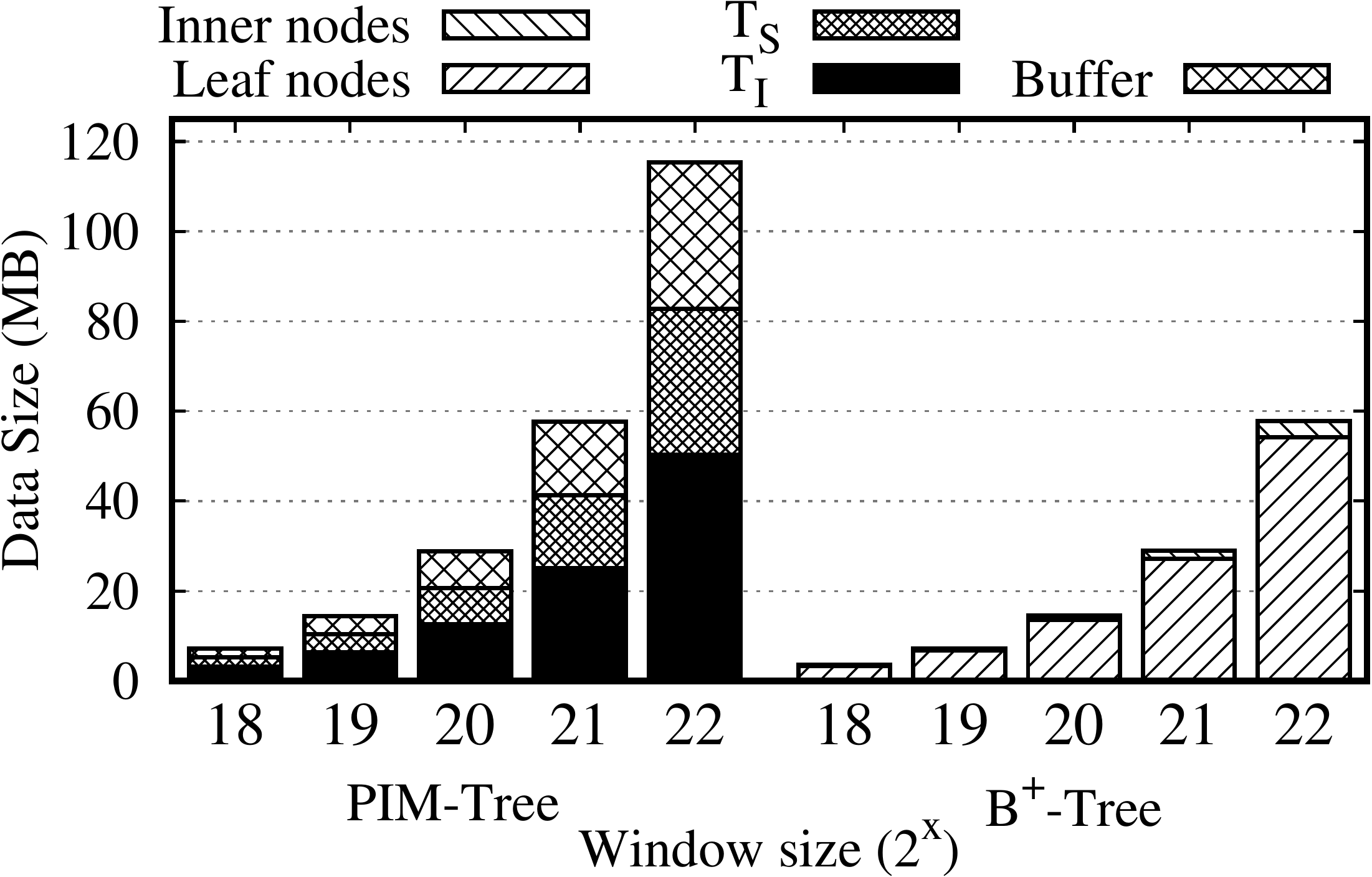}
		{\vspace*{-0.95cm} \caption{\hspace*{-3.9cm}\footnotesize(a)}}
	\end{minipage}
	\hfill
	\begin{minipage}[t]{0.24\textwidth}
		\captionsetup{labelformat=empty,labelsep=none}
		\centering
		\includepdfcrop{\columnwidth}{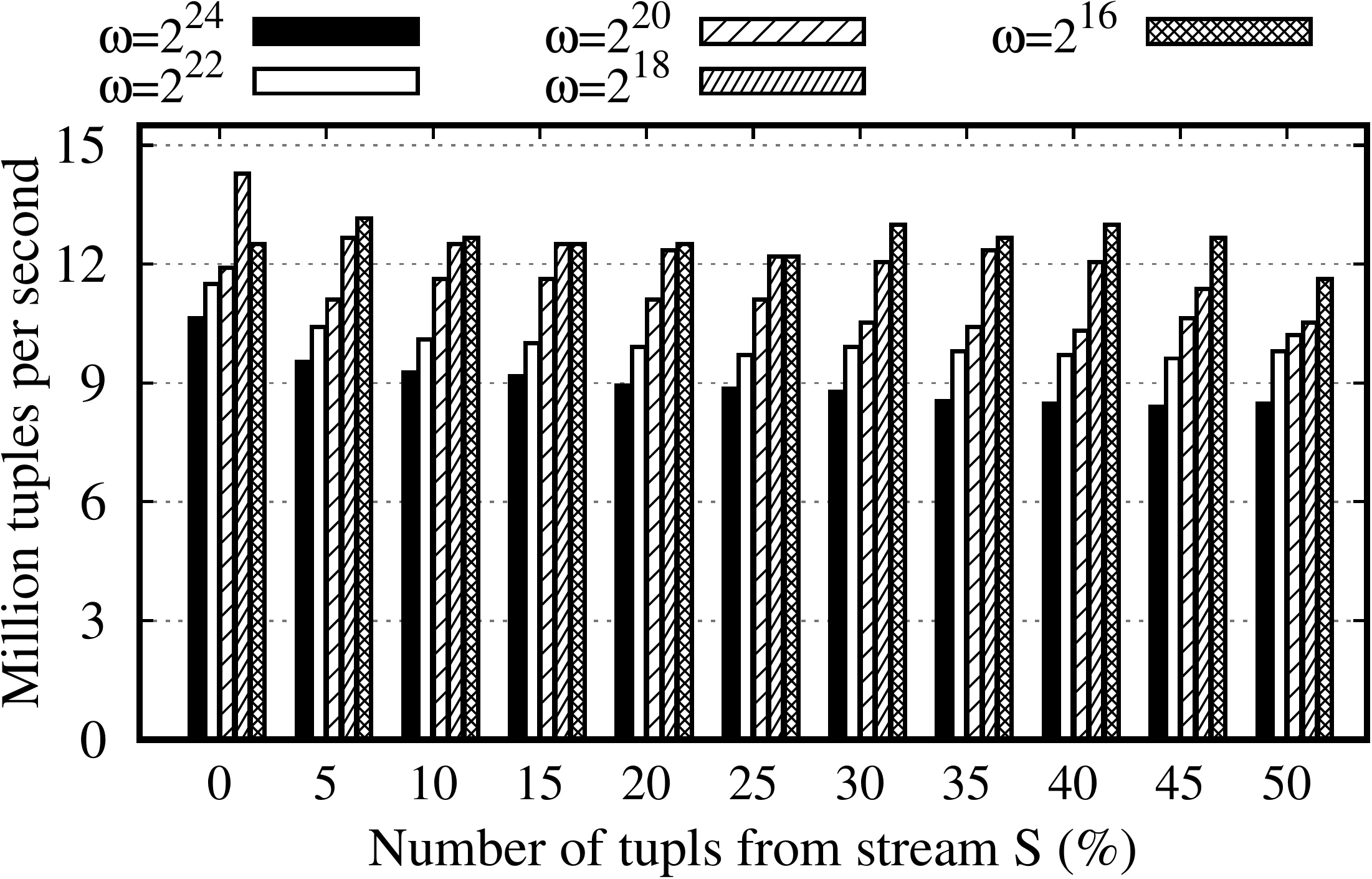}
		{\vspace*{-0.95cm} \caption{\hspace*{-3.9cm}\footnotesize(b)}}
	\end{minipage}
	\hfill
	\begin{minipage}[t]{0.24\textwidth}
		\captionsetup{labelformat=empty,labelsep=none}
		\centering
		\includepdfcrop{\columnwidth}{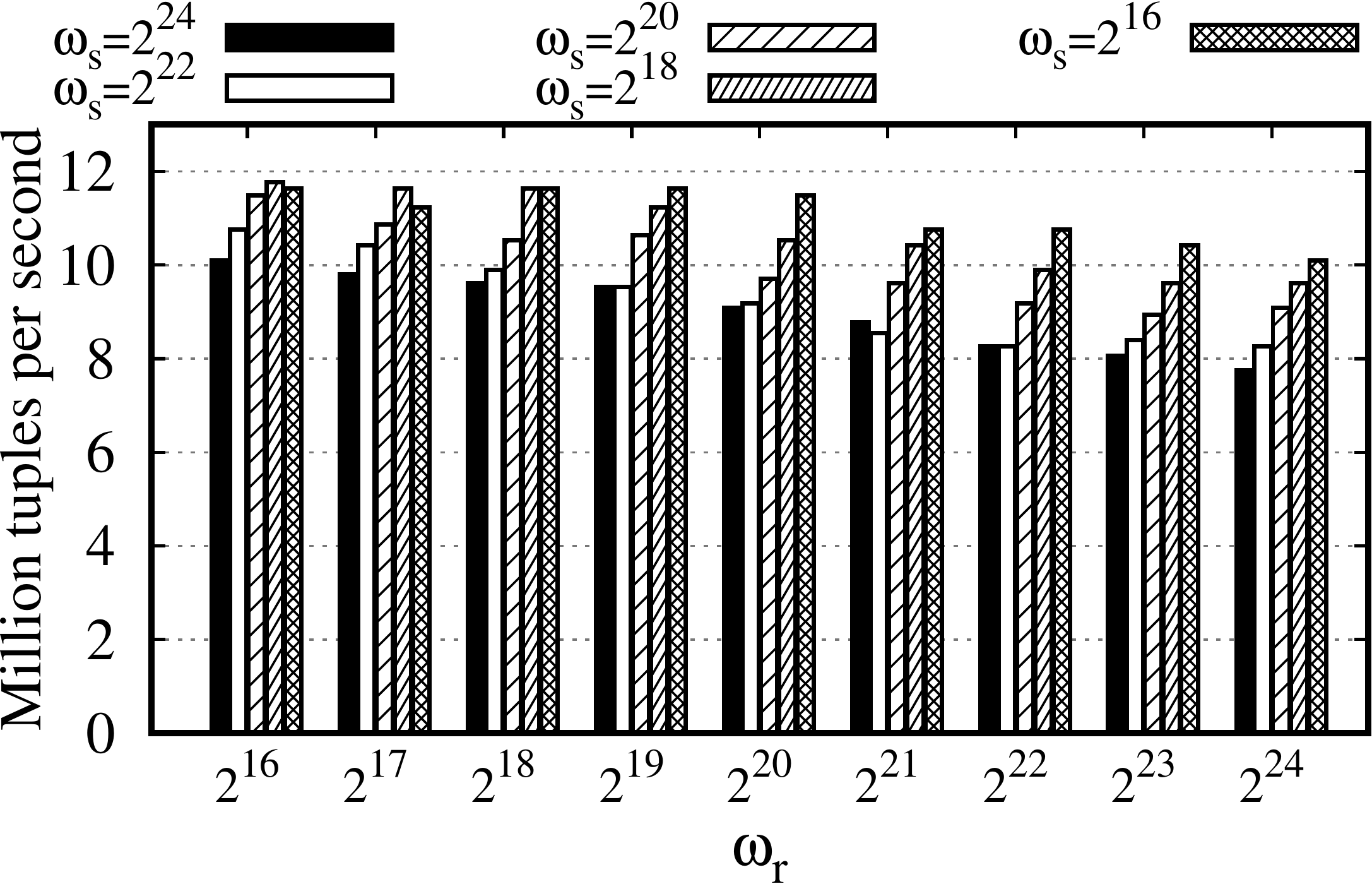}
		{\vspace*{-0.95cm} \caption{\hspace*{-3.9cm}\footnotesize(c)}}
	\end{minipage}
	\hfill
	\begin{minipage}[t]{0.24\textwidth}
		\captionsetup{labelformat=empty,labelsep=none}
		\centering
		\includepdfcrop{\columnwidth}{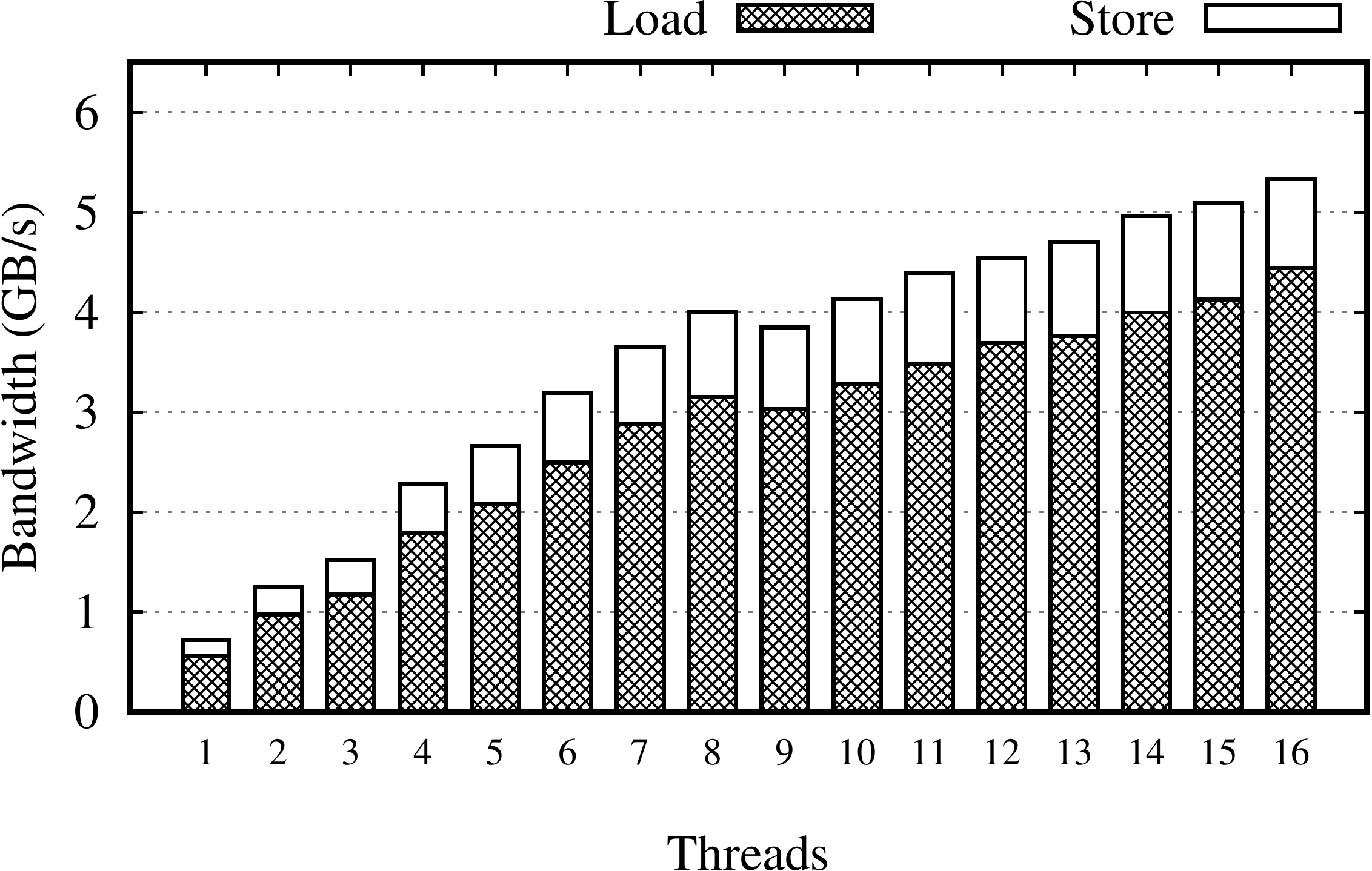}
		{\vspace*{-0.95cm} \caption{\hspace*{-3.9cm}\footnotesize(d)}}
	\end{minipage}
	\vspace{-0.2cm}
	\setcounter{figure}{10}
	\caption{
		a) Memory footprint comparison of \bptree{} and \dimtree.
		b) Evaluation of IBWJ using \dimtree\ with asymmetric input rate.
		c) Evaluation of IBWJ using \dimtree\ with asymmetric window sizes.
		d) Effective memory bandwidth of parallel IBWJ.
	}
	\label{plot:STWB}
	\vspace{-0.3cm}
\end{figure*}

\textbf{\bptree{} vs. \imtree\ vs. \dimtree} -- In this experiment, we
compare the performances of IBWJ using \bptree, \imtree\ and \dimtree.
For a more comprehensive comparison, we divide the process of finding
matching tuples into two steps: traversing the index tree for the
tuple with the lowest value, referred to as \textit{searching}, and
linearly checking tuples in leaf nodes, referred to as
\textit{scanning}. For each data structure, we measure the costs of the
different steps of performing IBWJ, including insert, delete,
search, scan, and merge. Figure~\ref{plot:HRR}b shows the results
for sliding window sizes of $2^{17}$ and $2^{23}$.

The merging overhead is almost identical for both \imtree\ and
\dimtree, and it constitutes 7\% and 11\% of the total processing for
$2^{17}$ and $2^{23}$ windows, respectively. Regarding the tuple
insertion performance, \dimtree\ and \imtree\ perform nearly
identically, and they are 1.5 and 2.6 times faster than \bptree{} for
$2^{17}$ and $2^{23}$ windows, respectively. For the smaller window
size ($2^{17}$), searching in \bptree{} is 75\% faster than searching
in \imtree\ and \dimtree. However, for the larger window size
($2^{23}$), the search performances corresponding to \dimtree\ and
\bptree{} are nearly identical, and both are slightly faster than
\imtree.

Figure~\ref{plot:SRS}a presents the throughput of single-threaded IBWJ using \bptree,
\imtree, and \dimtree\ for varying window sizes. We observe that
employing \dimtree\ and \bptree{} results in the best and the worst
performances, respectively. Considering IBWJ using \bptree{} as the
baseline, average improvements in system performance of 50\% and 63\%
in magnitude are achieved by employing \imtree\ and \dimtree,
respectively.

\textbf{Match rate ($\sigma_s$)} -- Figure~\ref{plot:SRS}b shows the
throughputs of four different implementations of IBWJ for the window
size of $2^{20}$ and match rates varying from $2^{-4}$ to $2^{10}$.
These implementations are three single-threaded IBWJ using \bptree,
\imtree\ and \dimtree\ and one multi-threaded IBWJ using \dimtree.
The join performance varies negligibly for the match rates between
$2^{-4}$ and $2^4$, which indicates that the join performance in this
range is bounded by index traversing rather than the linear leaf node
scans.  As the match rate increases beyond $2^4$, the join performance
noticeably decreases for all implementations.  This result implies that for
higher match rates, i.e., $2^5 \leq \sigma_s \leq 2^{10}$, the join
performance is bounded by system memory bandwidth due to extensive
leaf node scans.
Consequently, multithreading loses its advantage for IBWJ with high
selectivities, and its performance becomes closer to that of the
single-threaded implementations.  Additionally, the result indicates that
single-threaded IBWJ using \imtree\ and \dimtree\ for join with high selectivity results in
better performance than using \bptree, which is because of the more efficient leaf
node scan in immutable \bptree{} ($T_S$) than in regular \bptree.

\textbf{Task size} -- In this experiment, we study the influence of
the task size on our parallel window join algorithm.  Increasing the
task size decreases the overhead of task acquisition while
simultaneously increasing the system latency (task processing time).
Figures~\ref{plot:SRS}c and~\ref{plot:SRS}d illustrate the performance of IBWJ using
\dimtree\ over different task sizes ranging from 1 to 10 in terms of
throughput and latency, respectively. Increasing the task size to four
steadily improves the performance, which suggests that very small task
sizes lead to significant task scheduling overhead. For task sizes
from five to eight, a minor improvement is achieved, and for task
sizes larger than eight, the performance does not significantly vary.
The evaluation results shown in Figure~\ref{plot:SRS}c indicate that
the task size greatly influences the system latency: increasing the task
size leads to higher latencies. Additionally, we observe that the
latency of parallel IBWJ is higher for larger sliding windows. As the
window size increases, the \dimtree\ merge becomes more costly because
it leads to longer linear window scans during nonblocking merge
and consequently causes higher latency. In the remainder of the
evaluation, we use tasks of size eight.

\begin{figure*}
	\begin{minipage}[t]{0.3\textwidth}
		\captionsetup{labelformat=empty,labelsep=none}
		\centering
		\includepdfcrop{\columnwidth}{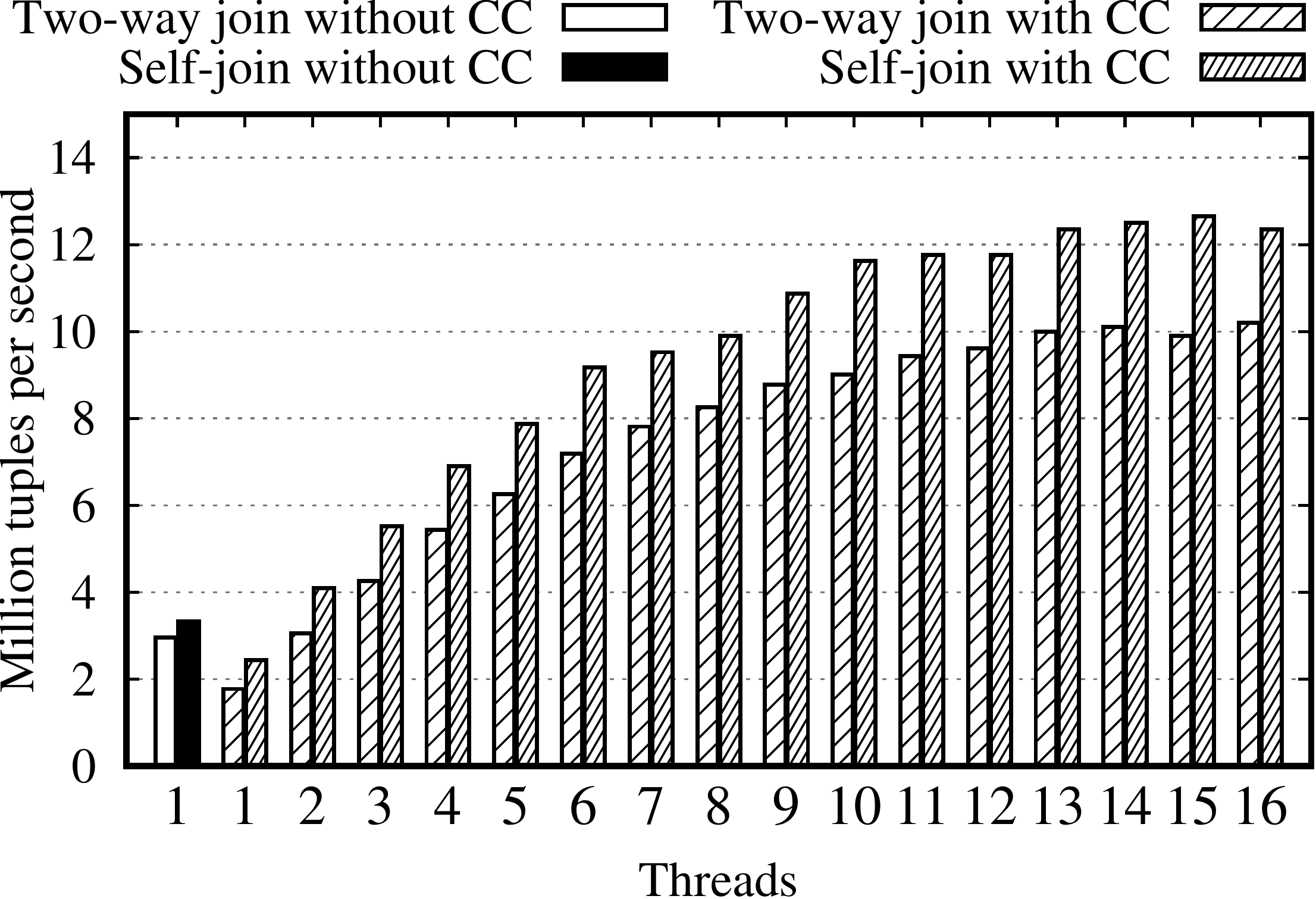}
		{\vspace*{-0.95cm} \caption{\hspace*{-4.1cm}\footnotesize(a)}}
	\end{minipage}
	\hfill
	\begin{minipage}[t]{0.3\textwidth}
		\captionsetup{labelformat=empty,labelsep=none}
		\centering
		\includepdfcrop{\columnwidth}{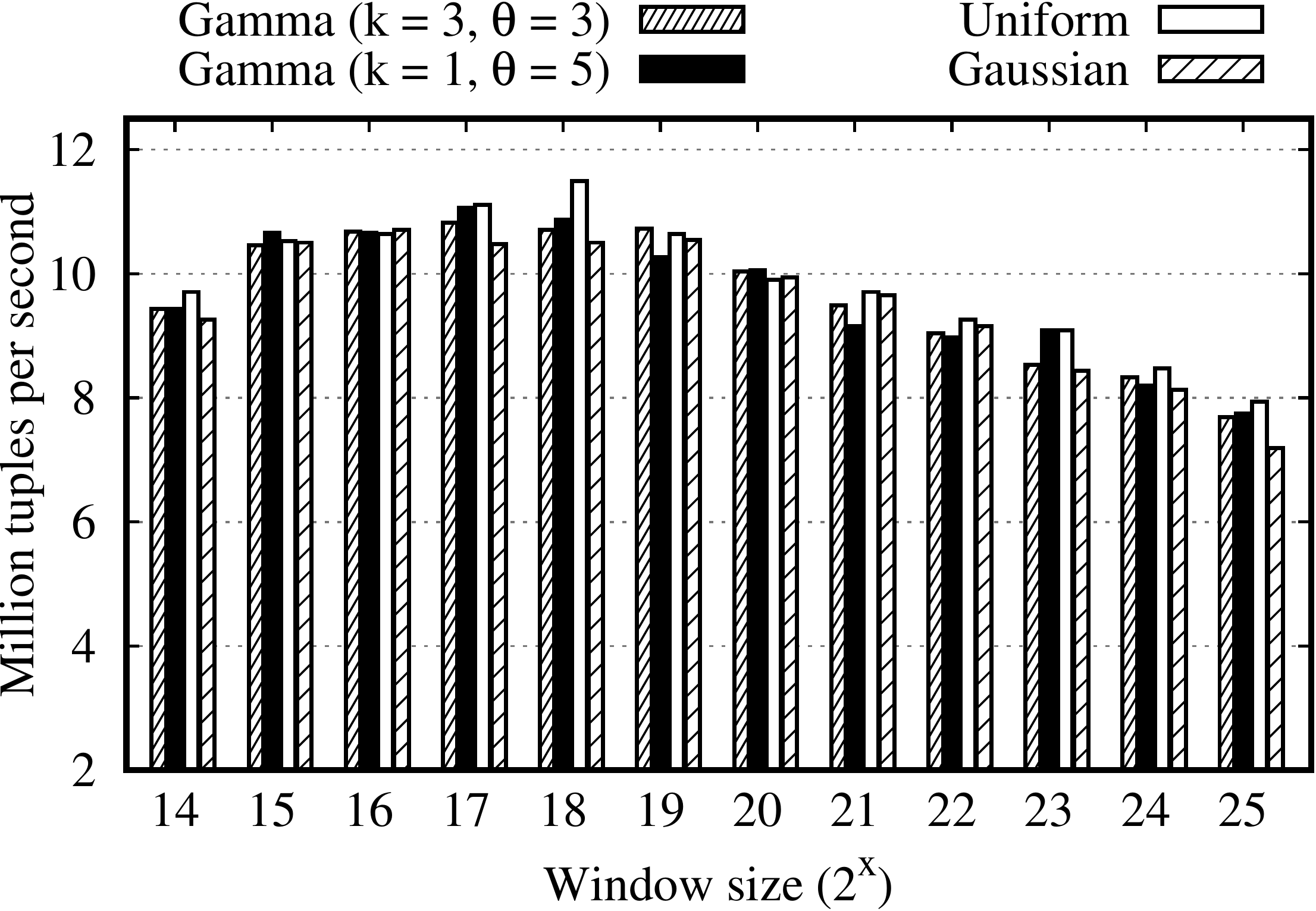}
		{\vspace*{-0.95cm} \caption{\hspace*{-4.1cm}\footnotesize(b)}}
	\end{minipage}
	\hfill
	\begin{minipage}[t]{0.38\textwidth}
		\captionsetup{labelformat=empty,labelsep=none}
		\centering
		\includepdfcrop{\columnwidth}{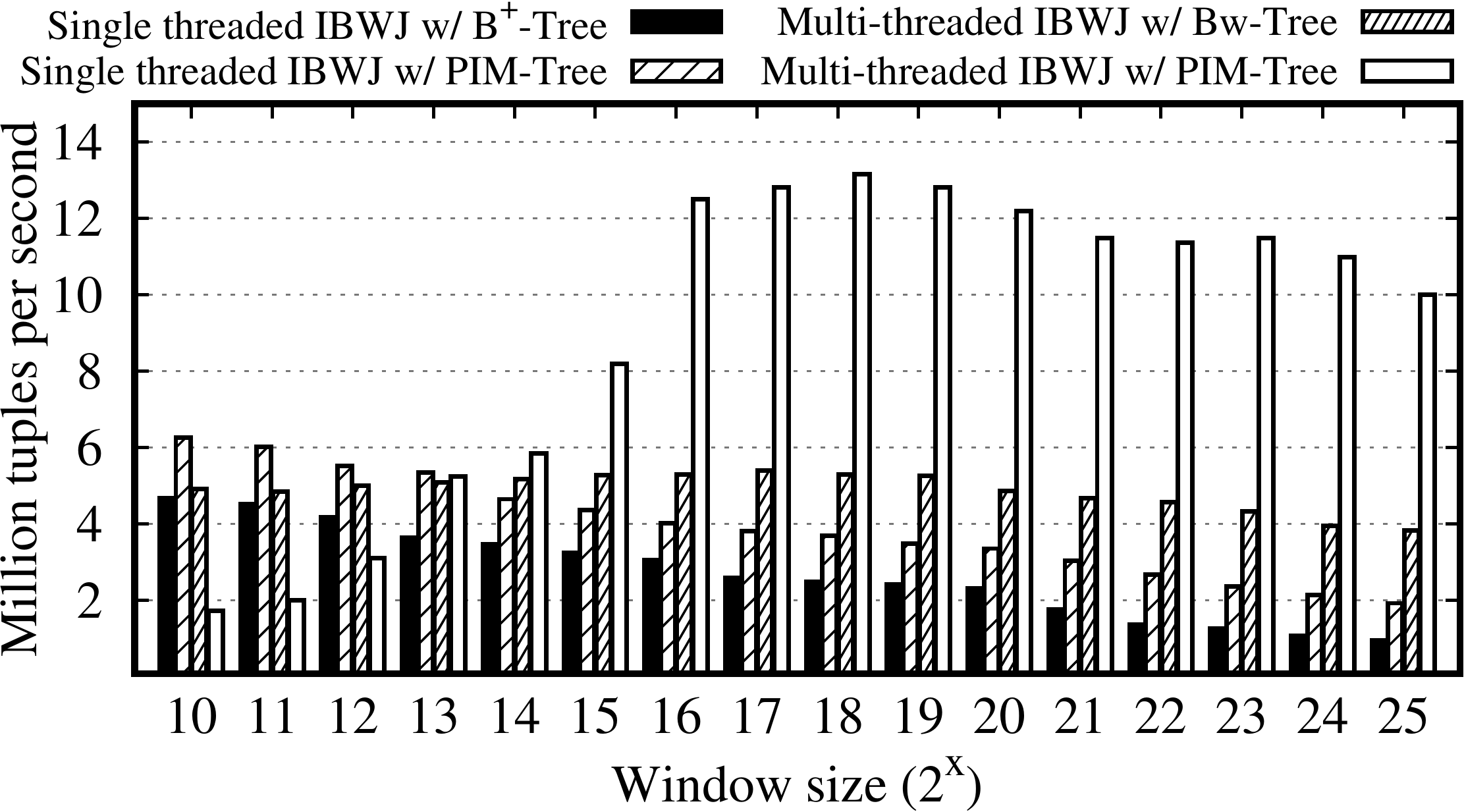}
		{\vspace*{-0.95cm} \caption{\hspace*{-5.5cm}\footnotesize(c)}}
	\end{minipage}
	\vspace{-0.1cm}
	\setcounter{figure}{11}
	\caption{
		a) Comparison of parallel IBWJ using \dimtree\ utilizing
		varying number of threads against the single-threaded implementation
		without concurrency control (CC) ($w = 2^{20}$).
		b) Evaluation of parallel IBWJ using \dimtree\ for different tuple value distributions.
		c) Performance comparison of single-threaded and multithreaded index-based self-join.
	}
	\label{plot:TLW}
	\vspace{-0.55cm}
\end{figure*}

\textbf{Memory consumption} -- Figure~\ref{plot:STWB}a compares the
memory space required for different components of \dimtree\ and
\bptree{} storing varying numbers of elements.  Each element is a pair
of 4 bytes for key and 4 bytes for sliding window reference.  The
storage required for \dimtree\ consists of the search-efficient
component ($T_S$), the insert-efficient component ($T_I$), and a
buffer that is required during nonblocking merge.  For this
experiment, the merge ratio is set to one such that $T_I$ is at the
largest possible size.  The results reveal that the space required for
\dimtree\ is almost double the space required for \bptree, regardless of
window size.

\textbf{Asymmetric sliding windows} -- In contrast to our other
experiments, where we considered the lengths of both sliding windows
to be equal, here we set different sizes for the sliding windows of streams
$R$ ($w_r$) and $S$ ($w_s$), and we examine whether asymmetric window
sizes impact the performance of IBWJ.
Figure~\ref{plot:STWB}c presents the throughout of parallel \IBWJ\
using \dimtree\ for various combinations of $w_r$ and $w_s$.  In
general, the system performance for asymmetric window sizes follows
the same pattern as for ones using symmetric window sizes.
Considering a fixed window size for one stream, increasing the size of
the other window decreases system performance, although the magnitude
of the performance decrease is less than when both sliding window sizes
increase.

\textbf{Asymmetric tuple distribution} -- Here, we examine 
the impact of asymmetric input rates on the performance of parallel
window join using \dimtree.  An asymmetric input rate skews the
distribution of search and insert operations among the two indexing
data structures (i.e., inserts in one index vs. search in the other
index). Thus, there are more insert operations in the index of the
stream with the higher input rate and, at the same time, more search
operations in its companion index.  Figure~\ref{plot:STWB}b illustrates
the throughput of parallel window join using \dimtree\ for various
input rates and window sizes.  The results show that the throughput
increases marginally as the input rate skew increases.  This indicates
that the parallel window join algorithm is resilient against input
rate fluctuations.

\textbf{Memory Bandwidth} -- The purpose of this experiment is to examine the 
impact of the system memory bandwidth on the performance of the
parallel window join. The maximum system memory bandwidth is 43~GB/s.
Figure~\ref{plot:STWB}d illustrates the effective system memory
bandwidth of parallel window join using \dimtree\ ($w =
2^{20}$).  The results indicate that 22\% of the total
memory bandwidth is due to store operations for the case of the 
single-threaded execution. This ratio decreases to 16\% as we increase the
number of threads.  The higher ratio of load operations for
multithreaded executions is the result of a less efficient sliding
window search during multithreaded window join compared with the
single-threaded execution.  The parallel window lookup consists of two
parts: (1) a linear scan between edge-tuple and sliding window head
and (2) querying the index for the remaining window portion.  In the
case of a single-threaded window join, there is no need for the linear
scan since the entire window content is always indexed; thus, the
sliding window lookup is at its most efficient operating point.  As we
increase the number of threads, the number of active tasks in the
system also increases. Consequently, the gap between edge-tuple
and sliding window head increases, which causes more costly linear
scans and consequently a less efficient window lookup. 

\textbf{Scalability} -- The objectives of this experiment are to first
study the overhead of the concurrency control mechanisms and to then
examine the scalability of our join algorithm using multiple
threads. Figure~\ref{plot:TLW}a compares the resulting throughputs
corresponding to self-join and two-way join using \dimtree\ under a varying number of
threads against the single-threaded implementation without concurrency
control (CC).

The results show that enforcing CC causes performance
degradations of nearly 40\% and 26\% for two-way join and self-join,
respectively, mainly as a result of the locking overhead. As we
increase the number of threads from one to eight, the performance of
both two-way join and self-join increase to 4.6 and 4 times of 
the single-threaded implementation with CC, respectively.
Moreover, the results reveal that enabling hyper-threading (16 threads) 
increases the throughput by 24\%, and the mentioned improvements increase 
to 5.7 and 5, respectively.

\textbf{Multithreading efficiency} -- In this experiment, we study
the efficiencies of our multithreading approach and nonblocking merge,
and we also compare \dimtree\ to the state-of-the-art parallel
indexing tree, Bw-tree.  Figure~\ref{plot:DSP}c shows the throughput
performances of five different implementations of the two-way IBWJ: (1)
single-threaded IBWJ using \bptree, (2) single-threaded IBWJ using
\dimtree, (3) parallel IBWJ using Bw-tree, (4) parallel IBWJ using
\dimtree, and (5) parallel IBWJ using \dimtree\ with blocking merge.

\begin{figure*}
	\begin{minipage}[t]{0.29\textwidth}
		\captionsetup{labelformat=empty,labelsep=none}
		\centering
		\includepdfcrop{\columnwidth}{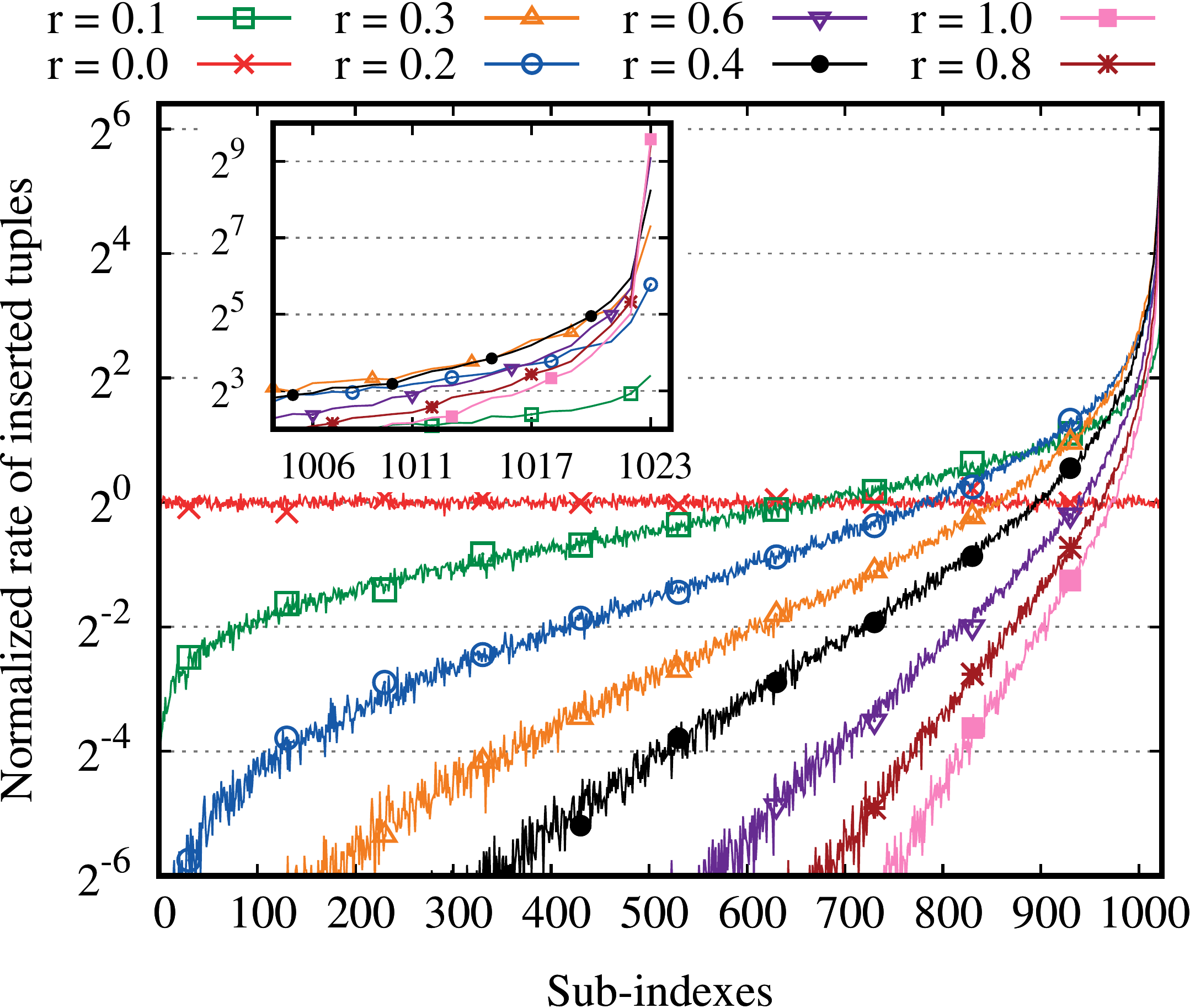}
		{\vspace*{-0.95cm} \caption{\hspace*{-4.3cm}\footnotesize(a)}}
	\end{minipage}
	\hfill
	\begin{minipage}[t]{0.28\textwidth}
		\captionsetup{labelformat=empty,labelsep=none}
		\centering
		\includepdfcrop{\columnwidth}{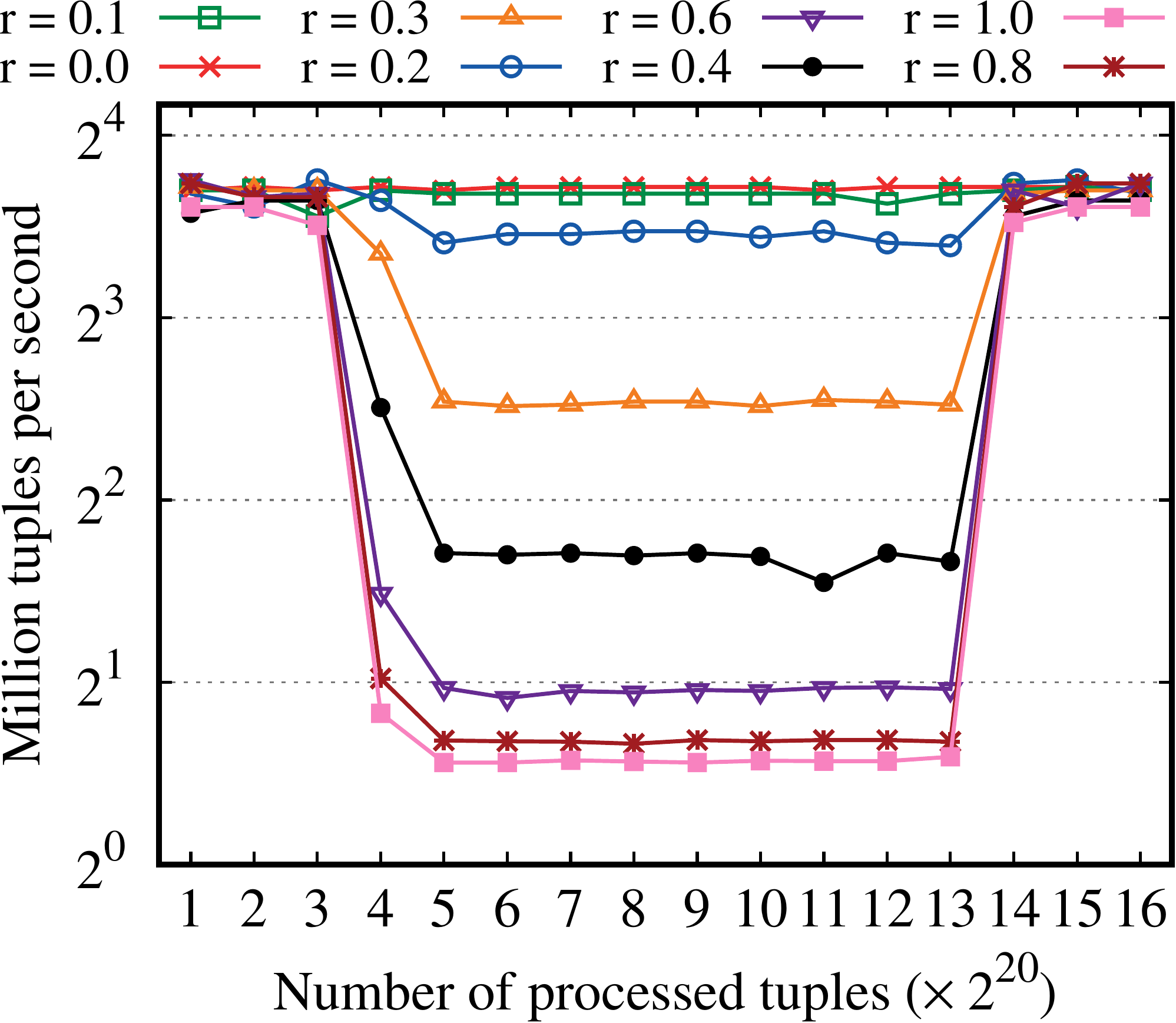}
		{\vspace*{-0.95cm} \caption{\hspace*{-4.1cm}\footnotesize(b)}}
	\end{minipage}
	\hfill
	\begin{minipage}[t]{0.4\textwidth}
		\captionsetup{labelformat=empty,labelsep=none}
		\centering
		\includepdfcrop{\columnwidth}{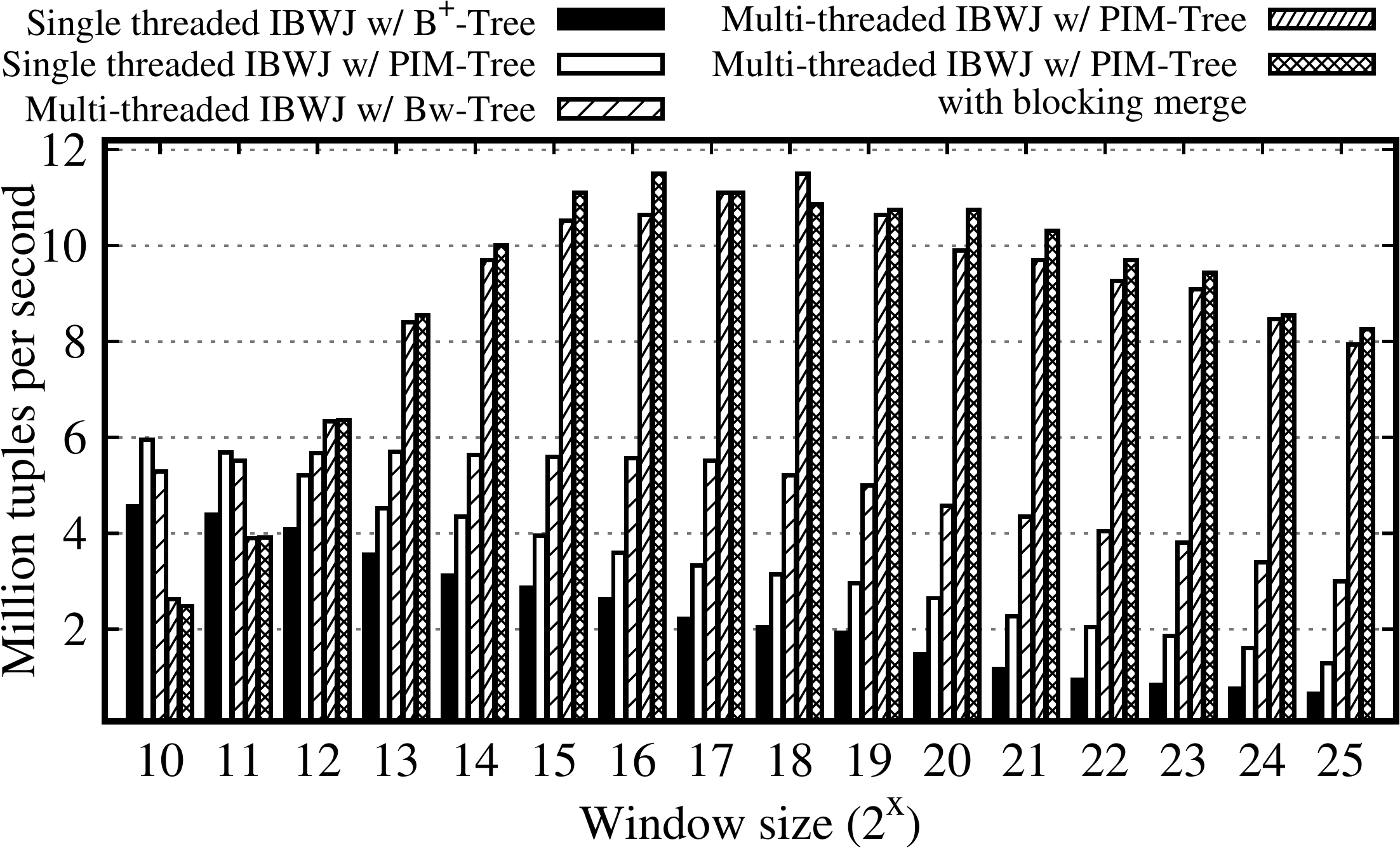}
		{\vspace*{-0.95cm} \caption{\hspace*{-6.5cm}\footnotesize(c)}}
	\end{minipage}
	\vspace{-0.2cm}
	\setcounter{figure}{12}
	\caption{
		a) Distribution of inserts among subindexes during drifting Gaussian distributions.
		b) Evaluation of multithreaded index-based self-join using \dimtree\ for shifting Gaussian distributions.
		c) Throughput comparison of single-threaded and multithreaded two-way join.
	}
	\label{plot:DSP}
	\vspace{-0.6cm}
\end{figure*}

The results of parallel IBWJ using \dimtree\ show that using 
blocking and non-blocking merge techniques result in similar performances 
while blocking merge is slightly faster than the non-blocking one which is 
because of its less complicated mechanism used to perform blocking merge operations. 
Moreover, the results reveal that our parallel approach is effective 
for window sizes larger than $2^{14}$. For the smaller evaluated 
window sizes ($2^{10}$ to $2^{13}$), merge operations occur very often 
which leads to frequent linear window scans during merge operations
and thus the system performance declines.
For window sizes between $2^{15}$ and $2^{25}$, our parallel IBWJ
using \dimtree\ results in on average 7.5 and 3.7 times higher throughput than 
the single-threaded IBWJ using \bptree{} and \dimtree, respectively.
The biggest improvement is achieved for the largest evaluated window 
size ($2^{25}$) that resulted in improvement increases of  
12 and 5.3 times, respectively.
The evaluation results of IBWJ using Bw-tree reveal that Bw-tree is 
also not effective for the smaller evaluated  window sizes ($2^{10}$ to $2^{13}$) which is 
because of the high conflict between threads during index operations.
For windows sizes between $2^{14}$ and $2^{25}$, parallel IBWJ using Bw-tree
results in 1.8 times higher throughput than our single threaded 
IBWJ using \dimtree , on average.
For the same range of window sizes, our parallel IBWJ using \dimtree\ 
outperforms the Bw-tree based implementation by a factor of 2.2 on average.
Although our \dimtree\ achieves better performance than Bw-tree, we do
not aim to challenge Bw-Tree in this work since Bw-tree is designed as
a generic parallel indexing tree that is highly efficient for OLTP
systems where the majority of queries are read accesses (more than
80\%~\cite{Krueger:2011}), whereas our design is specifically tuned
for highly dynamic systems such as data stream indexing with a
significantly higher rate of data modification.

Figure~\ref{plot:TLW}c presents the performance comparison of the
parallel and single-threaded IBWJ implementations for self-join. 
Similar to the experiment on two-way window joins, parallel self-join using 
\dimtree\ is not effective for the smaller evaluated window sizes ($2^{10}$ to $2^{15}$).
For window sizes between $2^{16}$ to $2^{25}$, parallel self-join using 
\dimtree\ achieves 7 and 4 times higher throughput than the single threaded 
self-join using \bptree{} and \dimtree, respectively.

\textbf{Impact of skewed data} -- We now study the impact of the tuple
value distribution on the performance of parallel \IBWJ\ using
\dimtree\ in two experiments.  First, we examine \IBWJ\ using three
differently skewed distributions, including a Gaussian distribution
($\mu$ = 0.5, $\sigma$ = 0.125) and two differently parameterized
Gamma distributions (k = 3, $\theta$ = 3 and k = 1, $\theta$ = 5), and
we compare them with the result of using a uniform distribution.  For
each evaluation, we adjust the band join predicate to keep the average
match rate equal to two. Figure~\ref{plot:TLW}b presents the
evaluation results ($w = 2^{20}$). The uniform distribution of the
join attributes always results in the highest throughput, although the
differences are not significant. On average, the resulting throughput
of \IBWJ\ using \dimtree\ for uniformly distributed join attributes is
between 2\% and 4\% higher than for Gaussian and Gamma distributions,
respectively.

In the second experiment, we examine the impact of a dynamic tuple
value distribution on the performance of \IBWJ\ using \dimtree.  In
the case of a fixed tuple value distribution, the insert operations
are spread uniformly across subindexes, even though the tuple value
distribution is skewed.  The reason is that \bptree{} nodes are
naturally adapting to the indexed values such that the subtrees of the
two inner nodes at the same depth have almost an equal number of
indexed values. Because $T_I$'s subindexes are adjusted according
to $T_S$'s inner nodes, the load among subindexes is uniformly
distributed regardless of the value distribution.  However, when the
distribution changes, the range assignment is no longer optimal and
causes skew in the insert operation among subindexes.

In contrast to the previous experiment where the distribution of values
was fixed, we now study \dimtree\ performance under a dynamic value
distribution, which results in a skewed distribution of inserts among
subindexes.  For this purpose, we create a tuple sequence in which
tuple values are generated based on a shifting Gaussian distribution,
and we then evaluate the performance of parallel index-based self-join
using \dimtree\ with this tuple sequence ($w = 2^{20}$).  The tuple
sequence consists of three phases.  In the first phase, the tuples are
generated according to the fixed Gaussian distribution
$\mathcal{N}(0.5, 0.125)$ ($\mu = 0.5, \sigma^2 = 0.125$).  During the
middle phase, the distribution of tuple values is linearly shifting
from $\mathcal{N}(0.5, 0.125)$ to $\mathcal{N}(r+0.5, 0.125)$, where
the constant value $r$ defines the speed of the distribution change;
thus, the larger $r$ is, the faster the mean value of the Gaussian
distribution shifts.  In the last phase, the tuples are generated
according to the Gaussian distribution $\mathcal{N}(r+0.5, 0.125)$.
We set the lengths of these three phases to 4M ($4\times2^{20}$), 10M
and 4M tuples, respectively.  $D_I$ is set to 4, which results in 1024
subindexes considering $f_{ib} = 32$ and $w = 2^{20}$.
Figure~\ref{plot:DSP}a illustrates the normalized distribution of
insert operations among $T_I$'s subindexes during distribution shifts
(second phase) for different values of $r$ ranging from 0 to 1.
It follows that inserts are spread among subindexes equally when the
tuple value distribution is fixed ($r = 0$), and as $r$ increases, the
distribution of inserts becomes more skewed.  For the highest value of
$r$ ($r = 1$), the insert distribution is highly skewed such that 77\%
of all inserts are assigned to a single subindex, and there are almost
no inserts assigned to the other 70\% of subindexes.
Figure~\ref{plot:DSP}b presents the evaluation results for multiple
values of $r$ ranging from 0 to 1. The join performance during the
distribution change depends on how fast the distribution shifts: slow,
moderate or fast.  During slow distribution shifts ($r = 0.1, 0.2$),
there is almost no decrease in the stream join performance, which indicates
that \dimtree\ is able to gracefully tolerate slow changes in the
tuple value distribution.  For moderate distribution shifts ($r = 0.4,
0.6$), the system performance decreases to 35\% on average, which is due to
high partition locking congestion.  The lowest performance results
from fast distribution shifts ($r = 0.8, 1.0$), where the performance
decreases to 16\%. The join performances for $r = 0.8$ and $r = 1.0$ are
nearly identical, which indicates that partition locking congestion is close
to its peak.  Additionally, the results imply that regardless of how fast the
distribution shifts during the second phase, as the distribution
becomes stationary again in the third phase, partitions in
\dimtree\ are adjusted accordingly, and stream join performance
recovers.

\vspace*{-0.25cm}
\section{Related Work} \label{Section:Related_Work}
 
Work related to our approach can be classified as follows: Tree indexing, 
Parallel \bptree, sliding window indexing, and parallel window join. 
We review these categories in this section.
 
\textbf{Tree indexing} -- Due to the advances in main memory
technology, many databases are currently able to store indexing
information in main memory and eliminate the expensive I/O overhead
arising from storage to disks. Consequently, a large body of work has
explored tree-based in-memory indexing.  \bptree{} is a popular
modification of B-Tree, which provides better range query
performance~\cite{elmasri2008fundamentals,
  Bayer:1970:OML:1734663.1734671}.  T-Tree is a balanced binary tree
specifically designed to index data for in-memory
databases~\cite{Lehman:1986:SIS:645913.671312}.  Although B-Tree was originally
designed as a disk-stored indexing data structure, when properly
configured, B-Tree outperforms T-Tree while enforcing concurrency control~\cite{lu2000t}.  
Rao et al.~\cite{Rao:2000:MBT:342009.335449}
extended CSS-Tree~\cite{Rao:1999:CCI:645925.671362} to the
cache-sensitive \bptree{} (CSB\textsuperscript{+}-Trees), which
supports update operations, although \bptree{} outperforms
CSB\textsuperscript{+}-Tree in applications that require incremental
updates.
{\color{black} LSM-Tree is a multi-level data structure which stores 
each component on a different storage medium~\cite{Neil1996}.
All new tuples are inserted into the lowest level component and 
whenever the size of each component exceeds a predefined threshold,
a part of the component merges into the higher level one. 
LSM-Tree improves system performance in write-intensive 
applications using delta merging, however, it does not provide a 
solution for multi-threaded indexing.}
Adaptive radix tree (ART) is a high-speed in-memory indexing
data structure that exhibits a better memory footprint than a
conventional radix tree and better point query performance than
\bptree ~\cite{leis2013adaptive}.  However, \bptree{} outperforms ART
in executing range queries~\cite{7113370}.  
We use \bptree{} as the
baseline in to evaluate our \dimtree{} since it supports incremental updates and
range queries better than other approaches.

 \textbf{Parallel \bptree} -- As we enter the multicore era,
 concurrent in-memory indexing is essential for databases to exploit
 the computational resources of a modern server.  Bayer and
 Schkolnick~\cite{Bayer1977} proposed a concurrency control method for
 supporting concurrent access in B-Trees based on \textit{coupled
   latching}, in which threads are required to obtain the associated
 latch for each index node in every tree traversal.  B-link is a
 \bptree{} with a relaxed structure that requires fewer latch
 acquisitions to handle concurrent
 operations~\cite{lehman1981efficient}. However, concurrency control
 methods based on coupled latching are known to suffer from high
 latching overhead and poor scalability for in-memory
 systems~\cite{cha2001cache}.
 
 PALM is a parallel latch-free \bptree{} based on bulk synchronous
 processing~\cite{sewall2011palm}. To avoid potential conflicts, PALM
 sorts all queries in bulk at each level of the tree traversal to
 guarantee that the operations on each node are assigned to a single
 thread. Although this approach is scalable and handles data
 distribution changes, it requires processing queries in large groups
 (the authors suggest groups of 8,000 queries to achieve a reasonable
 scale up). This requirement negatively affects the system response
 time, which is an important criterion for data stream processing
 applications. Although PALM could excel at supporting batch-oriented
 processing engines, such as Apache Spark~\cite{karau2015learning}, it
 does not meet the requirements of real-time, event-by-event stream
 processing frameworks, such as Apache Storm~\cite{toshniwal2014storm}
 or the sliding window indexing considered in this paper.  In contrast
 to query batching, Pandis et
 al.~\cite{Pandis:2011:PPL:2021017.2021019} proposed physiological
 partitioning (PLP) of indexing data structures on the basis of a
 multirooted \bptree. Using PLP, the index structure is partitioned
 into disjoint intervals, and each interval is assigned exclusively to
 a single thread. 
Although, both PLP and our \dimtree\ employ range partitioning
 to provide concurrent indexing operations, there the task distribution method 
 and concurrency control mechanism are different between these two methods.
PLP is a latch-free partitioning technique where only one dedicated thread accesses 
each sub-index, while, sub-indexes in \dimtree\ are uniformly accessible 
by all operating threads and concurrent accesses on each sub-index 
are synchronized using locks.
In PLP, a partition manager assigns queries to threads and
ensures that all work given to a thread involves only data that it
owns.  If a query execution requires accessing multiple intervals,
then the partition manger breaks the query into multiple subtasks
and assembles the subresults to finish the query.  Although the
overhead of the partition manager is negligible for transaction
processing in a database, in the case of processing a single
streaming tuple, this overhead causes a significant performance
decrease.  
 
 Rastogi et al.~\cite{Rastogi:1997:LPV:645923.671017} introduced a
 multiversion concurrency control and recovery method in which update
 transactions create a new version of a node to avoid conflicting with
 lookup transactions rather than using locks.  Since building a new
 node is required for every node modification, this method suffers
 from high node creation and garbage collection overhead for use cases
 with many update operations.  Optimistic latch-free index traversal
 (OLFIT) is based on node versioning to ensure data consistency during
 tree traversal, but it does not require the creation of a new
 physical node to avoid conflicts~\cite{cha2001cache}.  In this
 approach, each node is assigned with a version number and a lock.  To
 update a node, it is necessary to obtain the associated lock and
 increment the version before releasing the lock. Node lookups are
 performed in an optimistic fashion. The reader thread compares the
 node version before and after the read operation; if the versions are
 identical and the node lock is not obtained, the read operation is
 successful. Otherwise, it repeats the entire operation. However, this
 approach does not provide an efficient node-merging algorithm, which
 is critical for preserving an efficient tree structure when the data
 distribution of tuples in the sliding window changes.  Bw-Tree is
 another optimistic latch-free parallel indexing data structure that
 utilizes atomic compare and swap (CAS) operations to avoid race
 conditions~\cite{levandoski2013bw}.  Bw-Tree is designed to
 simultaneously exploit the computational power of multicore
 processors and the memory bandwidth of underlying storage, such as
 flash memories.
  Among the aforementioned approaches, Bw-Tree is the best choice for
 use cases with frequent incremental updates, which is why we use it
 as the baseline for our multithreaded indexing approach using \dimtree{}.

\textbf{Sliding window indexing} -- A class of related work proposes
accelerating window queries by utilizing an index. Golab et
al.~\cite{golab2004indexing} evaluated different sliding window
indexing approaches, such as hash-based and tree-based indexing, for
different types of stream operators. Kang et
al.~\cite{kang2003evaluating} evaluated the performance of an
asymmetric sliding stream join using different algorithms, such as
nested loop join, hash-based join, and index-based join.  Lin et
al.~\cite{Lin:2015:SDS:2723372.2746485} and Ya-xin et
al.~\cite{ya2006indexed} proposed the \textit{chained index} to
accelerate index-based stream join utilizing coarse-grained tuple
disposal.  However, all of these approaches considered only
single-threaded sliding window indexing, thus avoiding concurrency
issues resulting from parallel update processing, which is central to
the focus of our work.

 \textbf{Parallel window join} -- Window join processing has received
 considerable attention in recent years due to its computational
 complexity and importance in various data management
 applications. Several approaches explore parallel window join
 processing.  Cell-join is a parallel stream join operator designed to
 exploit the computing power of the cell
 processor~\cite{gedik2009celljoin}. Handshake join is a scalable
 stream join that propagates stream tuples along a linear chain of
 cores in opposing directions~\cite{teubner2011soccer}.  Roy et
 al.~\cite{roy2014low} enhanced the handshake join by proposing a
 fast-forward tuple propagation to attain lower latency. SplitJoin is
 based on a top-down data flow model that splits the join operation
 into independent store and process steps to reduce the dependency
 among processing units~\cite{196251}. Lin et
 al.~\cite{Lin:2015:SDS:2723372.2746485} proposed a real-time and
 scalable join model for a computing cluster by organizing processing
 units into a bipartite graph to reduce memory requirements and the
 dependency among processing units.
 
 All these approaches are based on context-insensitive window 
 partitioning. Although these methods are effective for using nested 
 loop join or for memory-bounded joins with high selectivity, 
 context-insensitive window partitioning causes redundant index 
 operations using IBWJ, which limits the system efficiency. 

\vspace*{-0.29cm}
\section{Conclusions}

In this paper, we presented a novel indexing structure called
\dimtree\ to address the challenges of concurrent sliding window
indexing. Stream join using \dimtree\ outperforms the well-known
indexing data structure \bptree{} by a margin of 120\%. Moreover, we
introduced a concurrent stream join approach based on \dimtree, which
is, to the best of our knowledge, one of the first parallel
index-based stream join algorithms. Our concurrent solution improved
the performance of IBWJ up to 5.5 times when using an octa-core (16
threads) processor.

The directions for our future work are twofold: (1) developing a
distributed stream band join and (2) extending \dimtree\ to support
the indexing of multidimensional data. In this paper, we focused on
parallelism within a uniform shared memory architecture.  A further
challenge, but altogether a different problem, is to develop a
parallel IBWJ algorithm for nonuniform memory
access (NUMA) architectures, which requires addressing two main
concerns. First, a range partitioning technique that distributes a
workload uniformly among operating cores is needed. Second, a
repartitioning scheme that alleviates the overhead of data transfer
between memory nodes in a NUMA system is needed.  Although
\dimtree\ discretizes tuples into disjoint intervals, these intervals
are adjusted only according to the number of input tuples, which does
not necessarily lead to a uniform distribution of the workload across
all intervals.  In the solution for uniform memory access presented in
this paper, we used a shared work queue to distribute the workload
among operating cores; in a NUMA system, however, an efficient range
partitioning scheme is needed, which considers the numbers of both
input and output tuples of each interval.  Such a partitioning is not
needed for the approach presented in this paper.  Moreover, with
respect to supporting multidimensional data, \dimtree\ is designed to
index one-dimensional data.  Multidimensional indexing is a vital
requirement for many applications, specifically those that utilize
spatiotemporal datasets. Thus, a further direction is the design of a
multidimensional \dimtree.

\vspace*{-0.cm}

{
	\setstretch{1.0}
	\bibliographystyle{IEEEtran}
	\bibliography{PIMTREE_tech.bib}  

\begin{thebibliography}{10}
\providecommand{\url}[1]{#1}
\csname url@samestyle\endcsname
\providecommand{\newblock}{\relax}
\providecommand{\bibinfo}[2]{#2}
\providecommand{\BIBentrySTDinterwordspacing}{\spaceskip=0pt\relax}
\providecommand{\BIBentryALTinterwordstretchfactor}{4}
\providecommand{\BIBentryALTinterwordspacing}{\spaceskip=\fontdimen2\font plus
\BIBentryALTinterwordstretchfactor\fontdimen3\font minus
  \fontdimen4\font\relax}
\providecommand{\BIBforeignlanguage}[2]{{%
\expandafter\ifx\csname l@#1\endcsname\relax
\typeout{** WARNING: IEEEtran.bst: No hyphenation pattern has been}%
\typeout{** loaded for the language `#1'. Using the pattern for}%
\typeout{** the default language instead.}%
\else
\language=\csname l@#1\endcsname
\fi
#2}}
\providecommand{\BIBdecl}{\relax}
\BIBdecl

\bibitem{gao2015parallel}
X.~Gao, E.~Ferrara, and J.~Qiu, ``Parallel clustering of high-dimensional
  social media data streams,'' in \emph{CCGrid}, 2015, pp. 323--332.

\bibitem{zhang2008detecting}
L.~Zhang and Y.~Guan, ``Detecting click fraud in pay-per-click streams of
  online advertising networks,'' in \emph{ICDCS}, 2008, pp. 77--84.

\bibitem{montana2008data}
G.~Montana, K.~Triantafyllopoulos, and T.~Tsagaris, ``Data stream mining for
  market-neutral algorithmic trading,'' in \emph{Proceedings of the 2008 ACM
  symposium on Applied computing}, pp. 966--970.

\bibitem{stonebraker20058}
M.~Stonebraker, U.~{\c{C}}etintemel, and S.~Zdonik, ``The 8 requirements of
  real-time stream processing,'' \emph{SIGMOD}, pp. 42--47, 2005.

\bibitem{publication-15241}
D.~Dell'~Aglio, E.~Della~Valle, F.~van Harmelen, and A.~Bernstein, ``Stream
  reasoning: A survey and outlook : A summary of ten years of research and a
  vision for the next decade,'' \emph{Data Science}, pp. 59--83, 2017.

\bibitem{babu2001continuous}
S.~Babu and J.~Widom, ``Continuous queries over data streams,'' \emph{ACM
  Sigmod Record}, pp. 109--120, 2001.

\bibitem{golab2004indexing}
L.~Golab, S.~Garg, and M.~T. {\"O}zsu, ``On indexing sliding windows over
  online data streams,'' in \emph{International Conference on Extending
  Database Technology}, 2004, pp. 712--729.

\bibitem{zhang2015memory}
H.~Zhang, G.~Chen, B.~C. Ooi, K.-L. Tan, and M.~Zhang, ``In-memory big data
  management and processing: A survey,'' \emph{IEEE Transactions on Knowledge
  and Data Engineering}, pp. 1920--1948, 2015.

\bibitem{gepner2006multi}
P.~Gepner and M.~F. Kowalik, ``Multi-core processors: New way to achieve high
  system performance,'' in \emph{PARELEC}, 2006, pp. 9--13.

\bibitem{trill}
B.~Chandramouli, J.~Goldstein, M.~Barnett \emph{et~al.}, ``Trill: A
  high-performance incremental query processor for diverse analytics,''
  \emph{VLDB}, pp. 401--412, 2014.

\bibitem{streambox}
H.~Miao, H.~Park, M.~Jeon, G.~Pekhimenko \emph{et~al.}, ``Streambox: Modern
  stream processing on a multicore machine,'' in \emph{USENIX ATC 17}, 2017,
  pp. 617--629.

\bibitem{saber}
A.~Koliousis, M.~Weidlich, R.~Castro~Fernandez \emph{et~al.}, ``Saber:
  Window-based hybrid stream processing for heterogeneous architectures,'' in
  \emph{SIGMOD}, 2016, pp. 555--569.

\bibitem{storm}
A.~Toshniwal, S.~Taneja, A.~Shukla, K.~Ramasamy \emph{et~al.},
  ``Storm@twitter,'' in \emph{SIGMOD}, 2014, pp. 147--156.

\bibitem{spark}
M.~Zaharia, R.~S. Xin, P.~Wendell, T.~Das \emph{et~al.}, ``Apache spark: A
  unified engine for big data processing,'' \emph{Communication of the ACM},
  pp. 56--65, 2016.

\bibitem{flink}
P.~Carbone, A.~Katsifodimos, S.~Ewen, V.~Markl \emph{et~al.}, ``Apache flink :
  Stream and batch processing in a single engine,'' \emph{Bulletin of the IEEE
  Computer Society Technical Committee on Data Engineering}, 2015.

\bibitem{Zeuch:2019:AES:3303753.3316441}
S.~Zeuch, B.~D. Monte, J.~Karimov \emph{et~al.}, ``Analyzing efficient stream
  processing on modern hardware,'' \emph{VLDB}, pp. 516--530, 2019.

\bibitem{levandoski2013bw}
J.~J. Levandoski, D.~B. Lomet, and S.~Sengupta, ``The bw-tree: A b-tree for new
  hardware platforms,'' in \emph{ICDE}, 2013, pp. 302--313.

\bibitem{gedik2009celljoin}
B.~Gedik, R.~R. Bordawekar, and S.~Y. Philip, ``Celljoin: a parallel stream
  join operator for the cell processor,'' \emph{The VLDB journal}, pp.
  501--519, 2009.

\bibitem{teubner2011soccer}
J.~Teubner and R.~Mueller, ``How soccer players would do stream joins,'' in
  \emph{Sigmod}, 2011, pp. 625--636.

\bibitem{Lin:2015:SDS:2723372.2746485}
Q.~Lin, B.~C. Ooi, Z.~Wang, and C.~Yu, ``Scalable distributed stream join
  processing,'' in \emph{SIGMOD}, 2015, pp. 811--825.

\bibitem{ya2006indexed}
Y.~Ya-xin, Y.~Xing-hua, Y.~Ge, and W.~Shan-shan, ``An indexed non-equijoin
  algorithm based on sliding windows over data streams,'' pp. 294--298, 2006.

\bibitem{roy2014low}
P.~Roy, J.~Teubner, and R.~Gemulla, ``Low-latency handshake join,''
  \emph{VLDB}, pp. 709--720, 2014.

\bibitem{196251}
M.~Najafi, M.~Sadoghi, and H.-A. Jacobsen, ``Splitjoin: A scalable, low-latency
  stream join architecture with adjustable ordering precision,'' in
  \emph{USENIX Annual Technical Conference}, 2016, pp. 493--505.

\bibitem{Neil1996}
P.~O'Neil, E.~Cheng, D.~Gawlick, and E.~O'Neil, ``The log-structured merge-tree
  (lsm-tree),'' \emph{Acta Informatica}, pp. 351--385, 1996.

\bibitem{lehman1981efficient}
P.~L. Lehman \emph{et~al.}, ``Efficient locking for concurrent operations on
  b-trees,'' \emph{ACM Transactions on Database Systems}, pp. 650--670, 1981.

\bibitem{bingmann2008stx}
T.~Bingmann, ``{STX} {B}+tree {C}++ template classes,'' \emph{URL
  http://panthema. net/2007/stx-btree}, 2008.

\bibitem{Rao:1999:CCI:645925.671362}
J.~Rao and K.~A. Ross, ``Cache conscious indexing for decision-support in main
  memory,'' in \emph{VLDB}, 1999, pp. 78--89.

\bibitem{Krueger:2011}
J.~Krueger, C.~Kim, M.~Grund \emph{et~al.}, ``Fast updates on read-optimized
  databases using multi-core cpus,'' \emph{VLDB}, pp. 61--72, 2011.

\bibitem{elmasri2008fundamentals}
R.~Elmasri, \emph{Fundamentals of database systems}.\hskip 1em plus 0.5em minus
  0.4em\relax Pearson Education India, 2008.

\bibitem{Bayer:1970:OML:1734663.1734671}
R.~Bayer and E.~McCreight, ``Organization and maintenance of large ordered
  indices,'' in \emph{SIGFIDET}, 1970, pp. 107--141.

\bibitem{Lehman:1986:SIS:645913.671312}
T.~J. Lehman and M.~J. Carey, ``A study of index structures for main memory
  database management systems,'' in \emph{VLDB}, 1986, pp. 294--303.

\bibitem{lu2000t}
H.~Lu, Y.~Y. Ng, and Z.~Tian, ``T-tree or {B}-tree: Main memory database index
  structure revisited,'' in \emph{ADC}, 2000, pp. 65--73.

\bibitem{Rao:2000:MBT:342009.335449}
J.~Rao and K.~A. Ross, ``Making {B}+-trees cache conscious in main memory,'' in
  \emph{SIGMOD}, 2000, pp. 475--486.

\bibitem{leis2013adaptive}
V.~Leis, A.~Kemper, and T.~Neumann, ``The adaptive radix tree: Artful indexing
  for main-memory databases,'' in \emph{ICDE}, 2013, pp. 38--49.

\bibitem{7113370}
V.~Alvarez, S.~Richter, X.~Chen, and J.~Dittrich, ``A comparison of adaptive
  radix trees and hash tables,'' in \emph{ICDE}, 2015, pp. 1227--1238.

\bibitem{Bayer1977}
R.~Bayer and M.~Schkolnick, ``Concurrency of operations on b-trees,''
  \emph{Acta Informatica}, pp. 1--21, 1977.

\bibitem{cha2001cache}
S.~K. Cha, S.~Hwang, K.~Kim \emph{et~al.}, ``Cache-conscious concurrency
  control of main-memory indexes on shared-memory multiprocessor systems,''
  \emph{VLDB}, pp. 181--190, 2001.

\bibitem{sewall2011palm}
J.~Sewall, J.~Chhugani, C.~Kim \emph{et~al.}, ``{PALM}: Parallel
  architecture-friendly latch-free modifications to {B}+ trees on many-core
  processors,'' \emph{VLDB}, pp. 795--806, 2011.

\bibitem{karau2015learning}
H.~Karau, A.~Konwinski, P.~Wendell, and M.~Zaharia, \emph{Learning spark:
  lightning-fast big data analysis}.\hskip 1em plus 0.5em minus 0.4em\relax "
  O'Reilly Media, Inc.", 2015.

\bibitem{toshniwal2014storm}
A.~Toshniwal, S.~Taneja, A.~Shukla, K.~Ramasamy \emph{et~al.}, ``Storm@
  twitter,'' in \emph{SIGMOD}, 2014, pp. 147--156.

\bibitem{Pandis:2011:PPL:2021017.2021019}
I.~Pandis, P.~T\"{o}z\"{u}n, R.~Johnson, and A.~Ailamaki, ``{PLP}: Page
  latch-free shared-everything {OLTP},'' \emph{VLDB}, pp. 610--621, 2011.

\bibitem{Rastogi:1997:LPV:645923.671017}
R.~Rastogi, S.~Seshadri, P.~Bohannon \emph{et~al.}, ``Logical and physical
  versioning in main memory databases,'' in \emph{VLDB}, 1997, pp. 86--95.

\bibitem{kang2003evaluating}
J.~Kang, J.~F. Naughton, and S.~D. Viglas, ``Evaluating window joins over
  unbounded streams,'' in \emph{Data Engineering, International Conference on},
  2003, pp. 341--352.

\end{thebibliography}
}

\appendix

\section{\dimtree\ Operations}

Here, we provide more detail about the implementation of immutable
\bptree\  and \dimtree.

\vspace*{-0.0cm}

\subsection{\dimtree\ insertion}

Algorithm~1 describes the process of inserting a new record into
\dimtree. The first part (Lines 1-7) is to search $T_S$ to depth $D_I$ 
to find the corresponding subindex in $T_I$ for the given
record.  For each inner node, the algorithm linearly searches its keys
(Lines 5-7) and then calculates the relative location of the next
inner node (Line 7).  

\vspace*{0.2cm}
\hspace*{-0.4cm}
\includegraphics[width=0.47\textwidth]{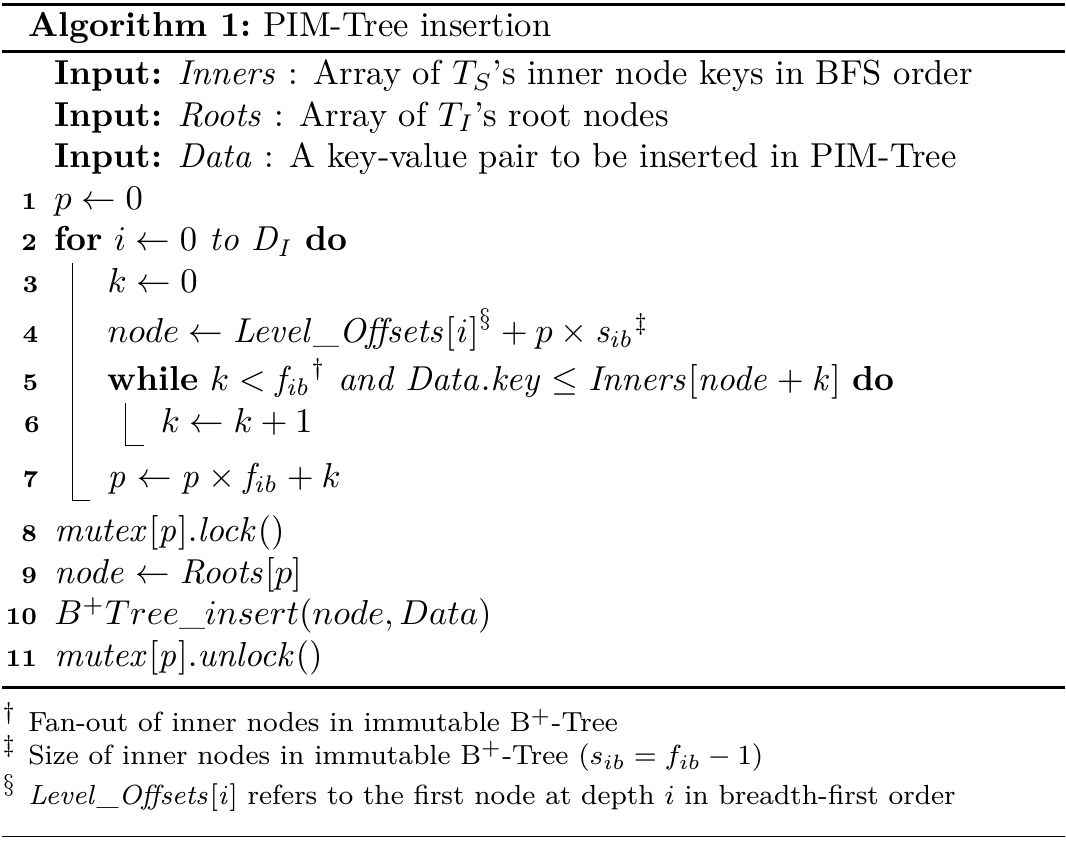}
\vspace*{0.2cm}

The second part (Lines 8-11) is to insert the
record into the corresponding subindex.  First, the insert routine
acquires the mutex that is associated with the targeted subindex
(Line 8), and then it fetches the root node of the subindex (Line 9).
Next, it inserts the record into the subindex using the \bptree\
insert algorithm.  The insert algorithm also takes care of correctly
setting the flag of the last leaf node in the case that the last leaf
node needs to be split.  Finally, the associated mutex is released and
the operation is terminated.

\subsection{\dimtree\ search}

Next, we describe search in \dimtree\ for a given range of values,
which is described in Algorithm~2.  The process starts with searching
$T_S$'s inner nodes for the minimum value of the given range
($Range.min$) (Lines 1-9).  At depth $D_I$, the search process detects
which $T_I$'s subindex stores records with value equal to
$Range.min$, which we refer to as $min\_sub\_index$.  At $T_S$'s leaf
nodes, the search process first searches for $Range.min$ (Lines 10-12)
and then linearly traverses through all matching records (Lines
13-15).

\vspace*{0.2cm}
\hspace*{-0.4cm}
\includegraphics[width=0.47\textwidth]{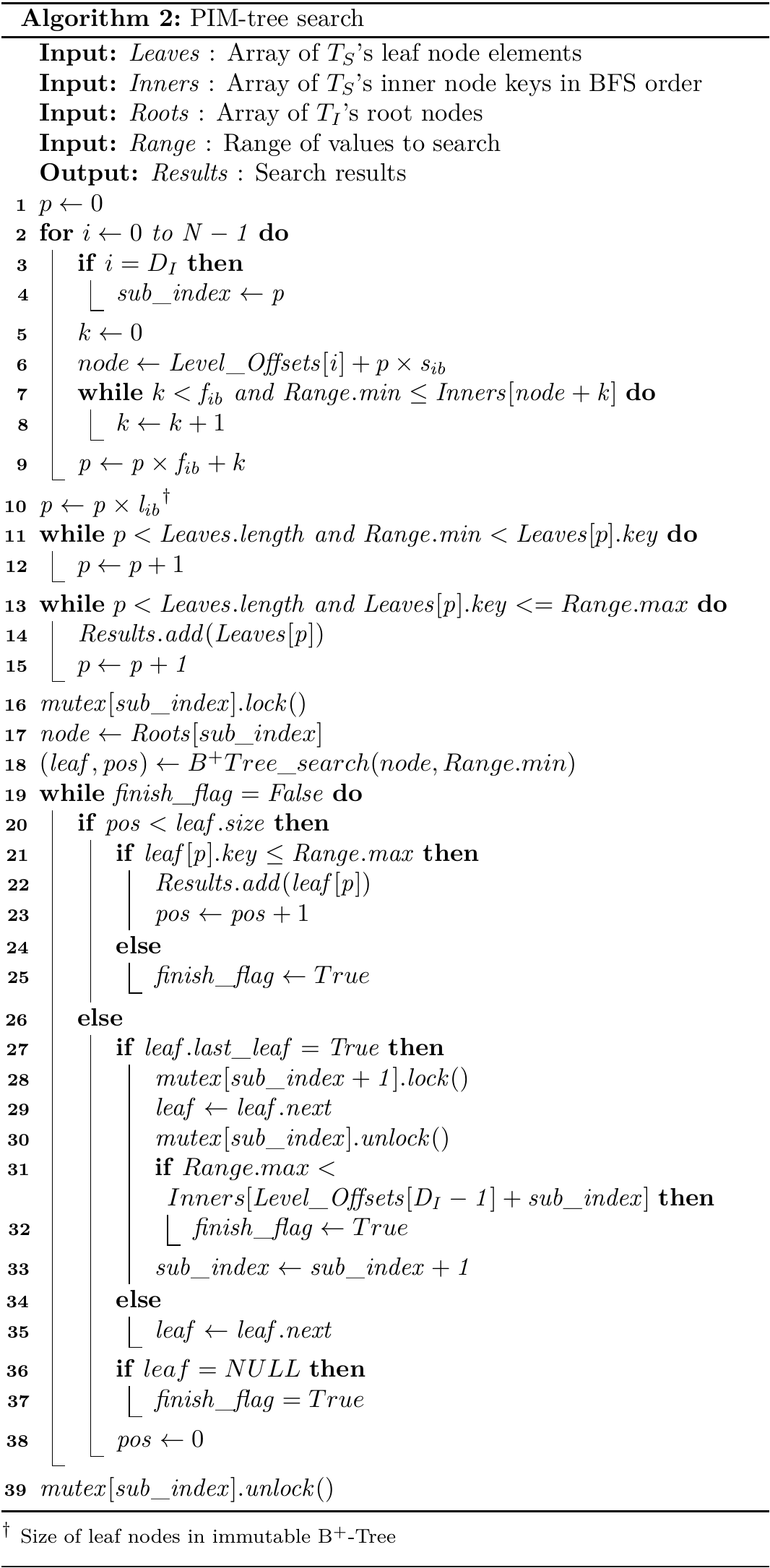}
\vspace*{0.3cm}

The next step is to look for matching tuples in $T_I$ (Lines 16-39).
The search process acquires the mutex associated with
$min\_sub\_index$ (Line 16), and then it fetches $min\_sub$ $\_index$'s
root node (Line 17).  Using the \bptree\ search routine, it locates
the first matching record in ascending order in $min\_$- $sub\_index$
that is equal to or greater than $Range.min$ (Line 18).  The return
values of the \bptree\ search are $leaf$ and $pos$, which indicates that the
desired record is the $pos$-th slot of node $leaf$.  If all records in
$min\_sub\_index$ are less than $Range.min$, then the search result is the
first empty slot of the last leaf node.

The final step is to scan $T_S$'s leaf nodes to find matching tuples
(Lines 19-38).  Whenever the search process switches from a node to
its successor, it checks whether it also switches a new subindex (Line 27).
If so, it acquires the mutex of the successor
subindex before switching to the successor leaf node and releases the
mutex for the current subindex afterwards (Lines 28-30).  Moreover,
in the case that the range of the new subindex does not overlap with
the given range, the search process terminates (Lines 31-32).  This
range checking is helpful to avoid searching through chains of empty
subindexes.  At the end, the process releases the current subindex's
mutex and terminates (Line 39).

\vspace*{-0.cm}

\subsection{Immutable \bptree\ creation}

Nodes in the immutable \bptree\ are arranged in a breadth-first
fashion. The relation among elements is deduced implicitly based on
their position rather than explicitly through pointers or references.
Using this node organization, if node $N$ is the $i_{th}$ node at
level $d$ in breadth-first order, then the $j_{th}$ child of $N$ is
at position $\operatorname{\mathit{Offset}}[d+1] + i \times f_{ib} + j$,
where $\operatorname{\mathit{Offset}}[d]$ is pointing to the beginning
of the $d_{th}$-level and $f_{ib}$ is the fan-out of inner nodes.  Since
it is not required to explicitly store
references to child nodes with this representation, it is possible to achieve a higher fan-out
using the same amount of space compared with the regular \bptree.
Consequently, the depth of the immutable \bptree\ is smaller than the
depth of the regular \bptree\ storing the same number of elements,
which results in better search performance.

\vspace*{0.3cm}
\hspace*{-0.4cm}
\includegraphics[width=0.47\textwidth]{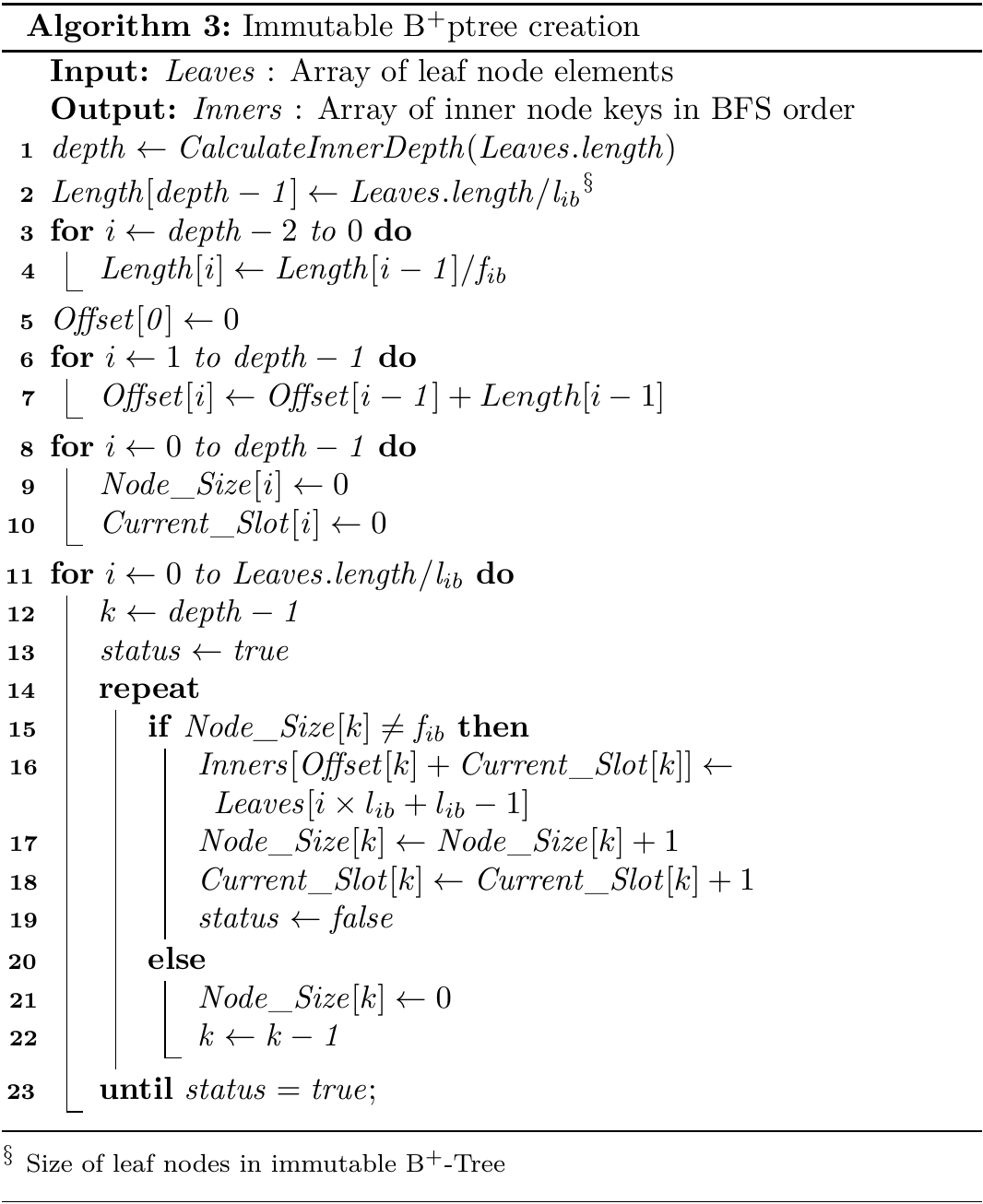}
\vspace*{0.1cm}

\begin{figure}[]
	\centering
	\includepdfcrop{0.7\columnwidth}{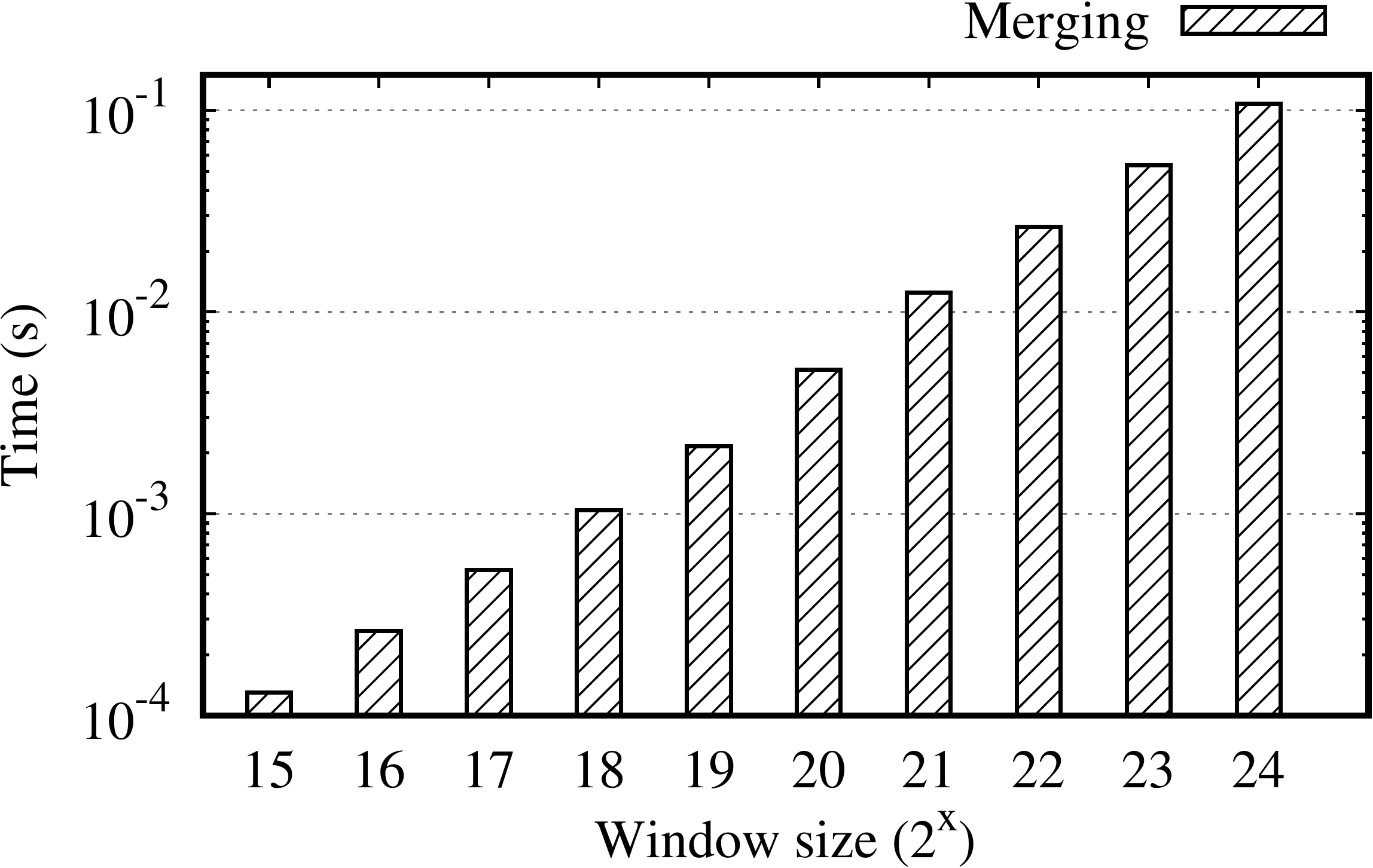}
	\setcounter{figure}{13}
	\vspace{-0.0cm}
	\caption{
		The cost of a \dimtree\ merge operation for various window sizes.}
	\label{plot:cost}
	\vspace{-0.3cm}
\end{figure}

Algorithm~3 describes how inner nodes are created for a given sorted
array of leaf node elements.  First, the algorithm calculates the tree
depth (Line~1) and the number of inner nodes at each depth
(Lines~2-4). In the next step (Lines~5-7), it determines the 
address of each tree level relative to the start of the inner nodes array
($\operatorname{\mathit{Offset}}[d]$).  
Then, the size of the current inner node and the current slot at each level 
are initialized to zero (Lines 8-10).  
For each leaf node, the algorithm starts
from the deepest level of inner nodes ($k =
\mathit{depth-1}$).  At each level, it checks whether there is an empty
slot in the current inner node.  If so, it assigns the largest key of
the current leaf node to the next available slot in the inner node and
increments the inner node size; then, it resumes tree creation by
advancing to the next leaf node.  If there is no empty slot left in
the inner node (Lines~21-22), the algorithm initializes a new node at
the current level (resetting the node size to zero), 
moves to the parent level, and repeats the same
procedure. Considering that $l$ is the total number of elements in 
leaf nodes and $d$ is the depth of the tree, Equation~\ref{equ:immutable} shows the
computational complexity of the immutable \bptree\ creation.

\vspace*{-0.2cm}
\begin{equation} \label{equ:immutable}
\overset{d}{\underset{k=1}{\sum}} k \cdot \dfrac{l}{{f_{ib}}^{k}} = O(l)
\end{equation}
\vspace*{-0.1cm}

Figure~\ref{plot:cost} illustrates the cost of the \dimtree\ merge
operation including merging nonexpired tuples of $T_S$ and $T_I$ into
a single sorted array and creating a new immutable \bptree.  As shown,
the cost of a merge operation increases linearly in the number of
elements in the tree.

\balance

\end{document}